%% file: funDecomp_arXiv_2024-01-14.tex
\documentclass[12pt]{article}

\usepackage[left=1in,right=1in,top=1in,bottom=1in]{geometry}
\usepackage{graphicx}
\usepackage{natbib}
\defcitealias{cea_2015}{DOT, CEA, and DOL, 2015}
\usepackage{setspace}
\usepackage{soul}               
\usepackage{caption}
\usepackage{booktabs}
\usepackage{threeparttable}
\usepackage{siunitx}            

\usepackage{scalerel}
\usepackage{nicefrac}

\usepackage[]{appendix}    

\usepackage[shortlabels]{enumitem}
\setlist[description]{leftmargin=0em,labelindent=\parindent}

\usepackage{amsmath,amsthm}

\theoremstyle{definition}

\newtheorem{result}{Result}

\newtheorem{prop}{Proposition}

\theoremstyle{remark}

\usepackage[T1]{fontenc}

\usepackage[scaled=.85]{beramono}
\usepackage[type1]{cabin}
\usepackage{amsmath,amsthm}
\usepackage[libertine]{newtxmath}

\usepackage[scr=rsfso]{mathalfa}
\usepackage{bm}
\usepackage[semibold]{libertine}

\usepackage{titlesec}
\titleformat{\section}[block]{\bf\Large}{\thesection\quad}{0pt}{}
\titleformat{\subsection}[block]{\bf\large}{\thesubsection\quad}{0pt}{}

\newenvironment{figurenotes}[1][Note]{\begin{minipage}[t]{\linewidth}\footnotesize{\itshape#1: }}{\end{minipage}}

\usepackage[dvipsnames]{xcolor}
\definecolor{csub-blue}{RGB}{0, 53, 148}
\definecolor{csub-blue-hl}{RGB}{0, 53, 148}
\usepackage[unicode,hyperfootnotes=true,linktocpage,pagebackref]{hyperref}
 \hypersetup{%
   colorlinks = true,
   pdfborder = {0 0 0},
   citecolor = csub-blue,
   linkcolor = csub-blue,
   urlcolor  = csub-blue,
 }%
\makeatletter
\def\@fnsymbol#1{\ensuremath{\ifcase#1\or \mathparagraph\or \|\or **\or \dagger\dagger \or \ddagger\ddagger \else\@ctrerr\fi}}
\makeatother

\usepackage{pdfpages}

\title{\textsf{\Large Unemployment Volatility:\\
  When Workers Pay Costs upon Accepting Jobs}}
\author{Rich Ryan\thanks{Email: \href{mailto:richryan@csub.edu}{richryan@csub.edu}. Department of Economics, California State University, Bakersfield,
  9001 Stockdale Highway, Bakersfield, CA 93311, USA.}} 
\date{January 14, 2024}

\newcommand{\estgamma}{0.103}
\newcommand{\gammahi}{1.27}
\newcommand{\etamuhi}{0.518}
\newcommand{\etamulo}{0.452}
\newcommand{\bndhi}{2.214}
\newcommand{\bndlo}{2}
\newcommand{\bndhihi}{11.862}
\newcommand{\bndhilo}{2.006}

\begin{document}
\maketitle

\begin{abstract}
  When a firm hires a worker,  
  adding the new hire to payroll is costly.
  These costs
  reduce the amount of resources that can go to recruiting workers
  and 
  amplify how unemployment responds to changes in productivity.
  Workers also incur up-front costs upon accepting jobs.
  Examples include moving expenses and regulatory fees.
  I establish that workers' costs
  lessen the response of unemployment to productivity changes and
  do not subtract from resources available for recruitment.
  The influence of workers' costs is bounded by properties of a matching function,
  which describes how
  job openings and unemployment produce hires.
  Using data on job finding that are adjusted
  for workers' transitions between employment and unemployment and 
  for how the Job Openings and Labor Turnover Survey records hires,
  I estimate a bound that ascribes limited influence to workers' costs. 
  The results demonstrate that costs paid by workers upon accepting jobs
  affect outcomes in the labor market
  (firms threaten workers with paying the up-front costs again if wage
  negotiations fail),
  but their influence on volatility is less important than firms' costs.
\end{abstract}

\vspace{3em}


\vspace{1em}


\vfill

\clearpage
\pagebreak

\section{Introduction}

Firms recruit workers by 
posting notices to online job boards and
taping help-wanted signs to storefront windows.
When the advertisement is answered and a firm decides to hire a worker,
the firm must pay the cost of adding the worker to payroll.
This one-off cost of creating a job reduces the amount of resources that can be allocated to recruitment.
Recruitment determines how job openings and therefore unemployment
respond
to changes in fundamentals like productivity  \citep{ljungqvist_sargent_2017}.

Imagine that
Acme Corporation wants to hire a worker and Don wants a job.
Acme posts their opening to an online job board,
which advertises the position for a monthly fee.
Acme expects to have trouble filling their position 
in periods when
there are many other firms looking to add workers and few workers looking for jobs.
Meanwhile,
Don searches for work, collects benefits from unemployment insurance, and
enjoys some leisure that in total amounts to $z$.
While scouring online job boards,
Don finds the Acme listing.
He applies, interviews, and accepts the job.
Don uses the technology at Acme to produce $y$.

The match between Don and Acme generates the surplus $y - z$.
But Acme must first pay the fixed cost of adding Don to payroll.
If this cost is properly accounted at $\xi$, 
then only $y - z - \xi$ can be allocated to vacancy creation.
Viewed as a fraction of output, 
potential resources for vacancy creation are increasing in $y$:
the derivative of $\left( y-z-\xi \right) / y$ is positive.
A change in productivity will generate a large increase in this fraction if
$\left( z+\xi \right)/y^{2}$ is large,
which occurs when $z + \xi$ is large.
The presence of costs paid by firms upon hiring workers
means resources allocated to job creation
will respond more to changes in productivity,
which will generate larger unemployment responses.
Importantly,
\citet{ljungqvist_sargent_2017} establish that
Acme's expected cost of posting the vacancy to the online job board is much less important than
the cost of adding Don to payroll.

One-off up-front costs paid by firms offer an answer to a widely acknowledged puzzle:
the failure of the standard search model to match the observed volatility of unemployment
\citep[see, for example,][]{hall_2005,shimer_2005,mortensen_nagypal_2007,costain_reiter_2008,pissarides_2009,gertler_trigari_2009,gomme_lkhagvasuren_2015,kehoe_etal_2023}.

That costs paid by firms 
can explain unemployment volatility
suggests these types of costs are worth exploring.
Surprisingly,
while firms' costs have been partially explored,
costs paid by workers upon taking a job have not,
despite evidence that
the prevalence of these costs has risen.

\begin{description}
\item[Costs paid by workers upon accepting a job.] In short,
  here is the main issue:
  Responses of unemployment to changes in productivity
  depend on resources that can be allocated to vacancy creation.
  These resources are reduced by
  costs paid by firms upon hiring a worker.
  The effect is larger responses of unemployment to changes in productivity.
  Do costs paid by workers upon accepting a job have a similar effect?
  Put another way using the example above,
  if Don has to relocate for the job at Acme,
  do Don's costs subtract from resources available for vacancy creation?
\end{description}

Examples of up-front costs paid by workers include
not only
any costs to relocate for work
but also 
any administrative fees associated with job regulation.
These costs are placed
in an economy that features unemployment.
In the standard Diamond--Mortensen--Pissarides environment
this feature is not without precedent.
\citet{mortensen_nagypal_2007}, for example, consider turnover costs paid by firms.
Turnover costs include the costs of adding a worker to payroll and taxes for firing a worker.
\citet{pissarides_2009} uses the cost a firm pays to add a worker
to establish that
their incorporation into a standard DMP model
implies unemployment responds realistically to observed changes in productivity.

In contrast to costs paid by firms,
costs paid by workers reduce
the response of unemployment to changes in productivity and
there is little scope for them to influence unemployment volatility.
To preview results,
this conclusion is based on two contributions of this paper.

\begin{enumerate}
\item\label{item:contrib:two-factor-decomposition} I 
  generalize \citeauthor{ljungqvist_sargent_2017}'s \citeyearpar{ljungqvist_sargent_2017}
  fundamental decomposition
  to show that
  fixed costs paid by workers are subsumed by a
  factor that is bounded from above by the elasticity of matching with respect to unemployment,
  severely limiting the scope
  for workers' costs to influence volatility.
\item\label{item:contrib:est-match} This would
  be the end of story,
  except that I show this elasticity is not in fact bounded by
  ``a consensus about reasonable parameter values''
  \citep[][2636]{ljungqvist_sargent_2017}.
  To reconcile the consensus,
  I estimate a matching function that does
  not exhibit constant elasticity of matching with respect to unemployment
  (like in the Cobb--Douglas case),
  using data on the labor force
  from the US Bureau of Labor Statistics and
  hires from the
  US Bureau of Labor Statistics' Job Openings and Labor Turnover Survey.
  These data are adjusted for
  time aggregation---workers can continuously find and separate from a job within a month---and
  how the JOLTS program records hires---all hires within a month are reported,
  not only hires that remain employed at the end of the month.
  Once I adjust for these potential sources of bias, I
  estimate a nonlinear matching technology and
  re-establish the ``consensus.''
\end{enumerate}

The result bolsters
the idea that \citeauthor{ljungqvist_sargent_2017}'s \citeyearpar{ljungqvist_sargent_2017}
fundamental surplus
refers to resources \textit{available to firms} that
the invisible hand can allocate to vacancy creation.
Fixed costs of job creation paid by workers do not reduce these resources.

Of course
up-front costs paid by workers affect labor-market outcomes.
According to the traditional account,
occupational regulation boosts workers' wages by 
restricting the supply of workers and increasing demand through higher-quality output.
In the search environment, however,
the firm--worker pair already enjoys a bilateral monopoly over
the surplus they generate.
The presence of fixed costs paid by workers
allows a firm to threaten a worker with
paying the fixed costs again if the wage negotiation fails.
As a consequence,
the negotiated wage is lower.
This makes posting a vacancy more valuable,
which lowers equilibrium unemployment.
While wages are often observed to be higher in regulated occupations (think doctors and lawyers),
there are few stories of certified nursing assistants getting rich.
In the \textit{Occupational Outlook Handbook}, for example,
the BLS reports that
``the median annual wage for nursing assistants was \$35,760 in May 2022,''
which is \$17.18 per hour.\footnote{This information is available
  in the \textit{Occupational Outlook Handbook} at
  \url{https://www.bls.gov/ooh/healthcare/nursing-assistants.htm}.
Accessed January 12, 2024.}
Going beyond the scope of this paper,
firms' gains from workers' costs could partly explain the 
increased prevalence of such costs.

The remainder of the paper is organized as follows.
Section \ref{sec:motivation-of-job-creation-costs} provides motivating examples of job-creation costs paid by workers.
Section \ref{sec:model-dmp-envir} describes the economic environment and
section \ref{sec:text:fund-decomp} presents the two-factor fundamental decomposition
of the elasticity of labor-market tightness with respect to productivity.
Because unemployment is a fast-moving state variable,
a good approximation of model dynamics is an analysis of steady states indexed by productivity \citep{hall_2005restat,hall_2005}.
For certain productivity levels 
the consensus that bounds the influence of job-creation costs paid by workers falls apart.
Section \ref{sec:calibration} presents a calibration that restores the consensus.
I calibrate the model to match the labor-market evidence in \citet{pissarides_2009}.
I then estimate a matching function that does not exhibit constant elasticity of matching with respect to unemployment.
The elasticities that vary by the state of the business cycle are
close to the single number provided by the Cobb--Douglas evidence.
Section \ref{sec:properties-model} presents properties of the model,
including the insight that workers' costs can reduce wages and unemployment volatility.
The model's properties are discussed within the context of the literature in section \ref{sec:discussion},
which includes a discussion of outstanding questions.
Section \ref{sec:conclusion} concludes.

\section{Two Examples of Job-Creation Costs Paid by Workers}
\label{sec:motivation-of-job-creation-costs}

When a worker accepts a job they may be required to pay fees associated with regulation.
For example,
a worker may have to pay administrative fees to a government agency
to file their name, address, and qualifications before starting work
\citep{kleiner_krueger_2013}. 
Administrative fees are often required
when a government agency maintains a register.
Variants of ``Registered Dietitian'' can be used
by workers meeting certain requirements in California, for example,
but anyone can provide nutritional advice.\footnote{The requirements
  to be a Registered Dietitian in California can be found
  on websites maintained by the
  Department of Nutritional Sciences \& Toxicology at
  the University of California in Berkeley and
  the Department of Family \& Consumer Sciences at
  California State University, Sacramento:
  \url{https://nst.berkeley.edu/mnsd/how-to-become-a-registered-dietitian-nutritionist}
  and \url{https://www.csus.edu/college/social-sciences-interdisciplinary-studies/family-consumer-sciences/nutrition/becoming-rdn.html}.
Accessed January 12, 2024.}
Registration may also require a worker to post a bond in order to practice \citep{kleiner_2006}.
A related example is testing fees.
Testing fees are sometimes required when
the government or a nonprofit agency issues a certificate
to workers who demonstrate skill or knowledge of some tasks.
Registration and certification 
fall under the rubric of occupational regulation
\citep{kleiner_2000}.

In addition to these two forms of regulation,
``the toughest form of regulation is licensure'' \citep[][S175]{kleiner_krueger_2013}.
A licensing policy means a worker cannot legally work in an occupation
unless they meet some standard.
Recent data from the US Census Bureau's Survey of Income and Program Participation
indicate that
one in four adults aged 18 through 64 had attained a license or certificate
\citep{gittleman_klee_kleiner_2018}.
In an analysis of 102 licensed occupations that
pay below the average income in the United States,
\citet{knepper_etal_2022} document
that
a worker can expect to pay \$$295$ in licensing fees.

A few notable facts accompany this statistic on workers' costs. 
First, the amount does not include lost wages from time spent
earning a degree or accumulating experience.
Second,
``licensing burdens often bear little relationship to
public health or safety---the purported rationale for much
licensing'' \citep[][8]{kleiner_vorotnikov_2018}.
For example,
only 12 percent of the 102 occupations analyzed by \citet[][37]{knepper_etal_2022}
are licensed universally across states,
``which means workers are safely practicing them in at least one state---and often
many more than one---without a government permission slip.''
Third,
the prevalence of occupational licensing is on the rise
(\citeauthor{kleiner_krueger_2010}, \citeyear{kleiner_krueger_2010}, \citeyear{kleiner_krueger_2013};
\citetalias{cea_2015};
\citeauthor{furman_giuliano_2016}, \citeyear{furman_giuliano_2016}).\footnote{Additional 
  work on occupational licensing includes
  \citet{kleiner_todd_2009},
  \citet{kleiner_vorotnikov_2017},
  \citet{johnson_kleiner_2020}, and
  \citet{kleiner_timmons_2020}.}

A worker may be required to purchase parts of their uniform.  
For example, a worker may be asked to wear steel-toed boots.
If the employer permits them to be worn off the job-site,
then the worker may be asked to purchase the boots out of their own pocket.\footnote{The
  policy on steel-toed boots comes from the
  Occupational Safety and Health Administration, OSHA,
  Occupational Safety and Health Standards,
  Standard No.~1910.132(h)(2),
  United States Department of Labor,
\url{https://www.osha.gov/laws-regs/regulations/standardnumber/1910/1910.132}.}

Regulatory fees are one example of costs borne by workers when a job is created.
Another example is the cost of relocation:
once a job is accepted, a worker may have to move to begin work.
A reasonable inference from the documented fall in worker mobility
is a rise in relocation costs.
Such a straightforward interpretation, though, may be incomplete.
\citet{amior_2024} provides evidence that
workers move in exchange for large salaries that justify the cost of moving.
Recent perspectives on mobility are provided by
\citet{kennan_walker_2011},
\citet{molloy_etal_2016},
\citet{notowidigdo_2020},
\citet{schmutz_sidibe_vidal-naquet_2021}, and
\citet{zabek_2024}.

Both examples are multifaceted.
I take a straightforward approach to
analyzing one-off job-creation costs paid by workers.
The representative experience for a worker is
payment of a one-off fixed amount upon accepting a job.

\section{Model: A DMP Environment with One-Off Costs When a Job Is Created}
\label{sec:model-dmp-envir}

The environment shares the features of a conventional DMP model,
including 
linear utility,
workers with identical skills,
random search,
exogenous separations,
wages determined as the outcome of Nash bargaining, and
competitive job creation that drives the value of posting a vacancy to zero.
In the model,
firms' costs to match with a worker include recruitment costs,
which are paid each period an ad for a vacancy is posted. 
When there are fewer unemployed workers, 
a vacancy will take longer to fill,
making the cost of recruitment proportional
to the ratio of vacancies to unemployment.
This ratio is commonly called labor-market tightness.\footnote{\citet{pissarides_2000} and
  \citet{petrosky-nadeau_wasmer_2017} provide excellent textbook treatments.
  Essential contributions in this area are
  \citet{pissarides_1985} and \citet{mortensen_pissarides_1994}.
  \citet{diamond_1982,diamond_1982b} made fundamental earlier contributions.
  \citet{nobel_2010} provides further background.}

Job creation may also involve fixed matching costs paid by firms. 
These costs include ``training, negotiation, and one-off administrative costs of adding a worker on the payroll'' \citep[][1363]{pissarides_2009}.
These costs are often ignored but once they are added to a standard DMP environment,
\citet{pissarides_2009} demonstrates how their addition can generate unemployment fluctuations in response to changes in productivity
that match the magnitudes observed in data.
Some evidence, though, suggests that workers may bear a significant fraction of one-off administrative costs,
including
relocation costs,
training costs,
tuition, foregone wages, and testing fees. 

Motivated by these features,
I add fixed job-creation costs paid by workers to a standard DMP model.
In addition to costs paid by workers,
firms are required to pay turnover costs: hiring and firing costs.
Hiring costs are emphasized by \citet{pissarides_2009} and
firing costs are studied by \citet{mortensen_nagypal_2007}.\footnote{\citet{mortensen_nagypal_2007} also consider hiring costs paid by firms.
  They present a version of the fundamental decomposition
  emphasized by \citet{ljungqvist_sargent_2017}.

  \citet{silva_toledo_2009} document costs of hiring a worker and
  present a dynamic, stochastic model
  where new workers are less productive than experienced employees,
  which reflects the costs of hiring a worker
  who does not know how a specific firm operates.
  The costs documented by \citet{silva_toledo_2009}
  are beyond the scope of the model presented here.
  They rely on numerical work to show how their model fits the facts and the data.}

A synthesis of these features is presented by \citet{ljungqvist_sargent_2017}.
And I build directly upon their work.
In section \ref{sec:text:fund-decomp},
I establish that their fundamental decomposition,
which reduces the elasticity of tightness with respect to productivity into two terms,
holds not only
for a Cobb--Douglas matching function but also
for any reasonable matching function.
One of the two terms subsumes costs paid by workers upon taking a job and
this term is bounded by properties of a function that determines how many jobs are created
when there are $v$ vacancies and $u$ unemployed workers.
When the matching function is not Cobb--Douglas,
there is no consensus on its properties.
I therefore estimate in section \ref{sec:calibration}
another matching function using data that are adjusted for
worker flows and
how the JOLTS program records hires.
Properties of this function
determine the maximal influence that workers' costs can have on
how unemployment responds to changes in productivity.

But first I describe the model environment. 

\subsection{Description of the Model Environment}

Time is discrete.
A continuum of identical workers populate the economy.
Workers are risk neutral and live forever.
They are endowed with an indivisible unit of labor and
they discount the future using the discount factor $\beta = \left( 1+r \right)^{-1}$.
They derive utility by consuming a single homogeneous good.

Workers are either unemployed or employed.
Employed workers produce the consumption good using a technology owned by firms. 
Unemployed workers meet recruiting firms randomly.
The matching process is summarized by a matching function $M \left( u,v \right)$,
where $u$ is the number of unemployed workers and $v$ is the number of posted vacancies.
The value $u$ can be interpreted as the unemployment rate because the labor force is normalized to one.

I assume that $M$
increases in both its arguments and
exhibits constant returns to scale in $\left( u,v \right)$.
The function $M$
determines the rate at which each worker meets a firm.
This rate is identical across workers because
it does not depend on characteristics of workers or firms.
The ratio of vacancies to unemployment is denoted by $\theta \equiv v / u$ and
called tightness.
The rate at which a recruiting firm meets an unemployed worker is
$q \left( \theta \right) \equiv M \left( u,v \right) / v$.
The rate at which an unemployed worker meets a recruiting firm is
$\theta q \left( \theta \right) \equiv M \left( u,v \right) / u$.
I assume that $q$ satisfies standard regularity assumptions that ensure that
$q \left( \theta \right)$ is decreasing in $\theta$,
$\theta q \left( \theta \right)$ is increasing in $\theta$,
$\lim_{\theta \rightarrow \infty} q \left( \theta \right) =0$,
$\lim_{\theta \rightarrow 0} q \left( \theta \right) = 1$,
$\lim_{\theta \rightarrow \infty} \theta q \left( \theta \right) = 0$, and
$\lim_{\theta \rightarrow 0} \theta q \left( \theta \right) = 1$.
I refer to
$q \left( \theta \right)$ as the job-filling rate and
$f \left( \theta \right) \equiv \theta q \left( \theta \right)$ as the job-finding rate. 

In general, $\theta = \theta \left( t \right)$ is time-varying as
both the number of unemployed workers and posted vacancies can vary over time.
But I
focus on steady-state equilibria and drop explicit reference to time.

\subsection{Job Creation by Firms}

The value of a productive firm satisfies the Bellman equation
\begin{equation}
J = y - w + \beta \left[s\left(V-\tau\right)+\left(1-s\right)J\right].\label{eq:text:J}
\end{equation}
The value of a productive firm equals flow output, $y$, less the flow wage payment, $w$, plus the continuation value.
The continuation value is
the value of a vacancy in the event of a separation,
which occurs with probability $s$,
in which case the firm must pay the layoff tax $\tau$, plus
the value of continued production if a separation does not occur.
The continuation value needs to be discounted.
The intuition for the form of this equation is standard.

The value of a vacancy is
\begin{equation}
V = - c + \beta\left[q\left(\theta\right)\left(J-h\right)+\left(1-q\left(\theta\right)\right)V\right].\label{eq:text:V}
\end{equation}
A recruiting firm becomes productive by posting a vacancy,
which entails a flow cost of $c$.
The following period the vacancy is
unfilled with probability  $1 - q \left( \theta \right)$ and
filled   with probability      $q \left( \theta \right)$.
Upon a match, the firm must pay the fixed cost $h$,
which summarizes one-off administrative costs.

Competitive efforts by the large measure of firms drives
the value of a vacancy to zero.
Using the competitive assumption in equation \eqref{eq:text:V} implies
\begin{equation}
J = \frac{c}{\beta q\left(\theta\right)}+h.\label{eq:text:value-J}
\end{equation}
The value of a productive job under competition is driven to the expected
recruitment cost, $c / \beta / q \left( \theta \right)$,
plus the fixed cost of job creation paid by firms.

Substituting the expression for $J$ in \eqref{eq:text:value-J} into \eqref{eq:text:J}
implies a job-creation condition for firms equal to
\begin{equation}
\label{eq:text:J-infinite}
  J = \frac{1+r}{r+s}\left(y-w-\beta s\tau\right)
  = \sum_{j=0}^{\infty}\beta^{j}\left(1-s\right)^{j}\left(y-w-\beta s\tau\right)=\frac{c}{\beta q\left(\theta\right)}+h.
\end{equation}
The value of a productive firm equals 
the present value of flow profit, $y-w$
less the expected present value of the layoff tax faced by the firm, $\beta s \tau$.
Discounting includes
the discount factor and
the job-retention rate, $1-s$.
The amount $\beta s \tau$ is subtracted from flow profits because
``the invisible hand can never allocate'' these resources to vacancy creation \citep[][2642]{ljungqvist_sargent_2017}.
The right side shows that a firm's expected gain equals the expected cost of job creation.

There are two components to this cost.
The first is a proportional cost $c$ that rises in expectation with how long the
firm expects the vacancy to be posted before it is filled.
The second is the fixed cost emphasized by \citet{pissarides_2009}.

In summary,
equations \eqref{eq:text:J} and \eqref{eq:text:value-J} along with the zero-profit condition imply
\begin{equation}
\label{eq:text:LS:5}
w=y-\beta s\tau-\frac{r+s}{q\left(\theta\right)}c-\frac{r+s}{1+r}h.
\end{equation}
In $\theta$--$w$ space, equation \eqref{eq:text:LS:5} is downward sloping \citep[][chapter 1]{pissarides_2000}.
A higher wage makes posting vacancies less attractive for a firm.
The relationship represents job creation by firms.
A higher cost of job creation, $h$, shifts this job-creation condition downward.

Appendix \ref{sec:app:deriv-dmp-envir} provides a derivation of these results.

\subsection{Job Creation by Workers}

The value of employment for a worker satisfies the Bellman equation
\begin{equation}
E = w+\beta\left[sU+\left(1-s\right)E\right].\label{eq:E}
\end{equation}
The value of employment equals the flow wage, $w$, plus the discounted continuation value,
which equals
the value of unemployment, $U$, if a separation occurs, which happens with probability $s$, and
the value of employment, $E$,   if the job remains productive, which happens with probability $1-s$.
The value of unemployment for a worker satisfies the Bellman equation
\begin{equation}
U=z+\beta\left[f\left(\theta\right)\left(E-\ell\right)+\left(1-f\left(\theta\right)\right)U\right].\label{eq:U}
\end{equation}
The value of unemployment equals the flow value of nonwork, $z$, plus the discounted continuation value,
which equals
the value of employment,   $E$, if a job is found,     which occurs with probability $f \left( \theta \right)$, and
the value of unemployment, $U$, if a job is not found, which occurs with probability $1 - f \left( \theta \right)$.

Equations \eqref{eq:E} and \eqref{eq:U} can be combined to
form an expression similar to the one in \eqref{eq:text:J-infinite}:
\begin{equation}
\label{eq:reservation-wage}
\sum_{j=0}^{\infty}\left(1-s\right)^{j}\beta^{j}w_{R} 
= z
+ \beta\left[f\left(\theta\right)\left(\sum_{j=0}^{\infty}\left(1-s\right)^{j}\beta^{j}w - \ell\right)
  + \left( 1 - f\left(\theta\right) \right)  U\right],
\end{equation}
where $w_{R} = r U / \left( 1+r \right)$ and $U = W\left(w_{R}\right)$,
which is established in result \ref{result:reservation-wage} in the appendix.
The expression for the reservation wage indicates that
a worker is indifferent between earning the expected value of taking a job that pays the reservation wage
starting from the current period and
receiving $z$ in unemployment and then, the following period, accepting the expected value of a job,
which occurs with probability $f \left( \theta \right)$, and the present value
of unemployment with probability $1 - f\left(\theta\right)$.
Upon taking the job, the worker pays the fixed cost $\ell$.

By convention, 
wages are determined by the outcome of asymmetric Nash bargaining.
The present value of match surplus is the gain from a productive match,
$S = \left(J-V\right)+\left(E-U\right)$.
The outcome of bargaining specifies
\begin{equation}
E-U=\phi S \; \text{ and } \; J=\left(1-\phi\right)S,\label{eq:8}
\end{equation}
where
$\phi\in\left[0,1\right)$ measures workers' bargaining power.

The expression for surplus along with the Nash sharing rule in \eqref{eq:8} and
expressions for $J$ and $E$ in \eqref{eq:text:J} and \eqref{eq:E} can be rearranged to
express the wage as
\begin{equation}
\label{eq:text:LS:9}
w = \left(1-\phi\right)w_{R}+\phi\left(y-\beta s\tau\right).
\end{equation}
A worker earns fraction $1-\phi$ of their reservation wage plus fraction $\phi$ of flow profit net of the appropriate layoff-tax payment.
The expression for wages in \eqref{eq:text:LS:9}
is analogous to equation (9) in \citet[][2634]{ljungqvist_sargent_2017} and
is derived in appendix \ref{sec:app:deriv-dmp-envir}.

Equation \eqref{eq:text:LS:9} contains the value of unemployment, $U$, through $w_{R}$.
Appendix \ref{sec:app:deriv-dmp-envir} shows how equation \eqref{eq:U} and the outcome of Nash bargaining in \eqref{eq:8}
can be combined to solve for $U$.
Using this result in \eqref{eq:text:LS:9} yields
\begin{equation}
\label{eq:text:LS:11}
w = z + \phi\left(y-z-\beta s\tau+\theta c\right) + \beta f \left(\theta\right) \left[\phi h-\left(1-\phi\right)\ell\right].
\end{equation}
The worker earns the value of nonemployment plus their share of
the flow gain generated by a productive match plus $c \theta = c \times v / u$,
which is the total cost of hiring, $c v$, divided by the number of unemployed workers, or
the average hiring cost per unemployed worker \citep[][17]{pissarides_2000}.

The fixed cost $h$ increases wages with coefficient $\phi \beta f\left(\theta\right)$.
Because if the negotiation fails,
then the firm has to pay $h$ when it meets another worker.
This event
takes place the following period with probability $f\left(\theta\right)$ and
needs to be discounted.
By staying in the match, the worker saves the firm an expected cost $\beta f \left(\theta\right) h$ and
the worker captures fraction $\phi$ of this amount through bargaining.
In contrast,
the fixed cost $\ell$ decreases wages with coefficient $(1-\phi) \beta f\left(\theta\right)$.
Because if the negotiation fails,
then the worker has to pay $\ell$ when it meets another firm.
This event takes place the following period with probability $f\left(\theta\right)$ and
needs to be discounted.
By staying in the match, the firm saves the worker an expected cost $\beta f \left( \theta \right) \ell$ and
the firm captures fraction $1-\phi$ of this amount through bargaining.\footnote{I have adopted this discussion from \citet[][1364]{pissarides_2009},
who provides a description of an analogous wage equation in a model set in continuous time,
where there are fixed costs paid by firms upon a match but not costs to workers.
Appendix \ref{sec:app:deriv-dmp-envir} derives the expression for the wage in equation \eqref{eq:LS:11}.}

Equation \eqref{eq:text:LS:11} is the condition for job creation by workers.
If $h = \ell = 0$,
then $w = z + \phi\left(y-z-\beta s\tau + \theta c\right)$.
In $\theta$--$w$ space the condition is increasing in tightness.
A tighter labor market raises the cost of hiring a worker,
part of which the worker captures through the outcome of bargaining.
When $h$ and $\ell$ are both positive,
for the condition to be upward sloping in $\theta$--$w$ space,
parameters must be such that $\phi c + \beta f^{\prime} \left(\theta\right) \left[\phi h-\left(1-\phi\right)\ell\right] > 0$.
As shown below,
this condition will guarantee a unique equilibrium.

\subsection{Equilibrium}

A steady-state equilibrium is a list of values $\left\langle u, \theta, w \right\rangle$ that satisfy
the Bellman equations \eqref{eq:text:J}, \eqref{eq:text:V}, \eqref{eq:E}, \eqref{eq:U} and
the sharing rule in \eqref{eq:8}
along with the free-entry condition that requires $V = 0$ and
the steady-state unemployment rate.
The unemployment rate remains constant when
the number of workers who separate from jobs, $s \left( 1-u \right)$, equals
the number of unemployed workers who find employment, $\theta q \left( \theta \right) u$.
The steady-state value of $u$ is therefore $s / \left( s + f \left( \theta \right) \right)$.
In $u$--$v$ space, there is a negative relationship between
these two variables:
``when there are more vacancies, unemployment is lower because the unemployed find jobs more easily'' \citep[][20]{pissarides_2000}.
The model generates a Beveridge curve.\footnote{\citet{elsby_michaels_ratner_2015}
  provide an overview.}

The equilibrium value of $\theta$ 
is jointly determined by the two expressions for the wage rate in \eqref{eq:text:LS:5} and \eqref{eq:text:LS:11}:
\begin{equation}
  \label{eq:text:LS:12}
y-z-\beta s\tau-\frac{\beta\left(r+s\right)h}{1-\phi}=\frac{r+s+\phi\theta q\left(\theta\right)}{\left(1-\phi\right)q\left(\theta\right)}c+\frac{\theta q\left(\theta\right)}{1+r}\left(\frac{\phi}{1-\phi}h-\ell\right).
\end{equation}
Conditions for existence and uniqueness of an equilibrium are summarized in proposition \ref{prop:unique-theta}.

\begin{prop}[Existence and uniqueness of $\theta$]
  \label{prop:unique-theta}
  Assume $y>z$, which says that workers produce more of the homogeneous consumption good at work than at home, and
  assume that the initial vacancy yields a positive value,
  a condition that can be stated as $\left(1-\phi\right)\left(y - z - \beta s \tau \right)/\left(r+s\right) > c + \beta h$.
  In the canonical DMP search model,
  which features
  random search, linear utility, workers with identical capacities for work,
  exogenous separations, no disturbances in aggregate productivity, and
  fixed costs paid by workers and firms when a job is created,
  there exists a $\theta \in \left( 0, \bar{\theta} \right)$ that solves the steady-state relationship in \eqref{eq:text:LS:12},
  where  
\begin{equation}
\label{eq:theta-hi}
\bar{\theta} \equiv \frac{1-\phi}{c\phi} \left[y-z-\beta s\tau-\frac{\beta\left(r+s\right)}{1-\phi}h+\beta\ell\right] > 0.
\end{equation}
If the workers' condition for job creation slopes upwards,
then the equilibrium is unique.

The steady-state level of tightness, $\theta$, determines steady-state unemployment,
defined as the measure of
job creation, $f \left( \theta \right) u$, equaling
job destruction, $s \left( 1-u \right)$,
which implies $u = s / \left( s + f \left( \theta \right) \right)$.
\end{prop}

A detailed proof that uses insights from \citet{ryan_2023arxiv} is found in appendix \ref{sec:proof-prop-unique}.

For the purposes of this paper,
the most important part of proposition \ref{prop:unique-theta} are 
conditions that permit an investigation into how tightness responds to productivity.
In an economy with some unemployment,
which is guaranteed with exogenous separations,
the value of an initial vacancy can be thought of as
the value of $\lim_{\theta \rightarrow 0} V$.
In this scenario,
a vacancy is immediately filled and 
$-c - \beta h + \beta \left( 1-\phi \right) \left( y - z - \beta s \tau \right) / \left[ 1-\beta \left( 1-s \right) \right] > 0$,
which is the condition given in proposition \ref{prop:unique-theta}.

Existence can be expressed as $\mathcal{T} \left( \theta \right) = 0$, where
\begin{equation}
\label{eq:text:def:T}
\mathcal{T}\left(x\right) \equiv
y-z-\beta s\tau - \frac{\beta\left(r+s\right)}{1-\phi}h
- \frac{c}{1-\phi} \left[\frac{r+s+\phi xq\left(x\right)}{q\left(x\right)} +
  \frac{\beta xq\left(x\right)}{c} \left( \phi h-\left(1-\phi\right)\ell \right)\right].
\end{equation}
Then
\begin{align*}
\mathcal{T} \left( 0 \right) = y-z-\beta s\tau-\frac{\beta\left(r+s\right)}{1-\phi}h-\frac{c}{1-\phi}\left(r+s\right)>0,
\end{align*}
where the inequality uses the positive value of posting an initial vacancy.

A firm that posts the initial vacancy pays $c$ in the current period.
The vacancy is filled by the following period as there is a sole vacancy and many unemployed workers.
Upon the match, the following period the firm pays the fixed cost $h$, which must be discounted to the present.
These two costs must be less than the expected value of a productive match:
\begin{equation}
\label{eq:text:first-vacancy-explained}
c + \beta h < \beta \sum\limits_{j=0}^{\infty} \beta^{j} \left( 1-s \right)^{j} \left( 1-\phi \right) \left( y - z - \beta s \tau \right).
\end{equation}
To understand the expected value of a productive match, note that
the firm--worker pair begins production the following period,
which requires discounting.
The pair generates a net flow equal to flow surplus, $y-z$, less the appropriate resources deducted for the layoff tax, $\beta s \tau$.
The firm's bargaining position allows it to collect fraction $1-\phi$ of the flow.
Net flow is discounted by both the discount factor and the job-retention rate, $1-s$.
The condition in \eqref{eq:text:first-vacancy-explained} is a rearrangement of the condition $\mathcal{T} \left( 0 \right) > 0$.

In addition, using the definition of $\bar{\theta}$ in proposition \ref{prop:unique-theta},
it is straightforward to establish that $\mathcal{T} \left( \bar{\theta} \right) > 0$.
Because $\mathcal{T}$ is the composition of continuous functions,
it is continuous on $\left[0,\bar{\theta}\right]$.
Thus, by the intermediate-value theorem,
there exists $\theta\in \left( 0,\bar{\theta} \right)$ such that
$\mathcal{T}\left(\theta\right)=0$, which establishes existence.

If $\mathcal{T}$ is everywhere decreasing on $\left( 0,\bar{\theta} \right)$,
then the equilibrium will be unique.
The derivative of $\mathcal{T}$ is
\begin{equation}
  \label{eq:text:T:prime}
\mathcal{T}^{\prime}\left(x\right) \equiv \frac{c\left(r+s\right)}{\left(1-\phi\right)\left[q\left(x\right)\right]^{2}}q^{\prime}\left(x\right)-\frac{c\phi}{1-\phi}-\beta\frac{f^{\prime}\left(x\right)}{1-\phi}\left[\phi h-\left(1-\phi\right)\ell\right].
\end{equation}
As $q^{\prime} < 0$, the first term is negative.
The remaining terms are also negative,
for example,
if the condition for job creation by workers in \eqref{eq:text:LS:11} is increasing in tightness in $\theta$--$w$ space.
Then only one $\theta$ satisfies $\mathcal{T} \left( \theta \right) = 0$.
But a larger set of parameters implies a unique equilibrium.

The existence and uniqueness of the equilibrium permits the computation of a comparative static,
which is the subject of section \ref{sec:text:fund-decomp}. 

\section{The Fundamental Decomposition}
\label{sec:text:fund-decomp}

While policymakers' interest lies in how unemployment responds to productivity,
$u$ is not a fundamental,
as \citet{ljungqvist_sargent_2017} point out.
It makes sense to understand how $\theta$, a fundamental, responds to changes in $y$.
A related decomposition of the response of $\theta$ to changes in $y$ is provided by
\citet[][332]{mortensen_nagypal_2007}.

Paralleling the steps in the cited research for determining
the implied volatility of the conventional model,
I arrive at the following proposition.
The proposition provides a minor but important generalization of some results in \citet{ljungqvist_sargent_2017},
who assume the matching function is Cobb--Douglas.

\begin{prop}[Fundamental decomposition]
  \label{prop:fundamental-decomposition}
  In the economic environment described in proposition \ref{prop:unique-theta},
  the elasticity of market tightness with respect to productivity, $\eta_{\theta,y}$, can be decomposed into
  two multiplicative factors
  \begin{equation}
    \label{eq:elasticity-tight-y}
      \eta_{\theta,y} = \Upsilon \frac{y}{y-z-\beta s\tau-\frac{\beta\left(r+s\right)h}{1-\phi}},
  \end{equation}
  where the second factor is the inverse of the fundamental surplus fraction and
  the first factor is 
\begin{equation}
\label{eq:text:upsi}
\Upsilon \equiv \frac{r+s+\theta q\left(\theta\right)\left[\phi+\beta q\left(\theta\right)\left(\frac{\phi h-\left(1-\phi\right)\ell}{c}\right)\right]}{\left(r+s\right)\eta_{M,u}+\theta q\left(\theta\right)\left[\phi+\beta\left(1-\eta_{M,u}\right)q\left(\theta\right)\left(\frac{\phi h-\left(1-\phi\right)\ell}{c}\right)\right]}.
\end{equation}
The factor $\Upsilon$ is bounded by
\begin{equation}
\label{eq:Upsilon-bound}
0 < \Upsilon<\max\left\{ \frac{1}{\eta_{M,u}}, \frac{1}{1-\eta_{M,u}} \right\},
\end{equation}
where $\eta_{M,u} \in \left( 0,1 \right)$ is the elasticity of matching with respect to unemployment.
The elasticity of unemployment with respect to $y$ is $\eta_{u,y} = - \left( 1-u \right) \left( 1-\eta_{M,u} \right) \eta_{\theta,y}$.
\end{prop}

Both parts of proposition \ref{prop:fundamental-decomposition} are important.
The decomposition implies that
fixed costs of job creation paid by workers affect volatility through $\Upsilon$.
A main message of \citet{ljungqvist_sargent_2017}, though, is that the upper bound for $\Upsilon$ is small.
For example, many parameterizations posit that the matching function is Cobb--Douglas,
which exhibits constant elasticity of matching with respect to unemployment.
The elasticity is commonly chosen to be around $0.5$,
implying an upper bound for $\Upsilon$ of $2$.
This choice is
consistent with empirical evidence in \citet{petrongolo_pissarides_2001} and
consonant with efficiency
if the bargaining parameter $\phi$ also equals $0.5$,
so that \citeauthor{hosios_1990}'s \citeyearpar{hosios_1990} condition holds.
Notwithstanding some empirical evidence covered in \citet{jaeger_etal_2020},
this choice of $\phi$ implies workers and firms have equal bargaining power,
a reasonable assumption given limited data to assess.
But if $\Upsilon$ is bounded in magnitude by $2$, then $\ell$ cannot matter much for labor-market dynamics.

In general, however, unlike in the Cobb--Douglas parameterization,
$\eta_{M,u}$ varies with tightness.
Another matching technology
adopted in the literature 
implies the influence of $\Upsilon$ on volatility may be meaningful \citep{ryan_2023arxiv}.
For values of tightness observed in US data since December 2000,
the bound reached $\bndhihi$.
This leaves open the question of whether $\ell$ affects volatility.

Section \ref{sec:calibration} takes up this question by estimating the alternative, nonlinear matching function.
The estimation 
uses readily available data that must be corrected for
the fact that workers can transition between unemployment and employment continuously and
the way hires are recorded. 

\section{Calibration}
\label{sec:calibration}

To gain insights into how costs of job creation affect labor-market volatility,
I explore how the unemployment rate responds to changes in productivity.
As a shortcut for analyzing model dynamics, I compare steady states,
appealing to \citet[][39--40]{shimer_2005},
who ``documented that comparisons of steady states described by [the expression for $\eta_{\theta,y}$]
provide a good approximation to average outcomes from simulations of an economy subject to aggregate productivity shocks'' \citep[][2636]{ljungqvist_sargent_2017}.

Steady-state comparisons require assigning values to parameters.
All except one are largely agreed upon by convention.
The exception is the parameter that determines the elasticity of matching with respect to unemployment, $\eta_{M,u}$.
Proposition \ref{prop:fundamental-decomposition} establishes that $\eta_{M,u}$ bounds $\Upsilon$,
one of the two multiplicative factors that determine volatility.

A matching function that allows the elasticity of matching to vary with tightness
is suggested by \citet{den-haan_ramey_watson_2000}:
\begin{equation}
\label{eq:match-fnc}
M \left( u,v \right) = \mu \frac{uv}{\left( u^{\gamma} + v^{\gamma} \right)^{1/\gamma}}.
\end{equation}
One motivation for this form is random contact among all agents.
Imagine that an unemployed worker contacts all agents, including firms posting vacancies, randomly.
The probability the other agent is a recruiting firm is $v / \left( u + v \right)$,
implying matches total $uv / \left( u+v \right)$.
The nonlinear term can capture externalities from thick and thin markets \citep{den-haan_ramey_watson_2000}.
The parameterization in \eqref{eq:match-fnc} implies that $\eta_{M,u}$,
the elasticity of matching with respect to unemployment,
is $\theta^{\gamma} / \left( 1+\theta^{\gamma} \right)$.

The nonlinear elasticity may well imply a large bound for $\Upsilon$.
\citet{ryan_2023arxiv} documents that taking $\gamma$ in \eqref{eq:match-fnc} equal to $\gammahi$,
a value found in the literature, implies that
the upper bound varies
between between $\bndhilo$ and $\bndhihi$
when $\theta$ takes on values observed in US data since December 2000.
The number $\bndhihi$
suggests there is scope for fixed costs of job creation paid by workers to affect volatility in the labor market. 
But the value $\gammahi$ was not estimated.
Because this value has implications for whether $\ell$ affects volatility,
I take up the task of estimating $\gamma$.

The estimation uses readily available data on hires from
the US Bureau of Labor Statistics' Job Openings and Labor Turnover Survey.
The value for hires records all hires made within a month,
even though the worker may not remain employed at the end of the month.
The unadjusted data will therefore bias job-finding higher.
I adjust the data to account for this bias.
As far as I know,
this adjustment to the measure of hires has not been done before and
this is the first estimate of the matching technology in \eqref{eq:match-fnc}.

Section \ref{sec:parameters-conventional} briefly discusses conventional parameter values,
section \ref{sec:paramters-match-fnc} discusses the bias adjustment, and
section \ref{sec:paramters-match-fnc} covers the estimation of $\gamma$.

\subsection{Standard Parameters Agreed upon by Convention}
\label{sec:parameters-conventional}

Except for the parameter that determines the elasticity of matching with respect to unemployment,
choices about parameter values are standard.
I adopt many of the values used by \citet{pissarides_2009},
who considers job-creation costs paid only by firms.

The value of output, $y$, produced by each firm's constant-returns-to-scale technology is $1$.
The value of nonwork, $z$, which includes leisure and compensation from unemployment insurance, is $0.71$.
I follow convention and set $\phi = 0.5$,
which is a common choice that specifies workers and firms split any surplus generated from a match.

The model period is one day.
This choice, as noted by \citet[][2639, FN 6]{ljungqvist_sargent_2017},
prevents job-finding and -filling rates from falling outside of $0$ and $1$.
The interest   rate is set so that the monthly interest rate is $0.004$ and
the separation rate is set so that the monthly separation rate is $0.036$.

The average level of tightness observed in the US between 1960 and 2006 is $0.72$ and
the average monthly job-finding probability observed over the same period is $0.594$.
With $\theta = 0.72$,
I target the monthly job-finding probability by adjusting the parameter for matching efficiency, $\mu$ in equation \eqref{eq:match-fnc},
using the estimate of $\gamma$ presented in section \ref{sec:paramters-match-fnc}.
The implied unemployment rate is $5.7$ percent.

The cost of advertising vacancies and recruiting, $c$, is implied by two features.
First,  its value reflects the normalization of output.
Second, its value is determined by the steady-state condition in \eqref{eq:text:LS:12}.

These values agree with \citet{pissarides_2009}.
In a baseline calibration, where $h = \ell = \tau = 0$,
the equilibrium wage is $0.988$.
\citet[1351]{pissarides_2009} points out that this represents a flow percentage gain of $\left( 0.988/0.71 - 1 \right) \times 100 \text{ percent } = 39 \text{ percent}$
when a worker transitions from unemployment into a job.
Which is substantially more than some parameterizations where $z$ nearly equals the wage.
These nearly-equal parameterizations rely on the story of competitive markets in which workers are indifferent between work and nonwork.

\subsection{A New Measure of Transition Probabilities That Corrects for Time Aggregation
  and How the JOLTS Program Records Hires}
\label{sec:new-meas-trans}

One way to estimate the matching function in \eqref{eq:match-fnc} uses the homogeneity of $M$ to write the model in terms of job finding and therefore tightness.
Measuring rates of job finding, though, is challenging.
One challenge is the data, which are available only at discrete, monthly intervals,
even though workers can transition between employment and unemployment continuously 
throughout the month.
Another challenge is the way matches or hires are reported.
Data on the number of hires are available from the US Bureau of Labor Statistics' Job Openings and Labor Turnover Survey or the JOLTS program.
The JOLTS program reports all hires within a month,
including hires who are fired before the month ends.

Using the unadjusted hires measure would bias the probability of finding a job upwards.
The rate of job finding in the model is $f\left(\theta\right) \equiv m \left(u,v\right) / u$.
Unlike in the model, however, hires recorded by the JOLTS program do not necessarily work the following period.
And even though the probability of separating from a job is low,
the number of people who find a job each month is large.

To account for this biasing feature of the data,
I model the process of job transitions,
using techniques developed by \citet{shimer_2012_RED},
to uncover instantaneous transition rates between employment and unemployment.
The adjusted probabilities of job finding for the month can then be uncovered.

\subsubsection{Description of the Data Environment}

The interval $[t,t+1)$, for $t\in\left\{ 0,1,2,\dots\right\}$, is referred to as ``period $t$.''
Within period $t$ 
workers neither exit or enter the laborforce.
I define the following quantities: 
\begin{itemize}
\item $f_{t}\in[0,1]$ is the probability of finding a job in period $t$,
  the probability that a worker who begins period $t$ unemployed finds at least one job during period $t$;
\item $s_{t}\in[0,1]$ is the probability a worker separates from a job in period $t$,
  the probability that a worker who begins period $t$ employed loses at least one job during period $t$;
\item $\varphi_{t}\equiv-\log\left(1-f_{t}\right)\geq0$ is the arrival rate of
  the Poisson process that changes a worker's state from unemployment to employment; and 
\item $\varsigma_{t}\equiv-\log\left(1-s_{t}\right)\geq0$ is the arrival rate of
  the Poisson process that changes a worker's state from employment to unemployment.
\end{itemize}
The relationship, for example, between $\varphi_{t}$ and $f_{t}$ is $f_{t} = 1-\exp \left( -\varphi_{t} \right)$,
or one minus the probability that no jobs are found during the period.
I am interested in uncovering $s_{t}$ and $f_{t}$,
using
hires data available from the JOLTS program and
unemployment data from the Current Population Survey.

\subsubsection{Intraperiod Evolution of Stocks}

Fix $t\in\left\{ 0,1,2,\dots\right\} $ and let $\tau\in\left[0,1\right]$ be the elapsed time since the start of the period.
The following intraperiod stocks are of interest: 
\begin{itemize}
\item $e_{t+\tau}\left(\tau\right)$ is the number of employed workers at time $t+\tau$, 
\item $u_{t+\tau}\left(\tau\right)$ is the number of unemployed workers at time $t+\tau$, and 
\item $e_{t+\tau}^{h}\left(\tau\right)$ is the number of workers who were unemployed at $t$ and employed at some time $t^{\prime}\in[t,t+\tau)$,
  which corresponds with the number of hires, where hires includes
``workers who were hired and separated during the month.''\footnote{\label{fn:jolts-def}This category of workers
  is listed in the US Bureau of Labor Statistics's \textit{Handbook of Methods} under
  ``Job Openings and Labor Turnover Survey: Concepts,''
  which is accessible at \href{https://www.bls.gov/opub/hom/jlt/concepts.htm}{https://www.bls.gov/opub/hom/jlt/concepts.htm}.
  Accessed October 5, 2023.}
\end{itemize}
The notation is overkill, but
the parenthesis emphasize that the stocks are functions of time within the period,
which is indexed by $\tau$, and
the subscripts emphasize that data are available for $e_{t}$, $e_{t}^{h}$, $e_{t+1}$, etc.

At the start of each period, there are no hires: $e_{t}^{h}\left(0\right)=0$ for all $t$.
And the number of hires at the end of the period is defined as
$e_{t+1}^{h} \equiv e_{t+1}^{h}\left(1\right)$,
the number of hires measured and reported by the JOLTS program for period $t$.
Employment within the period evolves according to the system of differential equations
\begin{align}
\dot{e}_{t+\tau}\left(\tau\right) &= \varphi_{t}u_{t+\tau}\left(\tau\right) - \varsigma_{t}e_{t+\tau}\left(\tau\right)\label{eq:text:JOLTS:adj-e}\\
\dot{e}_{t+\tau}^{h}\left(\tau\right) &= \varphi_{t}u_{t+\tau}\left(\tau\right).\label{eq:text:JOLTS:adj-eh}
\end{align}
Employment, as described in equation \eqref{eq:text:JOLTS:adj-e},
increases as unemployed workers find jobs at instantaneous rate $\varphi_{t}$ and
decreases as employed workers separate from jobs at instantaneous rate $\varsigma_{t}$.
Hires, as described in equation \eqref{eq:text:JOLTS:adj-eh},
cumulates all hires,
which corresponds to how the JOLTS program measures new hires and differs from
\citeauthor{shimer_2012_RED}'s \citeyearpar{shimer_2012_RED} model of short-term unemployment.

To make progress on solving the system of differential equations,
I 
assume that the labor force is constant within the period:
$l_{t}=e_{t+\tau^{\prime}}\left(\tau^{\prime}\right)+u_{t+\tau^{\prime}}\left(\tau^{\prime}\right)$ for all $t^{\prime}\in\left[0,1\right)$.
The constant-labor-force assumption can be substituted into the system of differential equations in
\eqref{eq:text:JOLTS:adj-e} and \eqref{eq:text:JOLTS:adj-eh} to yield
two differential equations for $e$, $e^{h}$, and $l$:
\begin{align}
\dot{e}_{t+\tau}\left(\tau\right) & =\varphi_{t}l_{t}-\left(\varphi_{t}+\varsigma_{t}\right)e_{t+\tau}\left(\tau\right)\label{eq:text:JOLTS:adj-e2}\\
\dot{e}_{t+\tau}^{h}\left(\tau\right) & =\varphi_{t}l_{t}-\varphi_{t}e_{t+\tau}\left(\tau\right).\label{eq:text:JOLTS:adj-eh2}
\end{align}

Using equation \eqref{eq:text:JOLTS:adj-eh2} to eliminate $\varphi_{t} l_{t}$ from equation \eqref{eq:text:JOLTS:adj-e2} yields
the differential equation $\dot{e}_{t}\left(\tau\right) = -\varsigma_{t}e_{t+\tau}\left(\tau\right)+\dot{e}_{t+\tau}^{h}\left(\tau\right)$.
Its general solution involves the integration of the unknown function $e_{t+\tau}^{h} \left( \tau \right)$.
Because I do not know of evidence about hires made within a month,
to make progress on the solution,
I assume new hires are added linearly: $e_{t+\tau}^{h}\left(\tau\right)=e_{t+1}^{h}\tau$.
The model for new hires is depicted in figure \ref{fig:fnc-new-hires}
in appendix \ref{sec:app:deriv-new-meas-find}.

Once the model for new hires is set,
the differential equation for $\dot{e}_{t}\left(\tau\right)$ can be solved as
\begin{equation}
\label{eq:text:JOLTS:nonlinear-s}
e_{t+1} = e_{t}\left(1-s_{t}\right)+e_{t+1}^{h}-e_{t+1}^{h}\left[1+\frac{s_{t}}{\ln\left(1-s_{t}\right)}\right].
\end{equation}
The evolution of employment in \eqref{eq:text:JOLTS:nonlinear-s}
approximates 
the evolution of employment in stock--flow models of the labor market:
\begin{equation}
  \label{eq:text:JOLTS:nonlinear-s-approx}
  e_{t+1}\approx e_{t}\left(1-s_{t}\right)+e_{t+1}^{h}-e_{t+1}^{h}\left[1+\frac{s_{t}}{-s_{t}}\right]=e_{t}\left(1-s_{t}\right)+e_{t+1}^{h},
\end{equation}
where I have used the approximation $\ln\left(1-s_{t}\right)\approx-s_{t}$.
Employment the following period is approximately the number of workers who remain employed because they did not separate
plus the number of new hires.
The approximation, however,
does not correct for the fact that JOLTS reports all hires,
including people who were hired and then let go within the month.
The difference between equations \eqref{eq:text:JOLTS:nonlinear-s} and \eqref{eq:text:JOLTS:nonlinear-s-approx} 
is a correction for time aggregation.
The correction may be quantitatively important.
Because the rate of job finding in the United States is high,
even a low separation rate could mean the number of hires who remain employed the following month is meaningfully
less than the number reported by the JOLTS program.
This feature of the data could induce a meaningful upward bias
in measures of job finding that rely on unadjusted hiring data from the JOLTS program. 

Equation \eqref{eq:text:JOLTS:nonlinear-s} is a nonlinear equation for $s_{t}$,
which can be solved for using data on employment and new hires.
Appendix \ref{sec:app:deriv-new-meas-find} provides a derivation.
The appendix also discusses the difference between this model of transitions and
\citeauthor{shimer_2012_RED}'s \citeyearpar{shimer_2012_RED} model,
which makes use of a neat cancellation.

Given $s_{t}$, I still need to recover $f_{t}$.
But this is achievable because equation \eqref{eq:text:JOLTS:adj-eh2} is a linear differential equation
for $\dot{e}_{t+\tau} \left( \tau \right)$ with constant coefficients.
Its solution is
\begin{equation}
  \label{eq:text:shimer2012:5}
e_{t+1}=\frac{\varphi_{t}l_{t}}{\varsigma_{t}+\varphi_{t}}\left(1-e^{-\left(\varsigma_{t}+\varphi_{t}\right)}\right)+e_{t}e^{-\left(\varsigma_{t}+\varphi_{t}\right)}.
\end{equation}
Equation \eqref{eq:text:JOLTS:nonlinear-s} provides a solution for $s_{t}$ and therefore $\varsigma_{t}$.
Given this value,
equation \eqref{eq:text:shimer2012:5} implicitly defines $\varphi_{t}$.
Appendix \ref{sec:app:deriv-new-meas-find} provides a derivation and
shows that equation \eqref{eq:text:shimer2012:5} implies $e_{t+1} \approx e_{t}$ in steady state,
when the number of jobs created equals the number of jobs destroyed.

\subsection{Estimate of a Matching Function}
\label{sec:paramters-match-fnc}

Section \ref{sec:new-meas-trans} provides a way to measure the probability of job finding from readily available data on employment and hires.
The counterpart in the model is the probability, $f \left( \theta \right) \equiv M\left(u,v\right)/u$.
The parameterization in \eqref{eq:match-fnc} implies 
$f \left( \theta \right) = \mu \theta \left(1+\theta^{\gamma}\right)^{-1/\gamma}$.

There are many concerns when estimating a relationship between job-finding and tightness.
One concern is that hires measure employment-to-employment transitions. I will ignore this issue
so that I can use readily available data.
Another concern is that matching efficiency, $\mu$, can vary over the business cycle  \citep{borowczyk-martins_jolivet_postel-vinay}.
For example, on uncharacteristically sunny days workers may be more optimistic
and apply for more jobs.
In addition,
matching efficiency may vary over distinct eras of the business cycle,
which could reflect different ways of organizing economic activity.

I account for these features by modeling $\mu$ with two components.
The first component is an independent and identically distributed factor $\exp\left(\varepsilon_{t}\right)$.
The second component allows matching efficiency to shift
after the Great Recession, which began in December 2012, and
after the COVID-19 recession, which began in February 2020.\footnote{There are other concerns when estimating a matching technology, including compositional effects
and firms and workers adjusting behavior around periods characterized by high or
low rates of job finding.
But these concerns fall outside the scope of this paper.}
These assumptions imply a statistical model for the probability that a worker finds a job in month $t$:
\begin{equation}
  \label{eq:text:match-estimation}
  \log f\left(\theta_{t}\right) = \alpha + \log\theta_{t} -\frac{1}{\gamma} \log\left(1+\theta_{t}^{\gamma}\right)
  + \psi \; G\left(t\right) +\xi \; C\left(t\right) +\varepsilon_{t},
\end{equation}
where
$\psi$ is a component of matching efficiency that captures shifts after the Great Recession,
$G\left(t\right)$ is a function that equals $1$ if the month occurs after November 2007 and
before February 2020,
$\xi$ is a component of matching efficiency that captures shifts after the COVID-19 recession,
$C\left(t\right)$ is a function that equals $1$ if the month occurs after January 2020, and
$\varepsilon_{t}$ is an unobserved, time-varying error capturing shocks to search and recruiting intensity along with other factors.

\begin{figure}[htbp]
\centerline{\includegraphics[width=\textwidth]{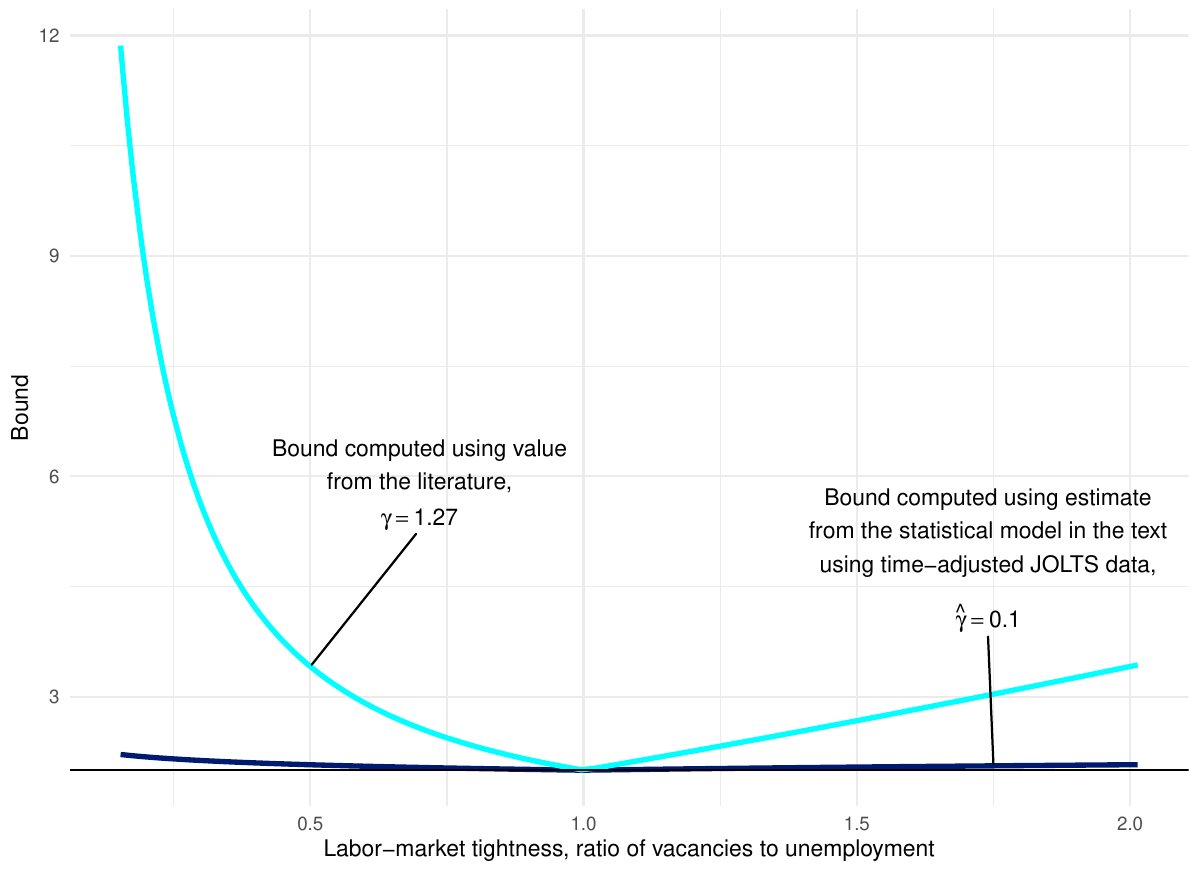}}
\caption[]{\label{fig:bound-tightness} Upper bounds for $\Upsilon$ computed for levels of tightness
  observed in the US economy after December 2000.}
\begin{figurenotes}[Notes]
  Tightness in the labor market is $\theta \equiv v / u$.
  The upper bound for $\Upsilon$ is given in \eqref{eq:Upsilon-bound}.
  Both series in blue depict the upper bound using
  the matching function given in \eqref{eq:match-fnc}
  for different values of $\gamma$.
  The horizontal, black line depicts the upper bound using the Cobb--Douglas matching function and
  elasticity parameter equal to $0.5$.
\end{figurenotes}
\begin{figurenotes}[Sources]
  Authors calculations that use data from FRED. 
  Data on vacancies are from the US Bureau of Labor Statistics' series
  Job Openings: Total Nonfarm [JTSJOL], retrieved from FRED, Federal Reserve Bank of St.~Louis; \url{https://fred.stlouisfed.org/series/JTSJOL}.
  Data on the level of unemployment are from the US Bureau of Labor Statistics' series
  Unemployment Level [UNEMPLOY], retrieved from FRED, Federal Reserve Bank of St. Louis;
  \url{https://fred.stlouisfed.org/series/UNEMPLOY}.
\end{figurenotes}
\end{figure}

Using nonlinear least squares on the statistical model in \eqref{eq:text:match-estimation},
I estimate $\hat{\gamma} = \estgamma$.
The estimate of $\gamma$ differs from the value used by \citet{den-haan_ramey_watson_2000} and \citet{petrosky-nadeau_wasmer_2017}. 
They take $\gamma$ to be $\gammahi$.
The two values imply meaningfully different bounds for $\Upsilon$,
which can be seen in figure \ref{fig:bound-tightness}.
The matching function in \eqref{eq:match-fnc} implies $\eta_{M,u} = \theta^{\gamma} / \left( 1+\theta^{\gamma} \right)$ and
the upper bound is given in \eqref{eq:Upsilon-bound}.

Figure \ref{fig:bound-tightness} depicts the upper bound in \eqref{eq:Upsilon-bound} for values of
tightness observed in the US economy after December 2000
when JOLTS data on vacancies became available.
The bound computed using $\gamma = \gammahi$ is always above the bound computed using $\hat{\gamma}$.
The difference is meaningful:
The bound computed using $\hat{\gamma}$ takes on values between $\bndlo$ and $\bndhi$, whereas
the bound computed using $\gammahi$ takes on values between $\bndhilo$ and $\bndhihi$.
The decomposition in \eqref{eq:elasticity-tight-y}
implies there is much more scope
when $\gamma = \gammahi$ for the features of the economic environment subsumed in $\Upsilon$, including $\ell$, to affect labor-market volatility.
But when $\hat{\gamma}$ is used,
\citeauthor{ljungqvist_sargent_2017}'s \citeyearpar{ljungqvist_sargent_2017} synopsis applies:
``A consensus about reasonable parameter values bounds [$\Upsilon$'s] contribution to the elasticity of market tightness. Hence,
the magnitude of the elasticity of market tightness depends mostly on the second
factor in expression \eqref{eq:elasticity-tight-y}, i.e., the inverse of what in the introduction we defined to
be the fundamental surplus fraction.''\footnote{In the quote, I changed the equation number to reflect the decomposition presented in this text.
  \citet{petrosky-nadeau_zhang_2017},
  in an exercise aimed at uncovering the model dynamics in \citet{hagedorn_manovskii_2008},
  take $\gamma = 0.407$; but,
  in an exercise aimed at uncovering the model dynamics in \citet{petrosky-nadeau_zhang_kuehn_2018}, which is seemingly more realistic,
  take $\gamma = 1.25$.
  A value of $1.553$ is used by \citet{silva_toledo_2009} who consider
  a DMP model with
  turnover costs paid by firms, as in \citet{mortensen_nagypal_2007}, and
  a boost to productivity from experience earned by workers who transition from inexperienced to experienced at constant rates.
  Figures \ref{fig:bound-hi} and \ref{fig:bound} in the appendix show the bound over time.}

In figure \ref{fig:bound-tightness} the black, horizontal line with y-intercept $2$ shows 
the bound in \eqref{eq:Upsilon-bound} when
the matching function is Cobb--Douglas and
the elasticity-of-matching parameter (the exponent) is $0.5$,
which is the value used by \citet{ljungqvist_sargent_2017} and consistent with estimates summarized by \citet{petrongolo_pissarides_2001}.
The Cobb--Douglas bound is much closer to the bound computed using the value for $\gamma$ estimated using data adjusted for sources of bias.

This makes sense:
The Cobb--Douglas parameterization is empirically successful \citep{bleakley_fuhrer_1997,petrongolo_pissarides_2001}.
For the Cobb--Douglas case,
the elasticity of matching with respect to unemployment is constant and equal to $0.5$.
When the matching function takes the form in \eqref{eq:match-fnc},
the elasticity of matching with respect to unemployment estimated here
falls between $\etamulo$ and $\etamuhi$.
In other words,
the implied elasticities that can vary by the state of the business cycle summarized by tightness
are close to the constant case of Cobb--Douglas matching.

In summary,
a literature suggested the bound for $\Upsilon$ in many cases was high.
But it is not.
This conclusion is based on an estimate of a nonlinear matching function using
data adjusted for time aggregation and how the JOLTS program collects data on hires.
The estimated elasticities are in line with what is found when matching is parameterized as Cobb--Douglas.

What does the estimate of $\gamma$ mean for volatility in the labor market when workers pay one-off hiring costs upon accepting a job?
Because the costs are subsumed in $\Upsilon$,
the result emphasizes that \citeauthor{ljungqvist_sargent_2017}'s \citeyearpar{ljungqvist_sargent_2017} fundamental surplus refers to recruitment resources available to firms.
There is limited scope for hiring costs paid by workers to affect labor-market volatility.
The magnitude of the elasticity of market tightness depends on the fundamental-surplus channel.
Nevertheless,
one-off costs paid by workers influence equilibrium tightness, unemployment, and wages.
These effects are investigated in the next section.
In addition,
while a full accounting of the business cycle is beyond the scope of this paper,
a comparative steady-state analysis is undertaken.
This is a shortcut to analyzing the fully dynamic model, but goes beyond the elasticity,
which is computed for a specific value of $\theta$.

\section{Properties of the Model}
\label{sec:properties-model}

The costs to create a job affect labor-market outcomes in two primary ways.
Costs affect the steady-state equilibrium and
they  affect labor-market dynamics.
Both effects are investigated in this section.
Section \ref{sec:prop:steady-state-equil} shows that
costs paid by workers upon accepting a job reduce wages.
Which makes recruiting more profitable for firms, so equilibrium unemployment is lower. 
Section \ref{sec:model-dynamics} illustrates how the unemployment rate would change if baseline productivity is perturbed.
Costs of job creation borne by workers 
reduce the volatility of unemployment.

\subsection{Steady-State Equilibrium}
\label{sec:prop:steady-state-equil}

\begin{figure}[htbp]
\centerline{\includegraphics[width=\textwidth]{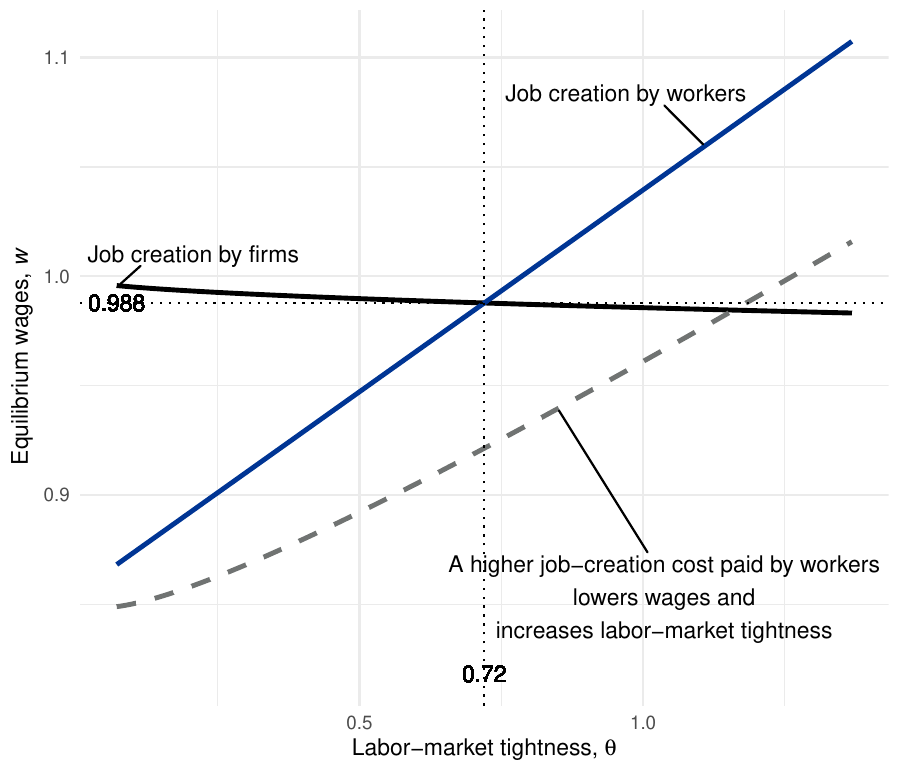}}
\caption[]{\label{fig:w-tight} Job creation by firms and workers in the presence of costs paid by workers upon accepting a job.}
\begin{figurenotes}[Notes]
  Job creation by firms   traces out the ordered pairs that satisfy equation \eqref{eq:text:LS:5} in black.
  Job creation by workers traces out the ordered pairs that satisfy equation \eqref{eq:text:LS:11} in blue.
  The blue curve represents an economy where $\tau = 0$, $h = 0$, and $\ell = 0$.
  Costs paid by workers upon accepting a job shift lower job creation by workers,
  which is represented by the dashed gray curve and is drawn for $\ell > 0$.
  Costs paid by workers upon accepting a job do not affect firms' hiring decisions independently 
  from the effect on wages, so the curve is fixed.  
  Because costs are paid by a worker upon a match,
  a firm can bargain for a lower wage by threatening the worker with
  paying the cost again upon a subsequent match if the wage negotiation fails.
  Lower wages encourage vacancy posting, which increases equilibrium tightness.
  Higher $\ell$ means a higher equilibrium $\theta$, lower equilibrium $w$, and lower equilibrium $u$.
  \end{figurenotes}
\end{figure}

In this section I am interested in how $\ell$ affects equilibrium.
To do this, I begin with a baseline.
I set the layoff tax, $\tau$, and costs of job creation paid by firms and workers, $h$ and $\ell$, to zero.
I then solve equation \eqref{eq:text:LS:12} for $c$ given the choice of tightness.
The choice of tightness implies a steady state job-finding and unemployment rate.
The steady-state unemployment rate is $5.7$ percent.
Given the solved-for $c$,
figure \ref{fig:w-tight}
depicts the two curves for job creation in $\theta$--$w$ space.
They intersect at the economy's equilibrium.
The curve that depicts job creation by firms   traces out the ordered pairs that satisfy equation \eqref{eq:text:LS:5} in black.
A higher wage reduces firms' willingness to post vacancies,
so the curve slopes downward.
The curve that depicts job creation by workers traces out the ordered pairs that satisfy equation \eqref{eq:text:LS:11} in blue.
Higher tightness increases the value of unemployment, which increases the bargaining position of workers.
The intersection of the curves shows steady-state tightness is $\theta = 0.72$,
as discussed in section \ref{sec:parameters-conventional}.
These results are standard \citep[][chapter 1]{pissarides_2000}.

Increasing $\ell$ does not affect firms' decisions to create jobs.
The costs are borne by workers and do not affect resources available for recruitment.
The costs do, however, lower wages for any given level of tightness through workers' job creation.
One-off costs paid by workers upon a match allow a firm to threaten a worker
with paying the costs again if the wage negotiation fails.
The equilibrium outcome is lower wages and increased tightness,
as depicted in figure \ref{fig:w-tight}. 

Firms are not directly affected by $\ell$. Instead,
firms are indirectly affected through a lower value of unemployment, $U$, and thus lower wages.
The indirect effect increases the value of a productive firm and encourages recruitment.
Unlike the case where $h > 0$, 
the cost $\ell$ does not subtract from resources that the invisible hand can allocation to vacancy creation.
The decomposition in \eqref{eq:elasticity-tight-y} makes clear that the fundamental surplus
applies to resources available \textit{to firms} for recruiting workers. 

\subsection{Model Dynamics}
\label{sec:model-dynamics}

To understand how costs affect economic dynamics,
I index four economies by costs of job creation.
In the steady state of each economy,
the level of productivity is $1$,
the level of tightness is $0.72$, the probability a worker finds a job is $0.594$, and the unemployment rate is $5.7$ percent.
The steady-state values imply different combinations of $c$, $h$, and $\ell$ that satisfy \eqref{eq:text:LS:12}.

At one extreme is a baseline economy where $h = \ell = 0$.
The only cost of job creation is the flow cost of posting a vacancy, $c$.
The value of $c$ is directly determined from equation \eqref{eq:text:LS:12}.
At the other extreme is an economy where costs of job creation are primarily paid by firms through a fixed cost,
instead of the proportional cost.
In this high-$h$ economy,
$c$ is nearly $0$ and $h$ is nearly $9$.
A value of $h$ a little higher than this amount would make any job posting unprofitable.
I then consider a middle-$h$ economy,
where the fixed cost of job creation paid by firms is half of what it is in the high-$h$ economy.
(To belabor the point, this implies a different flow cost of posting a vacancy.)
Finally,
I consider an economy where the high-$h$ value of
the fixed cost of job creation paid by firms is evenly split between firms and workers.

These economies are labeled in the first column of table \ref{tab:model-results}.
The second, third, and fourth columns report the values of $c$, $h$, and $\ell$ in each economy.
The fifth column reports the elasticity of tightness with respect to $y$ in each economy,
using the expression in \eqref{eq:elasticity-tight-y}.
Considering the period over which the steady-state values are calculated,
the data suggest that $\eta_{\theta,y}$ is around $7.56$.\footnote{The number comes from values reported by \citet{shimer_2005}.
  Table 1 of \citet[][28]{shimer_2005} reports that
  the correlation between $y$ and $\theta$ in the data is $0.396$,
  the standard deviation of $\theta$ is $0.382$, and
  the standard deviation of $y$ is $0.020$.
  All of these values are based on series that remove a low-frequency trend.
  Together, the values imply the correlation between $y$ and $\theta$ in the data is around $7.56$.
  \citet{pissarides_2009} notes that this number is the correct comparison and conducts a similar exercise that can be used as a comparison
  to my results.}

\input{tbl_04-plot-elasticity-experiment.tex}

Consistent with \citet{shimer_2005},
the baseline economy fails to generate the observed volatility in tightness.
In the baseline economy, $\eta_{\theta,y} = 3.602$.
This feature is sometimes referred to as the Shimer puzzle or the unemployment-volatility puzzle.
Turning to fixed costs,
as \citeauthor{ljungqvist_sargent_2017}'s \citeyearpar{ljungqvist_sargent_2017} decomposition proves,
shifting firms' costs from proportional costs to fixed costs raises volatility in the labor market.
In the high-$h$ economy, $\eta_{\theta,y}$ is $7.402$,
which is close to the data.
\citet{pissarides_2009} emphasizes this result.
The middle-$h$ economy, not surprisingly, lies between the two extremes.

Of interest is the effect $\ell$ has on volatility.
The fundamental decomposition in proposition \ref{prop:fundamental-decomposition} 
establishes that $\ell$ can have limited influence on labor-market dynamics
given the estimated matching technology, because
$\ell$ does not reduce firms' resources that can be allocated to vacancy creation.
The costs workers pay upon accepting jobs
enter only one term of the two-factor decomposition and this term is bounded by
a nonlinear function of $\eta_{M,u}$, which
in section \ref{sec:paramters-match-fnc} was estimated to range between $\bndlo$ and $\bndhi$.
The value of $\eta_{\theta,y}$ in the economy where costs are split corroborates these findings.
Labor-market volatility in the split economy is even less than in the middle-$h$ economy even though both
economies exhibit the same $h$ paid by firms,
since $\ell$ reduces $\Upsilon$ in \eqref{eq:elasticity-tight-y}.

The sixth column of table \ref{tab:model-results} reports the elasticity of wages with respect to $y$.
The values of $\eta_{w,y}$ in the last column of table \ref{tab:model-results}
demonstrate that the high-$h$ economy can generate realistic volatility without fixing wages.
\citet{hall_2005} fixes real wages in a DMP model to generate realistic volatility, but
data on real wages suggest wages are more procyclical.
For example, \citet{solon_barsky_parker_1994} document how wages are substantially procyclical once composition is accounted for,
consistent with the values reported in the sixth column of table \ref{tab:model-results}. 
\citet{gertler_trigari_2009} stagger wage bargaining, offering a waypoint between
period-by-period bargaining considered here and \citeauthor{hall_2005}'s \citeyearpar{hall_2005} model.
Some caution is called for when it comes to wages.
\citet{pissarides_2009} shows that wages in new jobs determine volatility and data on wages in new jobs are characterized by procyclicality.

The values of $\eta_{\theta,y}$ report how tightness responds to productivity at the steady state.
There are at least two concerns.
First, $\eta_{\theta,y}$ applies locally and responses may vary away from the steady state.
Second, there may be more interest in how unemployment responds as opposed to $\theta$.
To address these concerns,
I consider perturbations in productivity around each economy's steady-state value of $y = 1$.
I hold other parameters fixed and solve for the implied level of $\theta$ from \eqref{eq:text:LS:12},
which I use to compute the unemployment rate.
The comparative equilibrium analysis is a shortcut for modeling dynamics.

\begin{figure}[htbp]
\centerline{\includegraphics[width=\textwidth]{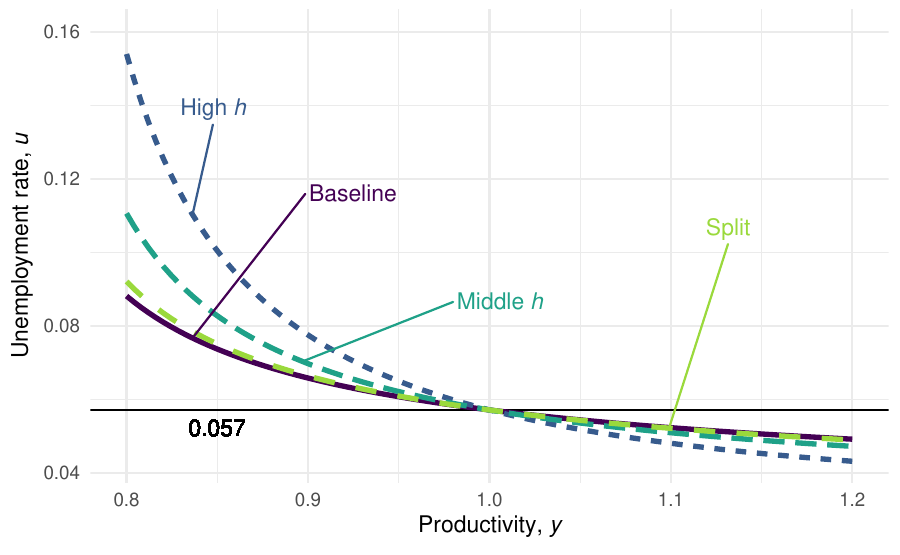}}
\caption[]{\label{fig:ur-dynamics} Steady-state unemployment dynamics for economies indexed by
  costs of job creation.}
\begin{figurenotes}[Notes]
  Economies are indexed by the job-creation costs listed in table \ref{tab:model-results}.
  For each economy,
  costs of job creation are adjusted so that tightness and the probability of finding a job equal their monthly targets,
  which produces a $5.7$ percent unemployment rate when $y$ is $1$.
  Steady-state dynamics are computed by solving for equilibrium tightness in \eqref{eq:text:LS:12}
  when productivity is perturbed around $y = 1$.
\end{figurenotes}
\end{figure}

Figure \ref{fig:ur-dynamics} depicts how each economy's unemployment rate responds to productivity perturbations.
The two extremes are the baseline and high-$h$ economies.
Unemployment in the baseline economy responds least to perturbations, whereas 
unemployment in the high-$h$ economy responds most. 
The difference in responses is especially pronounced for negative perturbations.
Responses of unemployment in the middle-$h$ economy are, not surprisingly, between these two.
Of some surprise, is the split economy,
where the fixed costs of job creation in the high-$h$ economy are split evenly
between workers are firms.
The split economy looks more like the baseline economy as opposed to the middle-$h$ economy,
even though $h$ is the same in the split and middle-$h$ economies.
The difference is explained by $\ell$.

Figure \ref{fig:ur-dynamics} shows that
responses of unemployment to productivity perturbations are asymmetric.
An increase in productivity generates a given fall in unemployment.
The magnitude of this fall is less than the increase in unemployment in response to a decrease in productivity of equal magnitude.
In steady state, where the unemployment rate is $5.7$ percent,
there is little way for the unemployment rate to fall, given exogenous separations.

\section{Discussion and Outstanding Questions}
\label{sec:discussion}

The model and numerical work predict that volatility in the labor market should decline as
the prevalence and magnitude of $\ell$-type costs increase.
Evidence supports this prediction.
The prevalence of such costs has increased over time
(\citetalias{cea_2015};
\citeauthor{kleiner_krueger_2013}, \citeyear{kleiner_krueger_2013},
figure 1, S177).
And \citet{barnichon_2010jme}, looking at data from the period 1948--2008, documents how
a positive productivity fluctuation
lowered unemployment on average early in the sample and
increased unemployment on average later in the sample.\footnote{How can a positive productivity fluctuation increase unemployment?
A positive innovation to productivity increases wages and thus demand.
But if firms' prices are stuck,
then the increase in demand is less than what firms can produce given the increase in productivity and size of the workforce.
The value of a worker is low and firms do not recruit workers,
which raises unemployment.
\citet{barnichon_2010jme} presents a model with these features.}
The model I presented offers a partial explanation for why
the cyclical component of unemployment and the cyclical component of productivity are less negatively correlated;
although, at this point,
the link is just suggestive and more evidence is needed.

In addition,
the perspective of costs paid by workers I have presented
raises a number of questions for future research.

Mechanisms that make
unemployment respond realistically to realistic changes in productivity will also make
unemployment respond meaningfully to changes in unemployment-insurance benefits.
[Equation \eqref{eq:text:LS:12} reveals that $\theta$ responds to
changes in $y$ and changes in $z$ symmetrically.]
Yet, data suggest
unemployment does not respond as much to the significant differences in benefits
observed across countries.
This dilemma was pointed out by \citet{costain_reiter_2008}.
\citet{rogerson_visschers_wright_2009} offer a solution:
a fixed factor of production like managerial talent or home production.
If the factor is abundant in certain states but scarce in others,
then unemployment responses can be muted
for large changes in benefits.
Does \citeauthor{rogerson_visschers_wright_2009}'s
\citeyearpar{rogerson_visschers_wright_2009} fixed-factor solution
interact with fixed costs of job creation?

In general,
unemployment insurance raises a trade-off between
smoothing consumption, which increases welfare, and
distorting search for work, which raises unemployment.
How does \citeauthor{andersen_2016}'s \citeyearpar{andersen_2016}
efficiency--equity locus change in the presence of job-creation costs?

The assumption of Nash bargaining is not essential.
As \citet{plotnikov_2019} emphasizes,
building on \citeauthor{farmer_2008}'s \citeyearpar{farmer_2008} insights,
any wage that divides the surplus is feasible.
\citet{plotnikov_2019} replaces the Nash sharing rule with
a rule that specifies agents' beliefs about their wealth
to determine aggregate demand and thus wages, output, and unemployment.
Do beliefs interact with turnover costs and job-creation costs paid by workers?    

Shifting the focus to costs of occupational regulation,
many features of registration, certification, and licensure are left unmodeled.
Work 
by \citet{tumen_2016} on job search using standard means versus social networks and
by \citet{florez_2019} on employment in an informal sector
demonstrates how selection matters.
In their models
composition across sectors is determined endogenously.
How would outcomes in the labor market change if workers
selected into sectors where work was regulated differently?

Several important features of the labor market are left unmodeled here too.
\citet{maury_tripier_2019} emphasize the importance large firms and
disruptions to productivity that induce firing.
Does intra-firm bargaining and firing change in the presence of fixed costs?
Do
policies that encourage participation in the labor market like
the earned income tax credit,
which \citet{regev_strawczynski_2019} study in the presence of risk-averse workers,
interact with costs of job creation?
Do these costs affect entrepreneurial decisions like
those modeled in \citet{gries_jungblut_naude_2016}?
Does the presence of a binding minimum wage,
which happens in the economies studied by
\citet{brecher_gross_2019} and
\citet{brecher_gross_2020}, 
interact with costs of job creation?
In contrast,
a minimum wage may not reduce labor demand in a model that assigns
heterogeneous workers to produce with heterogeneous capital,
as \citet{correa_parro_2020} demonstrate,
which raises the question:
how are workers assigned to tasks when creating a job costs a fixed amount?

\citet{guerrazzi_2023} takes up the question of optimal capital accumulation when
``hiring is a labor-intensive activity'' \citep[][footnote 1, 2]{guerrazzi_2023}.
Instead of posting vacancies,
large firms allocate part of their workforce to job creation \citep{shimer_2010}.
A wage rule is available that achieves optimal capital accumulation and labor allocation
in a decentralized equilibrium.
Capital and transitional dynamics are significant parts of the model.
How would transitional dynamics change
if workers paid a one-off cost upon taking a job or
if firms were forced to pay a cost to switch a worker from production to recruitment?

\section{Conclusion}
\label{sec:conclusion}

Creating jobs takes resources.
When workers pay a cost upon accepting a job,
wages are lower,
which increases the value of posting a vacancy and lowers unemployment.
The result exposes a previously unrecognized but important variable that affects wages.
Going beyond the scope of this paper,
this arrangement benefits firms and
could explain why workers more often have to pay costs like fees to take an exam before starting work.

Costs paid by workers
also reduce volatility in the labor market;
although, there is limited scope for this channel. 
The conclusion is based on a generalization of 
\citeauthor{ljungqvist_sargent_2017}'s \citeyearpar{ljungqvist_sargent_2017} fundamental
decomposition
that allows the matching function to take any reasonable form.
The decomposition reduces the elasticity of tightness with respect to productivity into two terms.
One of the two terms subsumes costs paid by workers upon taking a job.
This term is bounded by consensus, but
this judgment is based on 
the estimation of a matching function undertaken in this paper.
The estimation uses data that are adjusted for
worker flows and
how the JOLTS program records hires.
The estimated elasticity of matching with respect to unemployment
varies with tightness but the variation is small.
The updated bound limits the scope of influence that workers' costs can have on unemployment
volatility.
Nevertheless,
updated data used to assess the Shimer puzzle or unemployment-volatility puzzle
are consistent with the increased prevalence of costs paid by workers. 

Finally,
I close by suggesting that costs paid by workers upon accepting a job have
far-reaching implications.
The regulation of occupations is a growing ``phenomenon'' \citep[][S182]{kleiner_krueger_2013}.
By adding features of regulation into a DMP model,
I have taken a step towards understanding how such regulation affects dynamics in the labor market.
Likewise,
the flourishing of workers depends on their ability to take opportunities.
Workers may need help to overcome barriers like the cost of moving to accept jobs.

\bibliographystyle{../../../bibliography/bostonfed}
\bibliography{../../../bibliography/bibliography-org-ref}

\clearpage
\pagebreak

\section{Appendix}

Replication materials are available at:
\begin{center}
  \href{https://github.com/richryan/funDecomp}{https://github.com/richryan/funDecomp}.
\end{center}

\section{Derivations of Equations in Section \ref{sec:model-dmp-envir}}
\label{sec:app:deriv-dmp-envir}

\subsection{Bellman Equations for Firms}
\label{sec:app:deriv-dmp-envir-bellman-firms}

Key Bellman equations in the economy for firms are
\begin{equation}
J=y-w+\beta\left[s\left(V-\tau\right)+\left(1-s\right)J\right],\label{eq:J}
\end{equation}
\begin{equation}
V=-c+\beta\left[q\left(\theta\right)\left(J-h\right)+\left(1-q\left(\theta\right)\right)V\right].\label{eq:V}
\end{equation}
Imposing the zero-profit condition in equation \eqref{eq:V} implies
\begin{align*}
0 & =-c+\beta\left\{ q\left(\theta\right)\left(J-h\right)+\left[1-q\left(\theta\right)\right]\times0\right\} \\
\therefore c & =\beta q\left(\theta\right)\left(J-h\right)\\
\therefore\frac{c}{\beta q\left(\theta\right)} &= J-h
\end{align*}
or
\begin{equation}
J=\frac{c}{\beta q\left(\theta\right)}+h.\label{eq:value-J}
\end{equation}
Substituting this result into equation \eqref{eq:J} and using the
fact that vacancy creation exhausts all potential profits yields the
job-creation condition:
\begin{align}
J & =y-w+\beta\left[s\left(V-\tau\right)+\left(1-s\right)J\right]\nonumber \\
\therefore J & =y-w-\beta s\tau+\beta\left(1-s\right)J\nonumber \\
\therefore J\left[1-\beta\left(1-s\right)\right] & =y-w-\beta s\tau\nonumber \\
\therefore J & =\frac{y-w-\beta s\tau}{1-\beta\left(1-s\right)}\nonumber \\
 & =\frac{y-w-\beta s\tau}{\frac{1+r}{1+r}-\frac{1-s}{1+r}}\nonumber \\
 & =\frac{1+r}{r+s}\left(y-w-\beta s\tau\right).\label{eq:value-J-02}
\end{align}
Using the expressions for $J$ in \eqref{eq:value-J} and \eqref{eq:value-J-02},
the job-creation condition can be expressed as
\begin{align}
\label{eq:J-infinite}
  \begin{split}
    J &= \frac{1+r}{r+s}\left(y-w-\beta s\tau\right) \\
  &= \sum_{j=0}^{\infty}\left(\frac{1-s}{1+r}\right)^{j}\left(y-w-\beta s\tau\right) \\
  &= \sum_{j=0}^{\infty}\beta^{j}\left(1-s\right)^{j}\left(y-w-\beta s\tau\right)=\frac{c}{\beta q\left(\theta\right)}+h,
  \end{split}
\end{align}
where the second equality uses result \ref{result:1-plus-r}.
The expression in \eqref{eq:J-infinite} establishes equation \eqref{eq:text:J-infinite} in the main text.

What accounts for subtracting $\beta s\tau$ from flow profits?
The layoff tax represents resources that
``the invisible hand can never allocate'' to vacancy creation \citep[][2642]{ljungqvist_sargent_2017}.
\citeauthor{ljungqvist_sargent_2017}'s \citeyearpar{ljungqvist_sargent_2017} derivation of this result,
For a productive firm, the expected value of the layoff tax is
\begin{multline*}
\beta s\tau+\beta^{2}\left(1-s\right)s\tau+\beta^{3}\left(1-s\right)^{2}s\beta+\cdots \\
=s\tau\left[\beta+\beta\left(1-s\right)\beta+\beta^{2}\left(1-s\right)^{2}\beta+\cdots\right] \\
=\sum_{t=1}^{\infty}\beta^{t}\left(1-s\right)^{t-1}s\tau.
\end{multline*}
Because the firm is productive,
there is no chance of a layoff happening in the current period.
In the following period, with probability $s$ the layoff tax $\tau$ is paid, which must be discounted using $\beta$.
Two periods from now, with probability $s\left(1-s\right)$ the layoff tax $\tau$ is paid, which must be discounted using $\beta^{2}$.
The accounting of the layoff tax continues in a similar manner.
A flow amount each period must account for this expected amount each period the firm is in operation.
\citet[][2642]{ljungqvist_sargent_2017} call this amount $a$ and note that
\begin{align*}
\sum_{t=0}^{\infty}\beta^{t}\left(1-s\right)^{t}a=\sum_{t=1}^{\infty}\beta^{t}\left(1-s\right)^{t-1}s\tau,
\end{align*}
which implies $a = \beta s\tau$, using
\begin{align*}
\sum_{t=0}^{\infty}\beta^{t}\left(1-s\right)^{t}\beta s\tau & =\beta s\tau+\beta\left(1-s\right)\beta s\tau+\beta^{2}\left(1-s\right)^{2}\beta s\tau+\cdots\\
 & =s\tau\left[\beta+\beta\left(1-s\right)\beta+\beta^{2}\left(1-s\right)^{2}\beta+\cdots\right]\\
 & =\sum_{t=1}^{\infty}\beta^{t}\left(1-s\right)^{t-1}s\tau.
\end{align*}
The last line is
``the future layoff tax on the right side occurs after the initial period of operation.
Since the invisible hand can never allocate those resources to vacancy creation,
it is appropriate to subtract this annuity value when computing the surplus'' \citet[][2642]{ljungqvist_sargent_2017}.

Using equations \eqref{eq:J} and \eqref{eq:value-J} along with
the zero-profit condition imply
\begin{align*}
J & =y-w+\beta\left[s\left(V-\tau\right)+\left(1-s\right)J\right]\\
\therefore\frac{c}{\beta q\left(\theta\right)}+h & =y-w-\beta s\tau+\beta\left(1-s\right)\left(\frac{c}{\beta q\left(\theta\right)}+h\right)\\
\therefore w & =y-\beta s\tau-\left(\frac{c}{\beta q\left(\theta\right)}+h\right)+\beta\left(1-s\right)\left(\frac{c}{\beta q\left(\theta\right)}+h\right)\\
 & =y-\beta s\tau+\left[-1+\beta\left(1-s\right)\right]\left(\frac{c}{\beta q\left(\theta\right)}+h\right)\\
 & =y-\beta s\tau+\left[-\frac{1+r}{1+r}+\frac{1-s}{1+r}\right]\left(\frac{c}{\beta q\left(\theta\right)}+h\right)\\
 & =y-\beta s\tau-\frac{r+s}{1+r}\left(\frac{c}{\beta q\left(\theta\right)}+h\right)
\end{align*}
or
\begin{align*}
w & =y-\beta s\tau-\frac{r+s}{1+r}\left(\frac{c}{\beta q\left(\theta\right)}+h\right).
\end{align*}
This expression simplifies to
\begin{equation}
w=y-\beta s\tau-\frac{r+s}{q\left(\theta\right)}c-\frac{r+s}{1+r}h, \label{eq:LS:5}
\end{equation}
which is equation \eqref{eq:text:LS:5} in the main text.

\subsection{Workers' Bellman Equations}

\subsubsection{An Interpretation of $U$}

The key Bellman equations for workers are
\begin{equation}
E=w+\beta\left[sU+\left(1-s\right)E\right]\label{eq:6}
\end{equation}
\begin{equation}
U=z+\beta\left[\theta q\left(\theta\right)\left(E-\ell\right)+\left(1-\theta q\left(\theta\right)\right)U\right].\label{eq:7}
\end{equation}
These are equations \eqref{eq:E} and \eqref{eq:U} in the main text.

From equation \eqref{eq:6}
\begin{align}
E &= w+\beta\left[sU+\left(1-s\right)E\right]\nonumber \\
\therefore E\left[1-\beta\left(1-s\right)\right] & =w+\beta sU\nonumber \\
\therefore E &= \frac{w+\beta sU}{1-\beta\left(1-s\right)}.\label{eq:E-wR}
\end{align}
Using this result in \eqref{eq:7} yields
\begin{align*}
U & =z+\beta\left\{ \theta q\left(\theta\right)\left(E-\ell\right)+\left[1-\theta q\left(\theta\right)\right]U\right\} \\
 & =z+\beta\theta q\left(\theta\right)E-\beta\theta q\left(\theta\right)\ell+\beta\left[1-\theta q\left(\theta\right)\right]U\\
  & =z+\beta\theta q\left(\theta\right)\frac{w+\beta sU}{1-\beta\left(1-s\right)}-\beta\theta q\left(\theta\right)\ell+\beta\left[1-\theta q\left(\theta\right)\right]U.
\end{align*}
Using the definition of $\beta = \left( 1+r \right)^{-1}$ and result \ref{result:1-plus-r},
the latter can be written
\begin{align*}
U & =z+\beta\theta q\left(\theta\right)\frac{\left(1+r\right)w+sU}{r+s}-\beta\theta q\left(\theta\right)\ell+\beta\left[1-\theta q\left(\theta\right)\right]U\\
 & =z+\beta\theta q\left(\theta\right)\frac{\left(1+r\right)w}{r+s}+\frac{sU}{r+s}-\beta\theta q\left(\theta\right)\ell+\beta\left[1-\theta q\left(\theta\right)\right]U\\
  & =z+\beta\theta q\left(\theta\right)\sum_{j=0}^{\infty}\left(1-s\right)^{j}\beta^{j}w+\frac{sU}{r+s}-\beta\theta q\left(\theta\right)\ell+\beta\left[1-\theta q\left(\theta\right)\right]U.
\end{align*}
And collecting $U$ on the left side of the equation yields
\begin{align*}
U-\frac{sU}{r+s}-\beta\left[1-\theta q\left(\theta\right)\right]U & =z+\beta\theta q\left(\theta\right)\sum_{j=0}^{\infty}\left(1-s\right)^{j}\beta^{j}w-\beta\theta q\left(\theta\right)\ell\\
\therefore\left\{ 1-\frac{s}{r+s}-\beta\left[1-\theta q\left(\theta\right)\right]\right\} U & =z+\beta\theta q\left(\theta\right)\sum_{j=0}^{\infty}\left(1-s\right)^{j}\beta^{j}w-\beta\theta q\left(\theta\right)\ell\\
  \therefore\left\{ \frac{1}{r+s}\frac{1+r}{1}\frac{r}{1+r}U-\frac{1-\theta q\left(\theta\right)}{1}\frac{1}{1+r}U\right\}  & =z+\beta\theta q\left(\theta\right)\sum_{j=0}^{\infty}\left(1-s\right)^{j}\beta^{j}w-\beta\theta q\left(\theta\right)\ell.
\end{align*}
Collecting terms and prettying using result \ref{result:1-plus-r} yields 
\begin{align*}
\therefore\sum_{j=0}^{\infty}\left(1-s\right)^{j}\beta^{j}\frac{r}{1+r}U & =z+\beta\left[\theta q\left(\theta\right)\left(\sum_{j=0}^{\infty}\left(1-s\right)^{j}\beta^{j}w-\ell\right)+\left[1-\theta q\left(\theta\right)\right]U\right]\\
  \therefore \sum_{j=0}^{\infty}\left(1-s\right)^{j}\beta^{j}w_{R} &= z+\beta\left[\theta q\left(\theta\right)\left(\sum_{j=0}^{\infty}\left(1-s\right)^{j}\beta^{j}w-\ell\right)+\left[1-\theta q\left(\theta\right)\right]U\right],                                                                     
\end{align*}
where the last equality uses $r U / \left( 1+r \right) = w_{R}$,
which is established in result \ref{result:reservation-wage}.
This establishes equation \eqref{eq:reservation-wage} in the main text.

\subsubsection{Outcome of Asymmetric Nash Bargaining}

Wages, by convention in the economy, are determined by asymmetric Nash bargaining,
where the worker has bargaining parameter $\phi$.
As \citet{hall_2005} demonstrates in extreme by use of a fixed wage,
Nash bargaining is not essential,
but some form of rent sharing of the surplus generated from a productive match is essential.
The Nash-bargaining convention is adopted here to agree with the previous literature.
The outcome of asymmetric Nash bargaining is specified in \eqref{eq:8}.

Solving equation \eqref{eq:J} for $J$ yields
\begin{align}
J\left[1-\beta\left(1-s\right)\right] & =y-w-\beta s\tau\nonumber \\
\therefore J & =\frac{y-w-\beta s\tau}{1-\beta\left(1-s\right)}.\label{eq:J:w}
\end{align}
And solving \eqref{eq:6} for $E$ yields
\begin{align*}
E &= w+\beta\left[sU+\left(1-s\right)E\right]\\
\therefore E\left[1-\beta\left(1-s\right)\right] & =w+\beta sU\\
\therefore E & =\frac{w}{1-\beta\left(1-s\right)}+\frac{\beta sU}{1-\beta\left(1-s\right)}.
\end{align*}
Developing the expressions in \eqref{eq:8} yields
\begin{align*}
E-U &= \phi S\\
    &= \phi \frac{J}{1-\phi}
\end{align*}
and using the just-derived expressions for $J$ and $E$ yields
\begin{align*}
\underbrace{\left[\frac{w}{1-\beta\left(1-s\right)}+\frac{\beta sU}{1-\beta\left(1-s\right)}\right]}_{E}-U=\frac{\phi}{1-\phi}\underbrace{\left[\frac{y-w-\beta s\tau}{1-\beta\left(1-s\right)}\right]}_{J}.
\end{align*}

Developing this expression yields
\begin{align*}
w+\beta sU-\left[1-\beta\left(1-s\right)\right]U & =\frac{\phi}{1-\phi}\left(y-w-\beta s\tau\right)\\
\therefore w+\beta sU-U+\beta U-s\beta U & =\frac{\phi}{1-\phi}\left(y-w-\beta s\tau\right)\\
\therefore w & =\frac{\phi}{1-\phi}\left(y-w-\beta s\tau\right)+\left(1-\beta\right)U\\
\therefore\left(1-\phi\right)w & =\phi\left(y-w-\beta s\tau\right)+\left(1-\phi\right)\left(1-\beta\right)U\\
\therefore w & =\phi\left(y-\beta s\tau\right)+\left(1-\beta\right)U-\phi\left(1-\beta\right)U.
\end{align*}
Using the fact that
\[
1-\beta=1-\frac{1}{1+r}=\frac{1+r-1}{1+r}=\frac{r}{1+r},
\]
the latter expression can be written as
\begin{equation}
w = \frac{r}{1+r}U+\phi\left(y-\beta s\tau-\frac{r}{1+r}U\right), \label{eq:LS:9}
\end{equation}
which is equation (12) in \citet{ljungqvist_sargent_2017}.

\begin{align*}
w &= \frac{r}{1+r}U+\phi\left(y-\beta s\tau-\frac{r}{1+r}U\right)\\
\therefore w &= w_{R}+\phi\left(y-\beta s\tau\right)-\phi w_{R}\\
\therefore w &= \left(1-\phi\right)w_{R}+\phi\left(y-\beta s\tau\right).
\end{align*}
The worker earns a fraction of their reservation wages plus a fraction
of their output that accounts for the eventual layoff tax. 

To get an expression for the value $rU/\left(1+r\right)$, I solve
equation \eqref{eq:7} for $E-U$ and substitute this expression and
equation \eqref{eq:value-J} into equations \eqref{eq:8}. Turning
to equation \eqref{eq:7}:
\begin{align*}
U & =z+\beta\left\{ \theta q\left(\theta\right)\left(E-\ell\right)+\left[1-\theta q\left(\theta\right)\right]U\right\} \\
\therefore U & =z+\beta\theta q\left(\theta\right)E+\beta U-\beta\theta q\left(\theta\right)U-\beta\theta q\left(\theta\right)\ell\\
\therefore U & =z+\beta\theta q\left(\theta\right)\left(E-U\right)+\beta U-\beta\theta q\left(\theta\right)\ell\\
U\left(1-\beta\right)-z & =\beta\theta q\left(\theta\right)\left(E-U\right)-\beta\theta q\left(\theta\right)\ell\\
\therefore U\left(1-\beta\right)-z+\beta\theta q\left(\theta\right)\ell & =\beta\theta q\left(\theta\right)\left(E-U\right)\\
\therefore E-U & =\frac{1}{\beta\theta q\left(\theta\right)}\left[\left(1-\beta\right)U-z+\beta\theta q\left(\theta\right)\ell\right]\\
 & =\frac{1}{\beta\theta q\left(\theta\right)}\left[\left(1-\beta\right)U-z\right]+\ell\\
 & =\frac{1+r}{\theta q\left(\theta\right)}\left[\left(1-\beta\right)U-z\right]+\ell\\
 & =\frac{1+r}{\theta q\left(\theta\right)}\left[\left(1-\frac{1}{1+r}\right)U-z\right]+\ell\\
 & =\frac{r}{\theta q\left(\theta\right)}U-\frac{1+r}{\theta q\left(\theta\right)}z+\ell.
\end{align*}
Using the expression for $E-U$ in \eqref{eq:8} yields
\begin{align*}
E-U & =\phi S\\
\therefore\frac{r}{\theta q\left(\theta\right)}U-\frac{1+r}{\theta q\left(\theta\right)}z+\ell & =\phi S\\
 & =\phi\left(\frac{J}{1-\phi}\right)\\
 & =\frac{\phi}{1-\phi}\left(\frac{c}{\beta q\left(\theta\right)}+h\right),
\end{align*}
where the last equality uses the expression for $J$ in equation \eqref{eq:value-J}.
Developing this expression yields
\begin{align}
\frac{r}{\theta q\left(\theta\right)}U-\frac{1+r}{\theta q\left(\theta\right)}z & +\ell=\frac{\phi}{1-\phi}\left(\frac{c}{\beta q\left(\theta\right)}+h\right)\nonumber \\
\therefore rU-\left(1+r\right)z & +\theta q\left(\theta\right)\ell=\frac{\phi}{1-\phi}\frac{1}{\beta}c\theta+\theta q\left(\theta\right)\frac{\phi}{1-\phi}h\nonumber \\
\therefore rU-\left(1+r\right)z & =\frac{\phi}{1-\phi}\left(1+r\right)c\theta+\theta q\left(\theta\right)\left(\frac{\phi}{1-\phi}h-\ell\right)\nonumber \\
\therefore\frac{r}{1+r}U & =z+\frac{\phi c\theta}{1-\phi}+\frac{\theta q\left(\theta\right)}{1+r}\left(\frac{\phi}{1-\phi}h-\ell\right),\label{eq:LS:10}
\end{align}
which is comparable to equation (10) in \citet{ljungqvist_sargent_2017}.
Substituting equation \eqref{eq:LS:10} into equation \eqref{eq:LS:9}
yields
\begin{align*}
w & =\frac{r}{1+r}U+\phi\left(y-\beta s\tau-\frac{r}{1+r}U\right)\\
 & =\left[z+\frac{\phi c\theta}{1-\phi}+\frac{\theta q\left(\theta\right)}{1+r}\left(\frac{\phi}{1-\phi}h-\ell\right)\right]+\phi\left[y-\beta s\tau-z-\frac{\phi c\theta}{1-\phi}-\frac{\theta q\left(\theta\right)}{1+r}\left(\frac{\phi}{1-\phi}h-\ell\right)\right]\\
 & =\left(1-\phi\right)z+\frac{\phi c\theta}{1-\phi}\left(1-\phi\right)+\left(1-\phi\right)\frac{\theta q\left(\theta\right)}{1+r}\left(\frac{\phi}{1-\phi}h-\ell\right)+\phi\left(y-\beta s\tau\right)\\
 & =\left(1-\phi\right)z+\phi c\theta+\frac{\theta q\left(\theta\right)}{1+r}\left[\phi h-\left(1-\phi\right)\ell\right]+\phi\left(y-\beta s\tau\right)
\end{align*}
or
\begin{equation}
w=z+\phi\left(y-z-\beta s\tau+\theta c\right)+\frac{\theta q\left(\theta\right)}{1+r}\left[\phi h-\left(1-\phi\right)\ell\right].\label{eq:LS:11}
\end{equation}

Equation \eqref{eq:LS:11} can be re-arranged to read
\begin{align}
w & =z+\phi\left(y-z-\beta s\tau+\theta c\right)+\frac{\theta q\left(\theta\right)}{1+r}\left[\phi h-\left(1-\phi\right)\ell\right]\nonumber \\
\therefore w & =\left(1-\phi\right)\left[z-\beta\theta q\left(\theta\right)\ell\right]+\phi\left[y-\beta s\tau+\theta c+\beta\theta q\left(\theta\right)h\right].\label{eq:wage-01}
\end{align}

The two expressions for the steady-state wage rate in \eqref{eq:LS:5} and \eqref{eq:LS:11} jointly determine the equilibrium value of $\theta$.
Setting the two expressions for the steady-state wage rate equal to each other implies
\begin{align*}
  y-\beta s\tau-\frac{r+s}{q\left(\theta\right)}c-\frac{r+s}{1+r}h
  = z+\phi\left(y-z-\beta s\tau+\theta c\right)+\frac{\theta q\left(\theta\right)}{1+r}\left[\phi h-\left(1-\phi\right)\ell\right].
\end{align*}
Developing this expression yields
\begin{align*}
y-\beta s\tau-\frac{r+s}{q\left(\theta\right)}c-\frac{r+s}{1+r}h & =z+\phi\left(y-z-\beta s\tau+\theta c\right)+\frac{\theta q\left(\theta\right)}{1+r}\left[\phi h-\left(1-\phi\right)\ell\right]\\
\therefore\left(1-\phi\right)\left(y-z-\beta s\tau\right)-\frac{r+s}{q\left(\theta\right)}c-\frac{r+s}{1+r}h & =\phi\theta c+\frac{\theta q\left(\theta\right)}{1+r}\left[\phi h-\left(1-\phi\right)\ell\right]\\
\therefore\left(1-\phi\right)\left(y-z-\beta s\tau\right)-\frac{r+s}{1+r}h & =\left[\frac{r+s+\phi\theta q\left(\theta\right)}{q\left(\theta\right)}\right]c+\frac{\theta q\left(\theta\right)}{1+r}\left[\phi h-\left(1-\phi\right)\ell\right]\\
\therefore y-z-\beta s\tau-\frac{\beta\left(r+s\right)h}{1-\phi} & =\frac{c}{1-\phi}\left(\frac{r+s+\phi\theta q\left(\theta\right)}{q\left(\theta\right)}\right)+\frac{\theta q\left(\theta\right)}{1+r}\left(\frac{\phi}{1-\phi}h-\ell\right)
\end{align*}
Or
\begin{equation}
y-z-\beta s\tau-\frac{\beta\left(r+s\right)h}{1-\phi}=\frac{r+s+\phi\theta q\left(\theta\right)}{\left(1-\phi\right)q\left(\theta\right)}c+\frac{\theta q\left(\theta\right)}{1+r}\left(\frac{\phi}{1-\phi}h-\ell\right)\label{eq:LS:12}
\end{equation}
Expression \eqref{eq:LS:12} establishes equation \eqref{eq:text:LS:12} in the main text.

\section{Proof of Proposition \ref{prop:unique-theta}}
\label{sec:proof-prop-unique}

\subsection{Value of an Initial Vacancy}

In order for firms to post vacancies, it must the the case that the
initial vacancy is positive. Imagine an economy where the initial vacancy
is instantly filled. The hired worker is just like all other workers
productivity-wise. If the initial vacancy is not profitable, then
no other vacancies will be profitable. While intuitively clear within
the DMP class of models, the check yields a requirement for parameters.
\citet{ljungqvist_sargent_2021} in their discussion of \citet{christiano_eichenbaum_trabandt_2021,christiano_eichenbaum_trabandt_2020}
point out that failure to check may lead to downplaying key channels.

When the number of vacancies is vanishingly small compared to the
number of workers searching for a job, the job-filling probability
is $1$ and the job-finding probability is $0$:
\begin{align*}
\lim_{\theta\rightarrow0}q\left(\theta\right)=1\text{ and }\lim_{\theta\rightarrow0}f\left(\theta\right)=0.
\end{align*}

To compute the value of an initial vacancy, I need to know the 
wage rate paid to the initial worker and the value of a productive match
in the extreme initial case.
The value of the wage comes from \eqref{eq:wage-01}:
\begin{align*}
\lim_{\theta\rightarrow0}w & =\lim_{\theta\rightarrow0}\left(1-\phi\right)\left[z-\beta\theta q\left(\theta\right)\ell\right]+\phi\left[y-\beta s\tau+\theta c+\beta\theta q\left(\theta\right)h\right]\\
 & =\left(1-\phi\right)z+\phi\left(y-\beta s\tau\right)\\
 & =z+\phi\left(y-z-\beta s\tau\right).
\end{align*}
With the wage, the value of a productive firm, using \eqref{eq:J:w},
is 
\begin{align*}
\lim_{\theta\rightarrow0}J & =\lim_{\theta\rightarrow0}\frac{y-w-\beta s\tau}{1-\beta\left(1-s\right)}\\
 & =\lim_{\theta\rightarrow0}\frac{y-\left[z+\phi\left(y-z-\beta s\tau\right)\right]-\beta s\tau}{1-\beta\left(1-s\right)}\\
 & =\frac{\left(1-\phi\right)\left(y-z-\beta s\tau\right)}{1-\beta\left(1-s\right)}.
\end{align*}
The expression for $J$ in \eqref{eq:J:w} has used the idea that
once the initial vacancy yields a positive value, the recruitment
efforts of competitive firms will immediately drive the value of a
vacancy to $0$. The requirement for parameters will guarantee profitability.

The value of the initial vacancy is then
\begin{align*}
\lim_{\theta\rightarrow0}V & =\lim_{\theta\rightarrow0}-c+\beta\left[q\left(\theta\right)\left(J-h\right)+\left(1-q\left(\theta\right)\right)V\right]\\
 & =\lim_{\theta\rightarrow0}-c+\beta\left[q\left(\theta\right)\left(\frac{\left(1-\phi\right)\left(y-z-\beta s\tau\right)}{1-\beta\left(1-s\right)}-h\right)+\left(1-q\left(\theta\right)\right)V\right]\\
 & =-c+\beta\frac{\left(1-\phi\right)\left(y-z-\beta s\tau\right)}{1-\beta\left(1-s\right)}-\beta h\\
 & >0,
\end{align*}
where the third equality uses the fact that the opening is filled instantaneously.
The expression for the initial vacancy states that
\begin{equation}
\label{eq:first-vacancy-explained}
\beta \sum\limits_{j=0}^{\infty} \beta^{j} \left( 1-s \right)^{j} \left( 1-\phi \right) \left( y - z - \beta s \tau \right) > c + \beta h,
\end{equation}
which is equation \eqref{eq:text:first-vacancy-explained} in the main text.

This expression for $\lim_{\theta \rightarrow 0} V$ can be rearranged as
\begin{align*}
\beta\frac{\left(1-\phi\right)\left(y-z-\beta s\tau\right)}{1-\beta\left(1-s\right)} & >c+\beta h\\
\therefore\frac{1}{1+r}\frac{\left(1-\phi\right)\left(y-z-\beta s\tau\right)}{1-\frac{1}{1+r}\left(1-s\right)} & >c+\beta h
\end{align*}
or
\begin{equation}
\frac{\left(1-\phi\right)\left(y-z-\beta s\tau\right)}{r+s}>c+\beta h.\label{eq:parameter-requirement}
\end{equation}
The requirement in \eqref{eq:parameter-requirement} can be rearranged as
\begin{align*}
\frac{\left(1-\phi\right)\left(y-z-\beta s\tau\right)}{r+s} & >c+\beta h\\
\therefore\left(1-\phi\right)\left(y-z-\beta s\tau\right) & >\left(r+s\right)\left(c+\beta h\right)\\
\therefore y-z-\beta s\tau & >\frac{\left(r+s\right)c}{1-\phi}+\frac{\beta\left(r+s\right)h}{1-\phi}\\
\therefore y-z-\beta s\tau-\frac{\beta\left(r+s\right)h}{1-\phi} & >\frac{\left(r+s\right)c}{1-\phi},
\end{align*}
which establishes that
\begin{align*}
y-z-\beta s\tau-\frac{\beta\left(r+s\right)h}{1-\phi}>0.
\end{align*}
The above inequality guarantees that $\bar{\theta} > 0$,
which can be verified through the definition of $\bar{\theta}$ in \eqref{eq:theta-hi}.

\subsection{Existence and Uniqueness of Equilibrium Tightness $\theta$}

The equilibrium level of labor-market tightness in \eqref{eq:LS:12}
can be written $\mathcal{T}\left(\theta\right)=0$, where
\begin{equation}
\label{eq:def:T}
\mathcal{T}\left(x\right) \equiv y-z-\beta s\tau-\frac{\beta\left(r+s\right)}{1-\phi}h-\frac{c}{1-\phi}\left[\frac{r+s+\phi xq\left(x\right)}{q\left(x\right)}+\beta xq\left(x\right)\left(\frac{\phi h-\left(1-\phi\right)\ell}{c}\right)\right]
\end{equation}

An application of the intermediate value theorem establishes existence.
I will apply the theorem by establishing $\mathcal{T} \left( 0 \right) > 0$ and $\mathcal{T} \left( \bar{\theta} \right) > 0$,
where $\bar{\theta}$ is defined in \eqref{eq:theta-hi}.

To establish $\mathcal{T} \left( 0 \right) > 0$,
using $q \left( 0 \right) = 1$ and $0 \times q \left(0 \right) = 0$, I note that
\begin{align*}
\mathcal{T}\left(0\right)=y-z-\beta s\tau-\frac{\beta\left(r+s\right)}{1-\phi}h-\frac{c}{1-\phi}\left(r+s\right) > 0,
\end{align*}
where the inequality comes from the rearrangement of 
\begin{align*}
y-z-\beta s\tau-\frac{\beta\left(r+s\right)}{1-\phi}h-\frac{c}{1-\phi}\left(r+s\right) & >0\\
\therefore y-z-\beta s\tau & >\frac{\beta\left(r+s\right)}{1-\phi}h+\frac{c}{1-\phi}\left(r+s\right)\\
\therefore\frac{\left(1-\phi\right)\left(y-z-\beta s\tau\right)}{r+s} & >\beta h+c,
\end{align*}
which is true from \eqref{eq:parameter-requirement},
the requirement that the expected value of an initial vacancy yields a positive value.

To establish $\mathcal{T} \left( \bar{\theta} \right) < 0$,
the definition of $\mathcal{T}$, given in \eqref{eq:def:T}, implies
\begin{align*}
\mathcal{T}\left(\bar{\theta}\right) & =y-z-\beta s\tau-\frac{\beta\left(r+s\right)}{1-\phi}h-\frac{c}{1-\phi}\left[\frac{r+s}{q\left(\bar{\theta}\right)}+\phi\bar{\theta}+\beta\phi\bar{\theta}q\left(\bar{\theta}\right)\frac{h}{c}-\beta\left(1-\phi\right)\bar{\theta}q\left(\bar{\theta}\right)\frac{\ell}{c}\right]\\
                                     &= y-z-\beta s\tau-\frac{\beta\left(r+s\right)}{1-\phi}h-\frac{c}{1-\phi}\frac{r+s}{q\left(\bar{\theta}\right)}-\frac{\phi c}{1-\phi}\bar{\theta}-\frac{\beta}{1-\phi}\bar{\theta}q\left(\bar{\theta}\right)\left[\phi h-\left(1-\phi\right)\ell\right].
\end{align*}
The definition of $\bar{\theta}$, given in \eqref{eq:theta-hi}, implies the right side of the latter evaluates to
\begin{align*}                                       
\mathcal{T}\left(\bar{\theta}\right) &= y-z-\beta s\tau-\frac{\beta\left(r+s\right)}{1-\phi}h-\frac{c}{1-\phi}\frac{r+s}{q\left(\bar{\theta}\right)}-\left[y-z-\beta s\tau-\frac{\beta\left(r+s\right)}{1-\phi}h+\beta\ell\right] \\
  &\quad -\frac{\beta}{1-\phi}\bar{\theta}q\left(\bar{\theta}\right)\left[\phi h-\left(1-\phi\right)\ell\right]\\
 & =-\frac{c}{1-\phi}\frac{r+s}{q\left(\bar{\theta}\right)}-\frac{\beta}{1-\phi}\bar{\theta}q\left(\bar{\theta}\right)\left[\phi h-\left(1-\phi\right)\ell\right]-\beta\ell\\
 & =-\frac{c}{1-\phi}\frac{r+s}{q\left(\bar{\theta}\right)}-\frac{\beta}{1-\phi}\bar{\theta}q\left(\bar{\theta}\right)\phi h+\frac{\beta}{1-\phi}\bar{\theta}q\left(\bar{\theta}\right)\left(1-\phi\right)\ell-\beta\ell\\
 & =-\frac{c}{1-\phi}\frac{r+s}{q\left(\bar{\theta}\right)}-\frac{\beta}{1-\phi}\bar{\theta}q\left(\bar{\theta}\right)\phi h+\beta\bar{\theta}q\left(\bar{\theta}\right)\ell-\beta\ell\\
 & =-\frac{c}{1-\phi}\frac{r+s}{q\left(\bar{\theta}\right)}-\frac{\beta}{1-\phi}f\left(\bar{\theta}\right)\phi h-\beta\ell\left(1-f\left(\bar{\theta}\right)\right)\\
 & <0,
\end{align*}
where the last inequality uses the fact that $0<f\left(\bar{\theta}\right)<1$.

In addition, $\mathcal{T}$ is the composition of continuous functions
and therefore continuous on $\left[0,\bar{\theta}\right]$. Thus,
by the intermediate-value theorem, there exists $\theta\in \left( 0,\bar{\theta} \right)$
such that $\mathcal{T}\left(\theta\right)=0$, which establishes existence.
For any $\theta$ that satisfies this condition, the equilibrium level of unemployment is
\begin{equation}
u=\frac{s}{s+\theta q\left(\theta\right)}, \label{eq:LS:13}
\end{equation}
which is the steady state of
\begin{align*}
u_{t+1} = \left( 1-f \left( \theta \right) \right) u_{t} + s \left( 1-u_{t} \right),
\end{align*}
where $1-u_{t}$ is employment given the constant-labor-force assumption.
 
If $\mathcal{T}$ is everywhere decreasing,
then only one $\theta$ will satisfy $\mathcal{T}\left(\theta\right)=0$.
The derivative of $\mathcal{T}$ can be computed directly from \eqref{eq:def:T}.
It is stated in equation \eqref{eq:text:T:prime} in the main text and repeated here for convenience:
\begin{equation}
\mathcal{T}^{\prime}\left(x\right)\equiv\frac{c\left(r+s\right)}{\left(1-\phi\right)\left[q\left(x\right)\right]^{2}}q^{\prime}\left(x\right)-\frac{c\phi}{1-\phi}-\beta\frac{f^{\prime}\left(x\right)}{1-\phi}\left[\phi h-\left(1-\phi\right)\ell\right].\label{eq:T:prime}
\end{equation}
A tighter labor market makes it harder to fill a job and easier to find a job: $q^{\prime}<0$ and $f^{\prime}>0$.
The first term is therefore negative.
The term $c\phi/\left(1-\phi\right)$ is positive.
The sign of $\mathcal{T}^{\prime}$ therefore depends on the magnitudes of first two terms compared to the magnitude of
\begin{equation}
\label{eq:term-w-upward}
\frac{1}{1-\phi} \left[ c \phi + \beta f^{\prime} \left( x \right) \left[ \phi h - \left( 1-\phi \right) \ell \right] \right].
\end{equation}
The comparison can be easily checked for particular numerical values.

Yet,
other features of the economic environment can help rule out certain combinations of parameters.
If the job-creation condition of workers given in equation \eqref{eq:text:LS:11} is upward sloping in $\theta$--$w$ space,
then the term in \eqref{eq:term-w-upward} will be positive and
a unique equilibrium is guaranteed.
When there are no fixed costs of job creation, for example,
the job-creation condition of workers is upward sloping and this guarantees a unique equilibrium \citep[][chapter 1]{pissarides_2000}.

Table \ref{tab:model-results} describes sets of parameters guaranteeing a unique equilibrium.
Straightforward manipulation of the term listed in \eqref{eq:term-w-upward}
establish these cases.
The economies considered in section \ref{sec:properties-model}
are all characterized by a unique equilibrium.

\begin{table}
  \centering
  \caption{Sets of parameters guaranteeing a unique equilibrium.}
  \label{tab:params-guarantee-unique}
\begin{tabular}{llll}
  \toprule
                             & \multicolumn{3}{c}{Fixed cost of job creation} \\ \cmidrule(r){2-4}
Relative            & $h=\ell$ & $h>\ell$ & $h<\ell$ \\ 
bargaining strength &          &          &          \\ \midrule 
$\phi>1-\phi$ & Unique equilibrium & Unique equilibrium & Numerical check \\
$\phi=1-\phi$ & Unique equilibrium & Unique equilibrium & Numerical check \\
$\phi<1-\phi$ & Numerical check    & Numerical check    & Numerical check \\ \bottomrule
\end{tabular}
\end{table}

\section{Proof of Proposition \ref{prop:fundamental-decomposition}}

There are three parts of proposition \ref{prop:fundamental-decomposition}.
The first part establishes that the elasticity of tightness with respect to productivity can
be decomposed into two terms.
One of the terms establishes that the fundamental surplus is an essential object.
This part of the proof is established in appendix \ref{sec:fund-decomp}.
The second part establishes that the other term in the two-factor decomposition is bounded by
a nonlinear function of the elasticity of matching with respect to unemployment.
This part of the proof is established in appendix \ref{sec:bound-upsilon}.
The third part derives the elasticity of unemployment with respect to $y$.
This part of the proof is established in appendix \ref{sec:elast-unempl-wrt-y}.

\subsection{The Fundamental Decomposition: A Decomposition of the Elasticity of Market Tightness}
\label{sec:fund-decomp}

One way to understand how market tightness responds to changes in productivity is through
the elasticity of market tightness with respect to productivity, $\eta_{\theta,y} \equiv \left( d \theta / dy \right) \left( y / \theta \right)$.
The computation of  $\eta_{\theta,y}$ involves $d \theta / d y$.
The variable $\theta$ is defined implicitly by \eqref{eq:LS:12}.
The expression in \eqref{eq:LS:12} can be rearranged as
\begin{align}
  \label{eq:LS:53}    
  \begin{split}
y-z-\beta s\tau-\frac{\beta\left(r+s\right)h}{1-\phi} & =\frac{r+s+\phi\theta q\left(\theta\right)}{\left(1-\phi\right)q\left(\theta\right)}c+\frac{\theta q\left(\theta\right)}{1+r}\left(\frac{\phi}{1-\phi}h-\ell\right) \\
\therefore\frac{1-\phi}{c}\left[y-z-\beta s\tau-\frac{\beta\left(r+s\right)h}{1-\phi}\right] & =\frac{r+s+\phi\theta q\left(\theta\right)}{q\left(\theta\right)}+\frac{\theta q\left(\theta\right)}{1+r}\frac{\phi h-\left(1-\phi\right)\ell}{c} \\
 & =\frac{r+s}{q\left(\theta\right)}+\phi\theta+\frac{\theta q\left(\theta\right)}{1+r}\frac{\phi h-\left(1-\phi\right)\ell}{c}.
  \end{split}
\end{align}
This expression motivates defining the function $\digamma$ as
\begin{equation}
\label{eq:F}
\digamma \left(\theta,y\right) \equiv \frac{1-\phi}{c}\left(y-z-\beta s\tau-\frac{\beta\left(r+s\right)h}{1-\phi}\right)-\frac{r+s}{q\left(\theta\right)}-\phi\theta-\frac{\theta q\left(\theta\right)}{1+r}\frac{\phi h-\left(1-\phi\right)\ell}{c}.
\end{equation}
The implicit-function theorem then implies
\begin{align*}
\frac{d\theta}{dy} & =-\frac{\partial\digamma/\partial y}{\partial\digamma/\partial\theta}\\
 &= -\frac{\frac{1-\phi}{c}}{\frac{r+s}{\left[q\left(\theta\right)\right]^{2}}q^{\prime}\left(\theta\right)-\phi-\beta\left(\frac{\phi h-\left(1-\phi\right)\ell}{c}\right)\left(q\left(\theta\right)+\theta q^{\prime}\left(\theta\right)\right)}.
\end{align*}
Using the expression in \eqref{eq:LS:53} for $\left( 1-\phi \right) / c$, the latter can be written
\begin{align*}
\frac{d\theta}{dy} & =-\frac{\left[\frac{r+s}{q\left(\theta\right)}+\phi\theta+\beta\theta q\left(\theta\right)\frac{\phi h-\left(1-\phi\right)\ell}{c}\right]\frac{1}{y-z-\beta s\tau-\frac{\beta\left(r+s\right)h}{1-\phi}}}{\frac{r+s}{\left[q\left(\theta\right)\right]^{2}}q^{\prime}\left(\theta\right)-\phi-\beta\left(\frac{\phi h-\left(1-\phi\right)\ell}{c}\right)\left(q\left(\theta\right)+\theta q^{\prime}\left(\theta\right)\right)}\\
 & =-\frac{\left[\frac{r+s}{q\left(\theta\right)}+\phi\theta+\beta\theta q\left(\theta\right)\frac{\phi h-\left(1-\phi\right)\ell}{c}\right]\frac{1}{y-z-\beta s\tau-\frac{\beta\left(r+s\right)h}{1-\phi}}}{\frac{r+s}{\left[q\left(\theta\right)\right]^{2}}q^{\prime}\left(\theta\right)-\phi-\beta\left(\frac{\phi h-\left(1-\phi\right)\ell}{c}\right)\left(q\left(\theta\right)+\theta q^{\prime}\left(\theta\right)\right)}\times\frac{\theta q\left(\theta\right)}{\theta q\left(\theta\right)}.
\end{align*}
Developing this expression further yields
\begin{align*}
\frac{d\theta}{dy} & =-\frac{\left[\frac{r+s}{q\left(\theta\right)}+\phi\theta+\beta\theta q\left(\theta\right)\frac{\phi h-\left(1-\phi\right)\ell}{c}\right]\frac{1}{y-z-\beta s\tau-\frac{\beta\left(r+s\right)h}{1-\phi}}}{\frac{r+s}{\left[q\left(\theta\right)\right]^{2}}q^{\prime}\left(\theta\right)-\phi-\beta\left(\frac{\phi h-\left(1-\phi\right)\ell}{c}\right)\left(q\left(\theta\right)+\theta q^{\prime}\left(\theta\right)\right)}\times\frac{\theta q\left(\theta\right)}{\theta q\left(\theta\right)}\\
 & =\frac{\frac{r+s}{q\left(\theta\right)}+\phi\theta+\beta\theta q\left(\theta\right)\frac{\phi h-\left(1-\phi\right)\ell}{c}}{-\frac{r+s}{\left[q\left(\theta\right)\right]^{2}}q^{\prime}\left(\theta\right)+\phi+\beta\left(\frac{\phi h-\left(1-\phi\right)\ell}{c}\right)\left(q\left(\theta\right)+\theta q^{\prime}\left(\theta\right)\right)}\times\frac{\theta q\left(\theta\right)}{\theta q\left(\theta\right)}\frac{1}{y-z-\beta s\tau-\frac{\beta\left(r+s\right)h}{1-\phi}}\\
 & =\frac{\left(r+s\right)+\phi\theta q\left(\theta\right)+\beta\theta q\left(\theta\right)\left[q\left(\theta\right)\left(\frac{\phi h-\left(1-\phi\right)\ell}{c}\right)\right]}{-\left(r+s\right)\frac{\theta q^{\prime}\left(\theta\right)}{q\left(\theta\right)}+\phi\theta q\left(\theta\right)+\beta\theta q\left(\theta\right)\left(\frac{\phi h-\left(1-\phi\right)\ell}{c}\right)\left[q\left(\theta\right)+\theta q^{\prime}\left(\theta\right)\right]}\times\frac{\theta}{y-z-\beta s\tau-\frac{\beta\left(r+s\right)h}{1-\phi}}\\
 & =\frac{r+s+\phi\theta q\left(\theta\right)+\beta\theta q\left(\theta\right)q\left(\theta\right)\left(\frac{\phi h-\left(1-\phi\right)\ell}{c}\right)}{\left(r+s\right)\eta_{M,u}+\phi\theta q\left(\theta\right)+\beta\theta q\left(\theta\right)\left(1-\eta_{M,u}\right)q\left(\theta\right)\left(\frac{\phi h-\left(1-\phi\right)\ell}{c}\right)}\times\frac{\theta}{y-z-\beta s\tau-\frac{\beta\left(r+s\right)h}{1-\phi}},
\end{align*}
where the last line uses the expression for matching elasticities in \eqref{eq:eta-M-u} and \eqref{eq:1-eta-M-u}.
Tidying the latter expression up yields
\begin{equation}
\frac{d\theta}{dy}=\frac{r+s+\theta q\left(\theta\right)\left[\phi+\beta q\left(\theta\right)\left(\frac{\phi h-\left(1-\phi\right)\ell}{c}\right)\right]}{\left(r+s\right)\eta_{M,u}+\theta q\left(\theta\right)\left[\phi+\beta\left(1-\eta_{M,u}\right)q\left(\theta\right)\left(\frac{\phi h-\left(1-\phi\right)\ell}{c}\right)\right]}\times\frac{\theta}{y-z-\beta s\tau-\frac{\beta\left(r+s\right)h}{1-\phi}}.\label{eq:dtheta-dy}
\end{equation}
Thus
\begin{align*}
\eta_{\theta,y} = \frac{r+s+\theta q\left(\theta\right)\left[\phi+\beta q\left(\theta\right)\left(\frac{\phi h-\left(1-\phi\right)\ell}{c}\right)\right]}{\left(r+s\right)\eta_{M,u}+\theta q\left(\theta\right)\left[\phi+\beta\left(1-\eta_{M,u}\right)q\left(\theta\right)\left(\frac{\phi h-\left(1-\phi\right)\ell}{c}\right)\right]}\times\frac{y}{y-z-\beta s\tau-\frac{\beta\left(r+s\right)h}{1-\phi}},
\end{align*}
which establishes equation \eqref{eq:elasticity-tight-y} of proposition \ref{prop:fundamental-decomposition}.
Defining the term $\Upsilon$ as
\begin{equation}
\label{eq:upsi}
\Upsilon \equiv \frac{r+s+\theta q\left(\theta\right)\left[\phi+\beta q\left(\theta\right)\left(\frac{\phi h-\left(1-\phi\right)\ell}{c}\right)\right]}{\left(r+s\right)\eta_{M,u}+\theta q\left(\theta\right)\left[\phi+\beta\left(1-\eta_{M,u}\right)q\left(\theta\right)\left(\frac{\phi h-\left(1-\phi\right)\ell}{c}\right)\right]}.
\end{equation}
establishes \eqref{eq:text:upsi}  of proposition \ref{prop:fundamental-decomposition}.

\subsection{A Bound for $\Upsilon$}
\label{sec:bound-upsilon}

To establish a bound for $\Upsilon$, I start from
\begin{align*}
\frac{1}{\Upsilon}=\frac{\left(r+s\right)\eta_{M,u}+\theta q\left(\theta\right)\left[\phi+\beta\left(1-\eta_{M,u}\right)q\left(\theta\right)\left(\frac{\phi h-\left(1-\phi\right)\ell}{c}\right)\right]}{r+s+\theta q\left(\theta\right)\left[\phi+\beta q\left(\theta\right)\left(\frac{\phi h-\left(1-\phi\right)\ell}{c}\right)\right]}.
\end{align*}
There are two cases to consider.
The case where $\eta_{M,u} \geq 0.5$ and the case where $\eta_{M,u} < 0.5$.

I first consider the case where $\eta_{M,u} \geq 0.5$.
Thus $-\eta_{M,u}\leq-0.5$ and adding $1$ to both sides yields
\begin{equation}
1-\eta_{M,u}\leq\eta_{M,u}.\label{eq:case-eta-M-u-big}
\end{equation}
Starting from $1/\Upsilon$, I have
\begin{align*}
\frac{1}{\Upsilon} & =\frac{\left(r+s\right)\eta_{M,u}+\theta q\left(\theta\right)\left[\phi+\beta\left(1-\eta_{M,u}\right)q\left(\theta\right)\left(\frac{\phi h-\left(1-\phi\right)\ell}{c}\right)\right]}{r+s+\theta q\left(\theta\right)\left[\phi+\beta q\left(\theta\right)\left(\frac{\phi h-\left(1-\phi\right)\ell}{c}\right)\right]}\\
 & \geq\frac{\left(r+s\right)\left(1-\eta_{M,u}\right)+\theta q\left(\theta\right)\left[\phi+\beta\left(1-\eta_{M,u}\right)q\left(\theta\right)\left(\frac{\phi h-\left(1-\phi\right)\ell}{c}\right)\right]}{r+s+\theta q\left(\theta\right)\left[\phi+\beta q\left(\theta\right)\left(\frac{\phi h-\left(1-\phi\right)\ell}{c}\right)\right]},
\end{align*}
where the last inequality uses the assumption in \eqref{eq:case-eta-M-u-big}.
Developing the expression on the right side by adding and subtracting
$\left(1-\eta_{M,u}\right)\phi\theta q\left(\theta\right)$ to the numerator yields
\begin{gather*}
\frac{1}{\Upsilon}\geq\frac{\left(1-\eta_{M,u}\right)\left\{ r+s+\theta q\left(\theta\right)\left[\phi+\beta q\left(\theta\right)\left(\frac{\phi h-\left(1-\phi\right)\ell}{c}\right)\right]\right\} }{r+s+\theta q\left(\theta\right)\left[\phi+\beta q\left(\theta\right)\left(\frac{\phi h-\left(1-\phi\right)\ell}{c}\right)\right]}\\
+\frac{\theta q\left(\theta\right)\phi-\left(1-\eta_{M,u}\right)\phi\theta q\left(\theta\right)}{r+s+\theta q\left(\theta\right)\left[\phi+\beta q\left(\theta\right)\left(\frac{\phi h-\left(1-\phi\right)\ell}{c}\right)\right]}.
\end{gather*}
Developing the latter by canceling terms in the first expression and
collecting terms in the second yields
\begin{align*}
\frac{1}{\Upsilon} & \geq\left(1-\eta_{M,u}\right)+\frac{\phi\theta q\left(\theta\right)\eta_{M,u}}{r+s+\theta q\left(\theta\right)\left[\phi+\beta q\left(\theta\right)\left(\frac{\phi h-\left(1-\phi\right)\ell}{c}\right)\right]}\\
 & >1-\eta_{M,u}.
\end{align*}
Thus,
\begin{equation}
\label{eq:upsi-01}
\Upsilon < \left(1-\eta_{M,u}\right)^{-1}.
\end{equation}

Now consider the case where $\eta_{M,u}< 0.5$.
Thus $-\eta_{M,u}>-0.5$ and adding $1$ to both sides yields
\begin{equation}
1-\eta_{M,u}>0.5>\eta_{M,u}.\label{eq:case-eta-M-u-small}
\end{equation}
Starting from $1/\Upsilon$, I have
\begin{align*}
\frac{1}{\Upsilon} & =\frac{\left(r+s\right)\eta_{M,u}+\theta q\left(\theta\right)\left[\phi+\beta\left(1-\eta_{M,u}\right)q\left(\theta\right)\left(\frac{\phi h-\left(1-\phi\right)\ell}{c}\right)\right]}{r+s+\theta q\left(\theta\right)\left[\phi+\beta q\left(\theta\right)\left(\frac{\phi h-\left(1-\phi\right)\ell}{c}\right)\right]}\\
 & >\frac{\left(r+s\right)\eta_{M,u}+\theta q\left(\theta\right)\left[\phi+\beta\eta_{M,u}q\left(\theta\right)\left(\frac{\phi h-\left(1-\phi\right)\ell}{c}\right)\right]}{r+s+\theta q\left(\theta\right)\left[\phi+\beta q\left(\theta\right)\left(\frac{\phi h-\left(1-\phi\right)\ell}{c}\right)\right]},
\end{align*}
where the last inequality uses the assumption in \eqref{eq:case-eta-M-u-small}.
Developing the expression on the right side by adding and subtracting
$\eta_{M,u}\phi\theta q\left(\theta\right)$ to the numerator yields
\begin{gather*}
\frac{1}{\Upsilon}>\frac{\eta_{M,u}\left\{ r+s+\theta q\left(\theta\right)\left[\phi+\beta q\left(\theta\right)\left(\frac{\phi h-\left(1-\phi\right)\ell}{c}\right)\right]\right\} }{r+s+\theta q\left(\theta\right)\left[\phi+\beta q\left(\theta\right)\left(\frac{\phi h-\left(1-\phi\right)\ell}{c}\right)\right]}\\
+\frac{\theta q\left(\theta\right)\phi-\eta_{M,u}\phi\theta q\left(\theta\right)}{r+s+\theta q\left(\theta\right)\left[\phi+\beta q\left(\theta\right)\left(\frac{\phi h-\left(1-\phi\right)\ell}{c}\right)\right]}.
\end{gather*}
Developing the latter yields
\begin{align*}
\frac{1}{\Upsilon} & >\eta_{M,u}+\frac{\phi\theta q\left(\theta\right)\left(1-\eta_{M,u}\right)}{r+s+\theta q\left(\theta\right)\left[\phi+\beta q\left(\theta\right)\left(\frac{\phi h-\left(1-\phi\right)\ell}{c}\right)\right]}\\
 & >\eta_{M,u}.
\end{align*}
Thus, 
\begin{equation}
\label{eq:upsi-02}
\Upsilon<\left(\eta_{M,u}\right)^{-1}.
\end{equation}

The results in \eqref{eq:upsi-01} and \eqref{eq:upsi-02} establish \eqref{eq:Upsilon-bound} of proposition \ref{prop:fundamental-decomposition}.

The statement in \eqref{eq:Upsilon-bound} can be strengthened to
\begin{align*}
0 < \Upsilon < \max\left\{ \frac{1}{\eta_{M,u}}, \frac{1}{1-\eta_{M,u}} \right\} 
\end{align*}
if conditions are met for a unique equilibrium; namely,
if the term listed in \eqref{eq:parameter-requirement} is positive.
Result \ref{result:upward-slope} establishes the equivalency.
In other words, the result establishes the other direction too:
if $\Upsilon$ is unambiguously positive, then the equilibrium is unique.

\subsection{The Elasticity of Unemployment with Respect to Productivity}
\label{sec:elast-unempl-wrt-y}

The elasticity of unemployment with respect to $y$ is
$\eta_{u,y} = \left( du/dy \right) \left( y/u \right)$.
Starting from the expression for steady-state unemployment, $u = s / \left( s + f \left( \theta \right) \right)$,
it is true that
\begin{align*}
\frac{du}{dy} & =-s\left(s+f\left(\theta\right)\right)^{-2}f^{\prime}\left(\theta\right)\frac{d\theta}{dy}\\
 & =-\frac{u}{s+f\left(\theta\right)}\left(q\left(\theta\right)+\theta q^{\prime}\left(\theta\right)\right)\frac{d\theta}{dy},
\end{align*}
where the second equality uses the definition of steady-state $u$ and $f$.
Further development yields
\begin{align*}
\frac{du}{dy} & =-\frac{u}{s+f\left(\theta\right)}\left(q\left(\theta\right)+\theta q^{\prime}\left(\theta\right)\right)\frac{d\theta}{dy},\\
 & =-\frac{uq\left(\theta\right)}{s+f\left(\theta\right)}\left(1+\frac{\theta q^{\prime}\left(\theta\right)}{q\left(\theta\right)}\right)\frac{d\theta}{dy}\\
 & =-\frac{uq\left(\theta\right)}{s+f\left(\theta\right)}\left(1-\eta_{M,u}\right)\frac{d\theta}{dy},
\end{align*}
where the last line uses the definition of the elasticity of matching with respect to unemployment.
The elasticity of unemployment with respect to productivity is therefore
\begin{align*}
\frac{du}{dy}\frac{y}{u} & =-\frac{q\left(\theta\right)}{s+f\left(\theta\right)}\left(1-\eta_{M,u}\right)\frac{d\theta}{dy}y\\
 & =-\frac{\theta q\left(\theta\right)}{s+f\left(\theta\right)}\left(1-\eta_{M,u}\right)\frac{d\theta}{dy}\frac{y}{\theta}\\
 & =-\frac{f\left(\theta\right)}{s+f\left(\theta\right)}\left(1-\eta_{M,u}\right)\frac{d\theta}{dy}\frac{y}{\theta}\\
 & =-\left(1-u\right)\left(1-\eta_{M,u}\right)\frac{d\theta}{dy}\frac{y}{\theta}\\
 & =-\left(1-u\right)\left(1-\eta_{M,u}\right)\eta_{\theta,y},
\end{align*}
where the last equality uses the definition of $\eta_{\theta,y}$. 

A proof that $\eta_{M,u} \in \left( 0,1 \right)$ can be found in \citet{ryan_2023arxiv}.

\section{Auxiliary Results}

\begin{result}
\label{result:reservation-wage}The reservation wage, $U=W\left(w_{R}\right)$
is
\begin{align*}
w_{R}=\frac{r}{1+r}U.
\end{align*}
\end{result}

\begin{proof}
Starting from equation \eqref{eq:E-wR}:
\begin{align*}
E\left(w_{R}\right) & =\frac{w_{R}+\beta sU}{1-\beta\left(1-s\right)}=U\\
\therefore w_{R}+\beta sU & =U\left[1-\beta\left(1-s\right)\right]\\
\therefore w_{R} & =U\left[1-\beta\left(1-s\right)\right]-\beta sU\\
 & =U\left[1-\frac{1}{1+r}\left(1-s\right)-\frac{1}{1+r}s\right]\\
 & =U\left[\frac{1+r-1+s-s}{1+r}\right]\\
 & =U\frac{r}{1+r},
\end{align*}
establishing what was set out to be shown.
\end{proof}

\begin{result}
\label{result:1-plus-r}
The geometric series $ \left( 1+x \right) / \left( x+y \right)$ can be expressed as
\begin{align*}
\frac{1+x}{x+y} &= \sum_{j=0}^{\infty}\left(\frac{1-y}{1+x}\right)^{j} \\
  &= \frac{1}{1-\frac{1-y}{1+x}} \\
  &= \frac{1+x}{1+x-\left(1-y\right)} \\
  &=\frac{1+x}{x+y}.
\end{align*}
This result uses
\begin{align*}
\left|\frac{1-y}{1+x}\right| & <1 \; \text{ or } \;  -1 < \frac{1-y}{1+x} <1,
\end{align*}
which requiems
\begin{align*}
\frac{1-y}{1+x} <1 \; \text{  or  } \; 0 < x + y
\end{align*}
and
\begin{align*}
\frac{1-y}{1+x} > -1 \; \text{  or  } \; y - x < 2.
\end{align*}  
\end{result}

\begin{result}[Elasticity of matching with respect to unemployment]
  The elasticity of matching with respect to unemployment, $\eta_{M,u}$, is
  \begin{align}
    \label{eq:eta-M-u}
    \begin{split}
    \eta_{M,u} &\equiv \left(\frac{dM}{du}\right)\frac{u}{M} \\
    &= \left[\frac{d}{du}q\left(\theta\right)v\right]\frac{u}{M}=\left[\frac{d}{du}q\left(\theta\right)\right]\frac{vu}{M} \\
 & =\left[q^{\prime}\left(\theta\right)\frac{d\theta}{du}\right]\frac{vu}{M}=\left[q^{\prime}\left(\theta\right)\frac{-v}{u^{2}}\right]\frac{vu}{M} \\
 & =\left[q^{\prime}\left(\theta\right)\frac{-v}{u}\right]\frac{v}{M}=-q^{\prime}\left(\theta\right)\theta\frac{v}{M} \\
 & =-\frac{q^{\prime}\left(\theta\right)\theta}{q\left(\theta\right)}       
    \end{split}
\end{align}
where
the second line uses the fact that $d\theta/du = -v/u^{2}$ and
the last equality uses $q\left(\theta\right)=M/v$.
It follows that
\begin{equation}
  \label{eq:1-eta-M-u}
1-\eta_{M,u} = 1-\left[-\frac{q^{\prime}\left(\theta\right)\theta}{q\left(\theta\right)}\right]=\frac{q\left(\theta\right)+\theta q^{\prime}\left(\theta\right)}{q\left(\theta\right)}.
\end{equation}
\end{result}

\begin{result}
\label{result:upward-slope}
The slope of the job-creation condition for workers  in $\theta$--$w$ space is
\begin{align*}
c\phi+\beta f^{\prime}\left(x\right)\left[\phi h-\left(1-\phi\right)\ell\right] \text{ for } x \in \left( 0,\bar{\theta} \right),
\end{align*}
which can be derived from equation \eqref{eq:text:LS:11}.
A positive slope is equivalent to 
\begin{align*}
c\phi+\beta\left(q\left(x\right)+xq^{\prime}\left(x\right)\right)\left[\phi h-\left(1-\phi\right)\ell\right] &>0, \text{ using } f \left( x \right) = x q \left( x \right) \\
\therefore \frac{c\phi}{q\left(x\right)}+\beta\left(\frac{q\left(x\right)+\theta q^{\prime}\left(x\right)}{q\left(x\right)}\right)\left[\phi h-\left(1-\phi\right)\ell\right] &>0\\
\therefore\frac{c\phi}{q\left(x\right)}+\beta\left(1-\eta_{M,u}\right)\left[\phi h-\left(1-\phi\right)\ell\right] & >0, \text{ using result \eqref{eq:1-eta-M-u}},\\
\therefore\phi+q\left(x\right)\beta\left(1-\eta_{M,u}\right)\left[\frac{\phi h-\left(1-\phi\right)\ell}{c}\right] & >0,
\end{align*}
which is the denominator of $\Upsilon$.
\end{result}

\section{Derivation of a New Measure of Transition Probabilities That Corrects for Time-Aggregation and How JOLTS Records Hires}
\label{sec:app:deriv-new-meas-find}

Here is a derivation of transition probabilities that corrects for
data being available at discrete intervals even though workers can transition between employment and unemployment continuously. 
In addition, the adjustment corrects for how the JOLTS program records hires.
All hires are reported,
even
``workers who were hired and separated during the month.''
See footnote \ref{fn:jolts-def} in the main text.

Definitions of variables are included in the main text.
Some of the equations in the main text are reproduced here for convenience.

Section \ref{sec:derivations-transitions} contains the derivations and
section \ref{sec:hypoth-model-net} considers the hypothetical model
that would correspond to hew hires being measured as net hires.
Section \ref{sec:hypoth-model-net} highlights how this model differs from the model in \citet{shimer_2012_RED}.

\subsection{Derivations}
\label{sec:derivations-transitions}

In the model on transitions,
employment and new hires evolve according to 
\begin{align}
\dot{e}_{t+\tau}\left(\tau\right) &= \varphi_{t}u_{t+\tau}\left(\tau\right) - \varsigma_{t}e_{t+\tau}\left(\tau\right)\label{eq:JOLTS:adj-e}\\
\dot{e}_{t+\tau}^{h}\left(\tau\right) &= \varphi_{t}u_{t+\tau}\left(\tau\right).\label{eq:JOLTS:adj-eh}
\end{align}
These are equations \eqref{eq:text:JOLTS:adj-e} and \eqref{eq:text:JOLTS:adj-eh} in the main text.
With the assumption that
the labor force is constant within the period,
$l_{t}=e_{t+\tau^{\prime}}\left(\tau^{\prime}\right)+u_{t+\tau^{\prime}}\left(\tau^{\prime}\right)$ for all $t^{\prime}\in\left[0,1\right)$,
equations \eqref{eq:JOLTS:adj-e} and \eqref{eq:JOLTS:adj-eh} yield two differential equations for $e$ and $e^{h}$:
\begin{align}
\dot{e}_{t+\tau}\left(\tau\right) & =\varphi_{t}l_{t}-\left(\varphi_{t}+\varsigma_{t}\right)e_{t+\tau}\left(\tau\right)\label{eq:JOLTS:adj-e2}\\
\dot{e}_{t+\tau}^{h}\left(\tau\right) & =\varphi_{t}l_{t}-\varphi_{t}e_{t+\tau}\left(\tau\right).\label{eq:JOLTS:adj-eh2}
\end{align}
These are equations \eqref{eq:text:JOLTS:adj-e2} and \eqref{eq:text:JOLTS:adj-eh2} in the main text.

\begin{figure}[htbp]
\centerline{\includegraphics[width=0.8\textwidth]{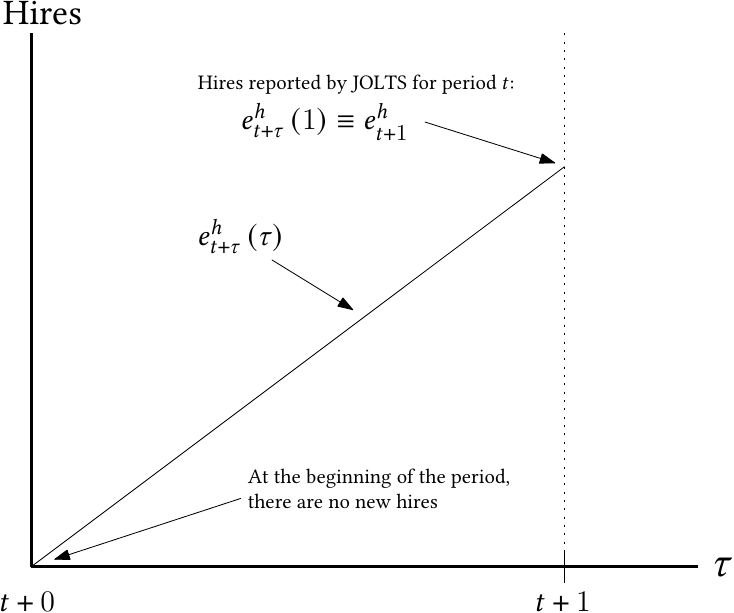}}
\caption[]{\label{fig:fnc-new-hires} Parameterized function for hires within a period.}
  \begin{figurenotes}[Notes]
  A the beginning of the period there are no new hires.
  At the end of the period there are $e_{t+1}^{h}$ new hires,
  a value that corresponds to data reported by the JOLTS program.
  I assume new hires are added linearly within the month.    
  \end{figurenotes}
\end{figure}

Using equation \eqref{eq:JOLTS:adj-eh2} to eliminate $\varphi_{t}l_{t}$
from equation \eqref{eq:JOLTS:adj-e2} yields the differential equation:
\begin{align*}
\dot{e}_{t}\left(\tau\right) & =\varphi_{t}l_{t}-\left(\varphi_{t}+\varsigma_{t}\right)e_{t+\tau}\left(\tau\right)\\
 & =\dot{e}_{t+\tau}^{h}\left(\tau\right)+\varphi_{t}e_{t+\tau}\left(\tau\right)-\left(\varphi_{t}+\varsigma_{t}\right)e_{t+\tau}\left(\tau\right)\\
 &= -\varsigma_{t}e_{t+\tau}\left(\tau\right)+\dot{e}_{t+\tau}^{h}\left(\tau\right).
\end{align*}
The general solution is \citep[923]{acemoglu_2009}:
\begin{align*}
e_{t}\left(\tau\right) & =\left[\varrho_{1}+\int_{0}^{\tau}\dot{e}_{t+x}^{h}\left(x\right)e^{\int_{0}^{x}\varsigma_{t}dv}dx\right]e^{\int_{0}^{\tau}-\varsigma_{t}dz}\\
 & =\left[\varrho_{1}+\int_{0}^{\tau}\dot{e}_{t+x}^{h}\left(x\right)e^{\varsigma_{t}x}dx\right]e^{-\varsigma_{t}\tau},
\end{align*}
where $\varrho_{1}$ is a constant of integration. The integral in
brackets can be evaluated using integration by parts:
\begin{align*}
\int_{0}^{\tau}\dot{e}_{t+x}^{h}\left(x\right)e^{\varsigma_{t}x}dx & =e^{\varsigma_{t}x}e_{t+x}^{h}\left(x\right)\bigg\vert_{x=0}^{x=\tau}-\int_{0}^{\tau}\varsigma_{t}e^{\varsigma_{t}x}e_{t+x}^{h}\left(x\right)dx\\
 & =e^{\varsigma_{t}\tau}e_{t+\tau}^{h}\left(\tau\right)-\varsigma_{t}\int_{0}^{\tau}e^{\varsigma_{t}x}e_{t+x}^{h}\left(x\right)dx.
\end{align*}
Therefore
\begin{equation}
e_{t+\tau}=\left[\varrho_{1}+e^{\varsigma_{t}\tau}e_{t+\tau}^{h}\left(\tau\right)-\varsigma_{t}\int_{0}^{\tau}e^{\varsigma_{t}x}e_{t+x}^{h}\left(x\right)dx\right]e^{-\varsigma_{t}\tau}.\label{eq:JOLTS:linear}
\end{equation}
\citet{shimer_2012_RED} at this point uses an immaculate cancellation.
As discussed below, this neat cancellation isn't available. To make
progress on the problem, I therefore assume hew hires are added linearly
within the period. Put another way, I posit a particular form for
the hires function within the period: $e_{t+\tau}^{h}\left(\tau\right)=e_{t+1}^{h}\tau$.
The model for new hires is depicted in figure \ref{fig:fnc-new-hires}.

With linearity, it is possible to make progress on the equation in
\eqref{eq:JOLTS:linear}:
\begin{align*}
e_{t+\tau} & =\left[\varrho_{1}+e^{\varsigma_{t}\tau}e_{t+\tau}^{h}\left(\tau\right)-\varsigma_{t}\int_{0}^{\tau}e^{\varsigma_{t}x}e_{t+1}^{h}xdx\right]e^{-\varsigma_{t}\tau}\\
 & =\left[\varrho_{1}+e^{\varsigma_{t}\tau}e_{t+\tau}^{h}\left(\tau\right)-\varsigma_{t}e_{t+1}^{h}\int_{0}^{\tau}e^{\varsigma_{t}x}xdx\right]e^{-\varsigma_{t}\tau}.
\end{align*}
The integral in the bracket is
\begin{align*}
\int_{0}^{\tau}e^{\varsigma_{t}x}xdx & =\frac{1}{\varsigma_{t}}xe^{\varsigma_{t}x}\bigg\vert_{x=0}^{x=\tau}-\frac{1}{\varsigma_{t}}\int_{0}^{\tau}e^{\varsigma_{t}x}dx\\
 & =\frac{1}{\varsigma_{t}}\tau e^{\varsigma_{t}\tau}-\frac{1}{\varsigma_{t}}\int_{0}^{\tau}e^{\varsigma_{t}x}dx\\
 & =\frac{1}{\varsigma_{t}}\tau e^{\varsigma_{t}\tau}-\frac{1}{\varsigma_{t}}\frac{1}{\varsigma_{t}}e^{\varsigma_{t}x}\bigg\vert_{x=0}^{x=\tau}\\
 & =\frac{1}{\varsigma_{t}}\tau e^{\varsigma_{t}\tau}-\frac{1}{\varsigma_{t}}\frac{1}{\varsigma_{t}}e^{\varsigma_{t}\tau}+\frac{1}{\varsigma_{t}}\frac{1}{\varsigma_{t}}\\
 & =\frac{1}{\varsigma_{t}}\left[\tau e^{\varsigma_{t}\tau}+\frac{1}{\varsigma_{t}}\left(1-e^{\varsigma_{t}\tau}\right)\right].
\end{align*}
Therefore
\begin{align*}
e_{t+\tau}=\left[\varrho+e^{\varsigma_{t}\tau}e_{t+\tau}^{h}\left(\tau\right)-e_{t+1}^{h}\left(\tau e^{\varsigma_{t}\tau}+\frac{1-e^{\varsigma_{t}\tau}}{\varsigma_{t}}\right)\right]e^{-\varsigma_{t}\tau}.
\end{align*}
When $\tau=0$, the latter express evaluates to
\begin{align*}
e_{t+0} & =\left[\varrho+e^{\varsigma_{t}0}e_{t+0}^{h}-e_{t+1}^{h}\left(0e^{\varsigma_{t}0}+\frac{1-e^{\varsigma_{t}0}}{\varsigma_{t}}\right)\right]e^{-\varsigma_{t}0}\\
\therefore e_{t} & =\varrho,
\end{align*}
using the fact that $e_{t+0}^{h}=0$. Therefore
\begin{align*}
e_{t+\tau} & =\left[e_{t}+e^{\varsigma_{t}\tau}e_{t+\tau}^{h}\left(\tau\right)-e_{t+1}^{h}\left(\tau e^{\varsigma_{t}\tau}+\frac{1-e^{\varsigma_{t}\tau}}{\varsigma_{t}}\right)\right]e^{-\varsigma_{t}\tau}\\
 & =e_{t}e^{-\varsigma_{t}\tau}+e_{t+\tau}^{h}\left(\tau\right)-e_{t+1}^{h}\left(\tau+\frac{e^{-\varsigma_{t}\tau}-1}{\varsigma_{t}}\right)\\
 & =e_{t}e^{-\varsigma_{t}\tau}+e_{t+\tau}^{h}\left(\tau\right)-e_{t+1}^{h}\left(\tau-\frac{1-e^{-\varsigma_{t}\tau}}{\varsigma_{t}}\right)
\end{align*}
Evaluated at $\tau=1$ yields
\begin{align*}
e_{t+1} & =e_{t}e^{-\varsigma_{t}}+e_{t+1}^{h}-e_{t+1}^{h}\left(1-\frac{1-e^{-\varsigma_{t}}}{s_{t}}\right)\\
 & =e_{t}\left(1-s_{t}\right)+e_{t+1}^{h}-e_{t+1}^{h}\left(1-\frac{s_{t}}{\varsigma_{t}}\right).
\end{align*}
Using $\varsigma_{t}=-\ln\left(1-s_{t}\right)$, the latter can be
written
\begin{align}
e_{t+1} &= e_{t}\left(1-s_{t}\right)+e_{t+1}^{h}-e_{t+1}^{h}\left[1+\frac{s_{t}}{\ln\left(1-s_{t}\right)}\right], \label{eq:JOLTS:nonlinear-s}
\end{align}
which is equation \eqref{eq:text:JOLTS:nonlinear-s} in the main text.

Equation \eqref{eq:JOLTS:nonlinear-s} is an equation only in the separation rate.
In addition, the approximation implies:
\begin{align}
e_{t+1} & \approx e_{t}\left(1-s_{t}\right)+e_{t+1}^{h}\nonumber \\
\therefore e_{t+1} & \approx e_{t}-e_{t}s_{t}+e_{t+1}^{h}\nonumber \\
\therefore e_{t}s_{t} & \approx e_{t}+e_{t+1}^{h}-e_{t+1}\nonumber \\
\therefore s_{t} & \approx1-\frac{e_{t+1}-e_{t+1}^{h}}{e_{t}}.\label{eq:JOLTS:approx-S}
\end{align}

As discussed in the main text,
I still need to recover the monthly job-finding rate.
To recover $f_{t}$,
I note that equation \eqref{eq:JOLTS:adj-e2} is a linear differential
equation for $\dot{e}_{t+\tau}\left(\tau\right)$ with constant coefficients.
The solution is 
\begin{align*}
e_{t+\tau} & =\frac{\varphi_{t}l_{t}}{\varsigma_{t}+\varphi_{t}}+\varrho_{2}e^{-\left(\varsigma_{t}+\varphi_{t}\right)\tau},
\end{align*}
where $\varrho_{2}$ is a constant of integration. Evaluating the
latter at $\tau=0$ yields
\begin{align*}
e_{t} & =\frac{\varphi_{t}l_{t}}{s_{t}+\varphi_{t}}+\varrho_{2}\\
\therefore\varrho_{2} & =e_{t}-\frac{\varphi_{t}l_{t}}{s_{t}+\varphi_{t}}.
\end{align*}
With the constant of integration, the solution is
\begin{align*}
e_{t+\tau} & =\frac{\varphi_{t}l_{t}}{\varsigma_{t}+\varphi_{t}}+\left(e_{t}-\frac{\varphi_{t}l_{t}}{s_{t}+\varphi_{t}}\right)e^{-\left(\varsigma_{t}+\varphi_{t}\right)\tau}\\
 & =\frac{\varphi_{t}l_{t}}{\varsigma_{t}+\varphi_{t}}\left(1-e^{-\left(\varsigma_{t}+\varphi_{t}\right)\tau}\right)+e_{t}e^{-\left(\varsigma_{t}+f_{t}\right)\tau}.
\end{align*}
Evaluated at $\tau=1$ yields
\begin{equation}
e_{t+1}=\frac{\varphi_{t}l_{t}}{\varsigma_{t}+\varphi_{t}}\left(1-e^{-\left(\varsigma_{t}+\varphi_{t}\right)}\right)+e_{t}e^{-\left(\varsigma_{t}+\varphi_{t}\right)}.\label{eq:shimer2012:5}
\end{equation}
To understand \eqref{eq:shimer2012:5},
note that in or near steady state,
the employment rate, $e_{t}/l_{t}$, is approximately $\varphi_{t}/\left(\varsigma_{t}+\varphi_{t}\right)$,
a standard formula derived from the number of jobs created totaling the number of jobs destroyed.
The approximation implies
\begin{align*}
l_{t}-\frac{u_{t}}{l_{t}}=\frac{e_{t}}{l_{t}}\approx1-\frac{\varsigma_{t}}{\varsigma_{t}+\varphi_{t}}=\frac{\varphi_{t}}{\varsigma_{t}+\varphi_{t}}.
\end{align*}
Using the latter in \eqref{eq:shimer2012:5} yields the steady-state approximation $e_{t+1} \approx e_{t}$:
\begin{align*}
e_{t+1} & \approx\frac{e_{t}}{l_{t}}l_{t}\left(1-e^{-\left(\varsigma_{t}+\varphi_{t}\right)}\right)+e_{t}e^{-\left(\varsigma_{t}+\varphi_{t}\right)}\\
 & =e_{t}\left[\left(1-e^{-\left(\varsigma_{t}+\varphi_{t}\right)}\right)+e^{-\left(\varsigma_{t}+\varphi_{t}\right)}\right]\\
 & =e_{t}\left\{ \left[1-\left(1-s_{t}\right)\left(1-f_{t}\right)\right]+\left(1-s_{t}\right)\left(1-f_{t}\right)\right\} \\
 & =e_{t}\left\{ \left[1-1+s_{t}+f_{t}-f_{t}s_{t}\right]+1-s_{t}-f_{t}+f_{t}s_{t}\right\} \\
 & =e_{t}\left\{ s_{t}+f_{t}-f_{t}s_{t}+1-s_{t}-f_{t}+f_{t}s_{t}\right\} \\
 & =e_{t}.
\end{align*}
But when the economy is away from steady state,
equation \eqref{eq:shimer2012:5} allows workers to transition between labor-market states continuously.
Equation \eqref{eq:shimer2012:5} ``captures the fact that a worker who loses
her job is more likely to find a new one without experiencing a measured
spell of unemployment'' \citep[131]{shimer_2012_RED}.

\subsection{Adjusted Data}
\label{sec:adjusted-data}

\begin{figure}[htbp]
\centerline{\includegraphics[width=\textwidth]{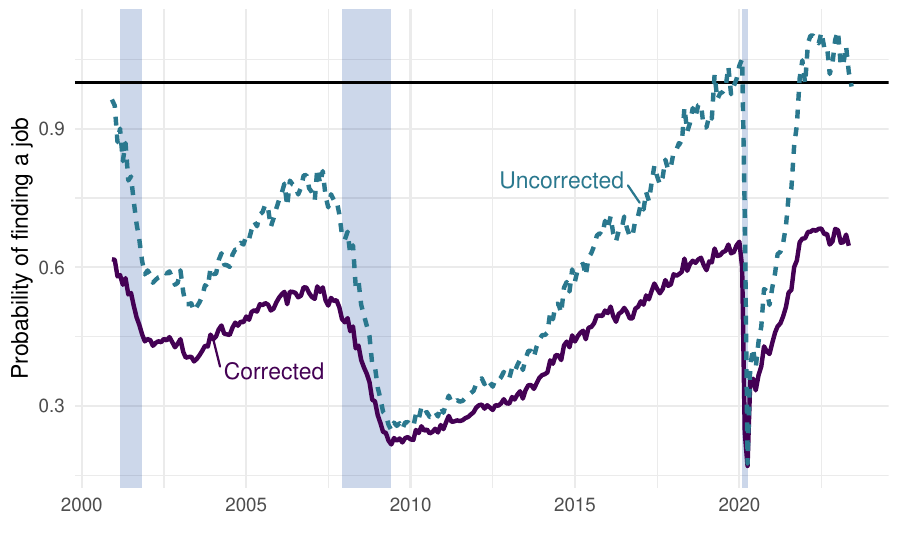}}
\caption[]{\label{fig:find-adj} Corrected and uncorrected probabilities of finding a job,
December 2000 through May 2023}
  \begin{figurenotes}[Notes]
    The uncorrected series
    reports the number of hires in the month divided by the level of unemployment,
    which corresponds to ``$M / u$'' in the model economy.
    The corrected series reports $f_{t} = 1-\exp \left( -\varphi_{t} \right)$,
    where $\varphi_{t}$ is implicitly defined in equation \eqref{eq:shimer2012:5}.
  \end{figurenotes}
\begin{figurenotes}[Sources]
  Authors calculations that use data from FRED.
  Data on hires are from the US Bureau of Labor Statistics' series Hires: Total Nonfarm [JTSHIL], retrieved from FRED, Federal Reserve Bank of St. Louis;
  \url{https://fred.stlouisfed.org/series/JTSHIL}.
  Data on the level of employment are from
  the US Bureau of Labor Statistics' series All Employees, Total Nonfarm [PAYEMS], retrieved from FRED, Federal Reserve Bank of St. Louis; \url{https://fred.stlouisfed.org/series/PAYEMS}.
  Data on the level of unemployment are from
  the US Bureau of Labor Statistics' series Unemployment Level [UNEMPLOY], retrieved from FRED, Federal Reserve Bank of St. Louis; \url{https://fred.stlouisfed.org/series/UNEMPLOY}.
  The labor force comprises
  employed and unemployed people.
\end{figurenotes}  
\end{figure}

\begin{figure}[htbp]
\centerline{\includegraphics[width=\textwidth]{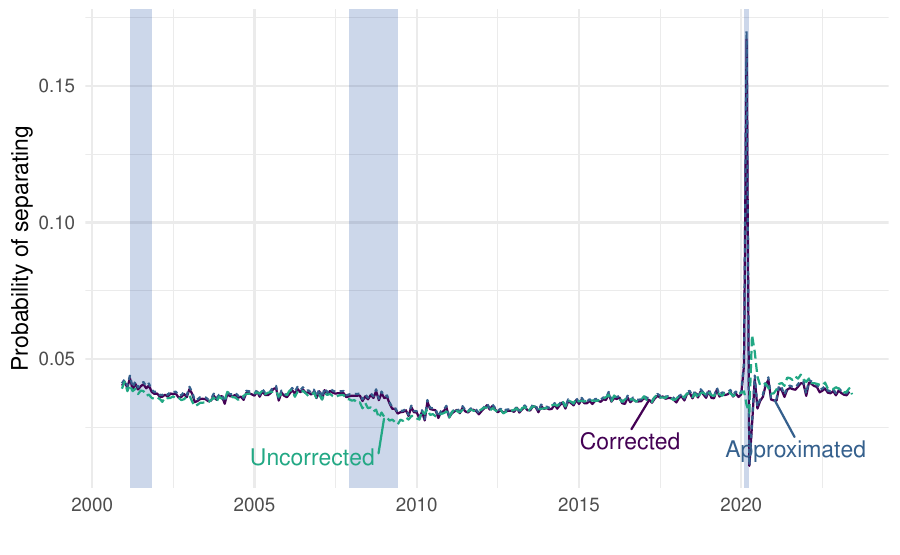}}
\caption[]{\label{fig:sep-adj} Corrected and uncorrected probabilities
  of separating from a job,
December 2000 through May 2023}
\begin{figurenotes}[Notes]
  The approximated series
  reports the separation rate in \eqref{eq:JOLTS:approx-S}.
  The uncorrected series uses
  an approximation of the unemployment rate.
  If $\tilde{s}_{t}$ is the uncorrected separation rate in month $t$ and
  $\tilde{f}_{t}$ is the uncorrected finding rate in month $t$,
  then the unemployment rate is roughly
  $\tilde{s}_{t} / \left( \tilde{s}_{t} + \tilde{f}_{t} \right)$.
  Given the monthly unemployment rate and $\tilde{f}_{t}$
  (from figure \ref{fig:find-adj}),
  I report the $\tilde{s}_{t}$ that would make the approximation hold with equality.
  The corrected series reports $s_{t} = 1-\exp \left( -\varsigma_{t} \right)$,
  where $\varsigma_{t}$ is implicitly defined in equation \eqref{eq:JOLTS:nonlinear-s}.
\end{figurenotes}
\begin{figurenotes}[Sources]
  Authors calculations that use data from FRED.
  The series for $\tilde{f}_{t}$ is described in the notes to figure \ref{fig:find-adj}.
  Data on the unemployment rate are from the US Bureau of Labor Statistics' series
  Unemployment Rate [UNRATE], retrieved from FRED, Federal Reserve Bank of St. Louis;
  \url{https://fred.stlouisfed.org/series/UNRATE}.
\end{figurenotes}  
\end{figure}

\begin{figure}[htbp]
\centerline{\includegraphics[width=\textwidth]{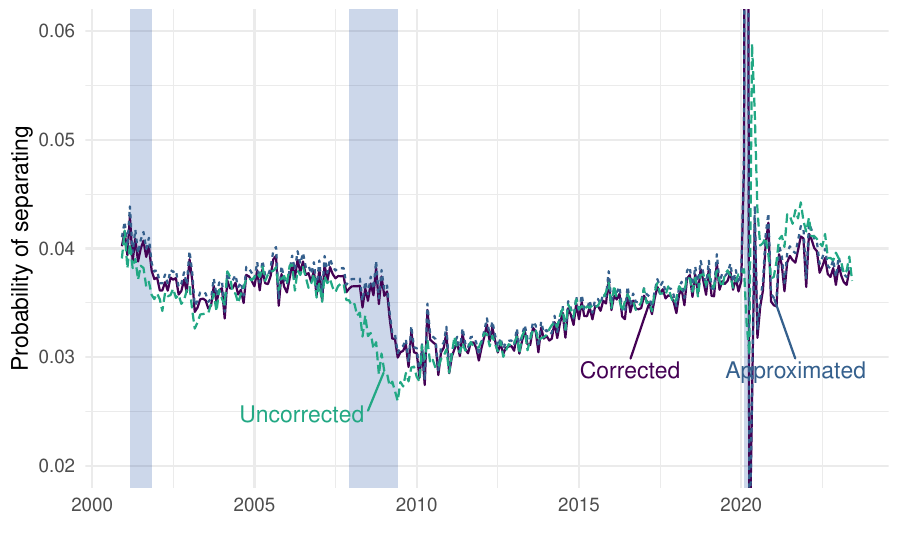}}
\caption[]{\label{fig:sep-adj-zoomed-in}  Corrected and uncorrected probabilities
  of separating from a job,
  December 2000 through May 2023.
  The series are the same as in figure \ref{fig:sep-adj}.
  The vertical axis is reduced to facilitate inspection.}
\end{figure}

The adjusted series are reported in figures
\ref{fig:find-adj}, \ref{fig:sep-adj}, and \ref{fig:sep-adj-zoomed-in}.
The data used in the figures are readily accessible
using \href{https://fred.stlouisfed.org/}{FRED}.
All the series are monthly.

Figure \ref{fig:find-adj} shows that
the correction for the probability of finding a job is meaningful.
The uncorrected series sometimes goes above $1$;
whereas,
the corrected series is closer to $0.7$.
As discussed in section \ref{sec:derivations-transitions},
the correction adjusts the probability of finding a job lower.

Figure \ref{fig:sep-adj} shows the
the approximate probability of separating from a job in \eqref{eq:JOLTS:approx-S} as well as
the corrected probability of separating from a job,
defined implicitly in \eqref{eq:JOLTS:nonlinear-s}.
The approximate and corrected series are close.
Equation \eqref{eq:JOLTS:nonlinear-s}
shows that the correction involves the separation rate,
which is small in magnitude.
The difference may be easier to see in figure \ref{fig:sep-adj-zoomed-in},
which shows the same series but truncates the vertical axis.

Figures \ref{fig:sep-adj} and \ref{fig:sep-adj-zoomed-in}
also report an uncorrected probability of separating from a job.
This series
corresponds to the separation rate
that would make the approximation to the unemployment rate,
$\tilde{s}_{t} / \left( \tilde{s}_{t} + \tilde{f}_{t} \right)$,
hold with equality.
The unemployment rate comes from the monthly BLS series and
I use the uncorrected finding rate depicted in figure \ref{fig:find-adj} for $\tilde{f}_{t}$.

\subsection{A Hypothetical Model of Net Hires and the Immaculate Cancellation in \citet{shimer_2012_RED}}
\label{sec:hypoth-model-net}

The JOLTS program measures cumulative hires.
If the JOLTS program measured net hires instead, then equation \eqref{eq:JOLTS:adj-eh} would be
\begin{align}
\dot{e}_{t+\tau}^{h} & =\varphi_{t}u_{t+\tau}-\varsigma_{t}e_{t+\tau}^{h}.\label{eq:adj-eh:immaculate}
\end{align}
Solving equation \eqref{eq:adj-eh:immaculate} for $\varphi_{t}u_{t+h}$
and substituting this into equation \eqref{eq:JOLTS:adj-e} yields
\begin{align*}
\dot{e}_{t+\tau} & =\varphi_{t}u_{t+\tau}-\varsigma_{t}e_{t+\tau}\\
 & =\left(\dot{e}_{t+\tau}^{h}+\varsigma_{t}e_{t+\tau}^{h}\right)-\varsigma_{t}e_{t+\tau}\\
 & =\dot{e}_{t+\tau}^{h}+\varsigma_{t}e_{t+\tau}^{h}-\varsigma_{t}e_{t+\tau}.
\end{align*}
This is a linear first-order equation for $\dot{e}_{t+\tau}$ with
$\tau\in\left[0,1\right)$. The general solution is
\begin{align}
e_{t+\tau} & =\left[c_{0}+\int_{0}^{\tau}\left(\dot{e}_{t+z}^{h}+\varsigma_{t}e_{t+z}^{h}\right)e^{-\int_{0}^{z}-\varsigma_{t}d\nu}dz\right]e^{\int_{0}^{\tau}-\varsigma_{t}dz}\nonumber \\
 & =\left[c_{0}+\int_{0}^{\tau}\left(\dot{e}_{t+z}^{h}+\varsigma_{t}e_{t+z}^{h}\right)e^{\varrho_{t}z}dz\right]e^{-\varsigma_{t}\tau},\label{eq:adj-t1:immaculate}
\end{align}
where $c_{0}$ is a constant of integration. Expanding the integral
in the bracketed term in \eqref{eq:adj-t1:immaculate} yields
\begin{align}
\int_{0}^{\tau}\left(\dot{e}_{t+z}^{h}+\varsigma_{t}e_{t+z}^{h}\right)e^{\varsigma_{t}z}dz & =\int_{0}^{\tau}e^{\varsigma_{t}z}\dot{e}_{t+z}^{h}dz+\int_{0}^{\tau}e^{\varsigma_{t}z}\varsigma_{t}e_{t+z}^{h}dz\nonumber \\
 & =\int_{0}^{\tau}e^{\varsigma_{t}z}\dot{e}_{t+z}^{h}dz+\varsigma_{t}\int_{0}^{\tau}e^{\varsigma_{t}z}e_{t+z}^{h}dz.\label{eq:adj-t2:immaculate}
\end{align}
The first integral on the right-hand size of the latter equation can be integrated by parts:
\begin{align*}
\int_{0}^{\tau}e^{\varsigma_{t}z}\dot{e}_{t+z}^{h}dz & =e^{\varsigma_{t}z}e_{t+z}^{h}\bigg\vert_{z=0}^{z=\tau}-\int_{0}^{\tau}\varsigma_{t}e^{\varsigma_{t}z}e_{t+z}^{h}dz\\
 & =\left(e^{\varsigma_{t}\tau}e_{t+\tau}^{h}-e^{\varsigma_{t}0}e_{t}^{h}\right)-\varsigma_{t}\int_{0}^{\tau}e^{\varsigma_{t}z}e_{t+z}^{h}dz\\
 & =e^{\varsigma_{t}\tau}e_{t+\tau}^{h}-\varsigma_{t}\int_{0}^{\tau}e^{\varsigma_{t}z}e_{t+z}^{h}dz,
\end{align*}
where the last equality uses $e_{t}^{h}=e_{t}^{h}\left(0\right)=0$.
Substituting this result into equation \eqref{eq:adj-t2:immaculate} yields
\begin{align*}
\int_{0}^{\tau}\left(\dot{e}_{t+z}^{h}+\varsigma_{t}e_{t+z}^{h}\right)e^{\varsigma_{t}z}dz & =e^{\varsigma_{t}\tau}e_{t+\tau}^{h}-\varsigma_{t}\int_{0}^{\tau}e^{\varsigma_{t}z}e_{t+z}^{h}dz+\varsigma_{t}\int_{0}^{\tau}e^{\varsigma_{t}z}e_{t+z}^{h}dz\\
 & =e^{\varsigma_{t}\tau}e_{t+\tau}^{h}.
\end{align*}
The cancellation yields a simpler expression than the one in \eqref{eq:JOLTS:linear}.
The result is an expression for next period's employment that does
not account for the hires number reporting cumulative hires. This
can be seen by substituting the result into equation \eqref{eq:adj-t1:immaculate},
which yields
\begin{align*}
e_{t+\tau} & =\left[c_{0}+\int_{0}^{\tau}\left(\dot{e}_{t+z}^{h}+\varsigma_{t}e_{t+z}^{h}\right)e^{\varsigma_{t}z}dz\right]e^{-\varsigma_{t}\tau}\\
 & =\left[c_{0}+e^{\varsigma_{t}\tau}e_{t+\tau}^{h}\right]e^{-\varsigma_{t}\tau}\\
 & =c_{0}e^{-\varsigma_{t}\tau}+e_{t+\tau}^{h}.
\end{align*}
The determination of $c_{0}$ comes from evaluating the latter at
$\tau=0$:
\[
e_{t}=c_{0}+e_{t}^{h}\left(0\right)=c_{0},
\]
as there are no new hires at the beginning of the period and $e_{t}^{h}\left(0\right)=0$.
Therefore
\[
e_{t+\tau}=e_{t}e^{-\varsigma_{t}\tau}+e_{t+\tau}^{h}.
\]
Evaluating the latter at $\tau=1$ yields
\begin{align}
e_{t+1} & =e_{t}e^{-\varsigma_{t}}+e_{t+1}^{h}\nonumber \\
 & =e_{t}\left(1-s_{t}\right)+e_{t+1}^{h},\label{eq:adj-t4:immaculate}
\end{align}
which indicates that the level of employment in the following survey
period equals the employed who do not separate from their jobs plus
new hires. Comparing \eqref{eq:adj-t4:immaculate} with \eqref{eq:JOLTS:approx-S}
and \eqref{eq:JOLTS:nonlinear-s} makes this point. In addition, the
comparison shows how
\citeauthor{shimer_2012_RED} \citeyearpar{shimer_2012_RED}'s adjustment for time
aggregation, which is meant for use with short-term unemployment data
from the CPS, differs from the adjustment for JOLTS data.

\subsection{Details of the Estimation of the Matching Function}

From the parameterization in \eqref{eq:match-fnc},
the implied job-finding probability, $M\left(u,v\right)/u$, is
\begin{align*}
f\left(\theta\right) &= \mu \frac{v}{\left(u^{\gamma}+v^{\gamma}\right)^{1/\gamma}}\times\frac{1/u}{1/\left( u^{\gamma} \right)^{1/\gamma}}=\mu\frac{\theta}{\left(1+\theta^{\gamma}\right)^{1/\gamma}}\\
\therefore\log\left(f\left(\theta\right)\right) &= \log\left(\mu\right)+\log\left(\theta\right)-\log\left(\left(1+\theta^{\gamma}\right)^{1/\gamma}\right) \\
 &= \log\left(\mu\right)+\log\left(\theta\right)-\frac{1}{\gamma}\log\left(1+\theta^{\gamma}\right).
\end{align*}
This derivation is used to produce the estimating equation in the main text in \eqref{eq:text:match-estimation},
which I repeat here for convenience:
\begin{equation}
  \label{eq:match-estimation}
  \log f\left(\theta_{t}\right) = \alpha + \log\theta_{t} -\frac{1}{\gamma} \log\left(1+\theta_{t}^{\gamma}\right)
  + \psi \; G\left(t\right) +\xi \; C\left(t\right) +\varepsilon_{t}.
\end{equation}
A description of the statistical model is provided in the main text.
I assume that $\varepsilon_{t}$ is independently and identically distributed according to the distribution $F$.

The untransformed expectation is 
\[
E\left[f\left(\theta_{t_{0}}\right)\right]=\int\exp\left(\alpha+\log\theta_{t_{0}}-\frac{1}{\gamma}\log\left(1+\theta_{t_{0}}^{\gamma}\right)+\psi_{G\left(t_{0}\right)}+\xi_{C\left(t_{0}\right)}+\varepsilon_{t_{0}}\right)dF\left(\varepsilon\right).
\]
The empirical cdf is $\widehat{F}\left(x\right)=T^{-1}\sum_{t=1}^{T}\boldsymbol{1}\left\{ \hat{\varepsilon}_{t}\leq x\right\} $,
where $\boldsymbol{1}\left\{ A\right\} $ is an indicator function
that takes the value $1$ if the event $A$ is true and $0$ otherwise.
\citet{duan_1983} shows how the unknown cdf $F$ can be replaced
by its empirical estimate, $\widehat{F}_{T}$:
\begin{align}
\widehat{E}\left[f\left(\theta_{t_{0}}\right)\right] & =\int\exp\left(\hat{\alpha}+\log\theta_{t_{0}}-\frac{1}{\hat{\gamma}}\log\left(1+\theta_{t_{0}}^{\hat{\gamma}}\right)+\hat{\psi}_{G\left(t_{0}\right)}+\hat{\xi}_{C\left(t_{0}\right)}+\varepsilon\right)d\widehat{F}_{T}\left(\varepsilon\right)\nonumber \\
 & =\frac{1}{T}\sum_{t=1}^{T}\int\exp\left(\hat{\alpha}+\log\theta_{t_{0}}-\frac{1}{\hat{\gamma}}\log\left(1+\theta_{t_{0}}^{\hat{\gamma}}\right)+\hat{\psi}_{G\left(t_{0}\right)}+\hat{\xi}_{C\left(t_{0}\right)}+\hat{\varepsilon}_{t}\right)\nonumber \\
 & =\hat{\mu}\frac{\theta_{t_{0}}}{\left(1+\theta_{t_{0}}^{\hat{\gamma}}\right)^{1/\hat{\gamma}}}\frac{1}{T}\sum_{t=0}^{T}\exp\left(\hat{\varepsilon}_{t}\right).\label{eq:smear-estimate}
\end{align}
\citet{duan_1983} establishes the result for a transformation that
yields a linear model and refers to the result as the smearing estimate.
I posit that the smearing estimate works for the nonlinear model in
\eqref{eq:match-estimation}. \citet{miller_1984} provides an example
where $F$ is lognormal.
The smearing estimate in \eqref{eq:smear-estimate}
is visually indistinguishable from the naive estimate
$\hat{\mu}\theta_{t_{0}}\left(1+\theta_{t_{0}}^{\hat{\gamma}}\right)^{-1/\hat{\gamma}}$.

\begin{figure}[htbp]
\centerline{\includegraphics[width=\textwidth]{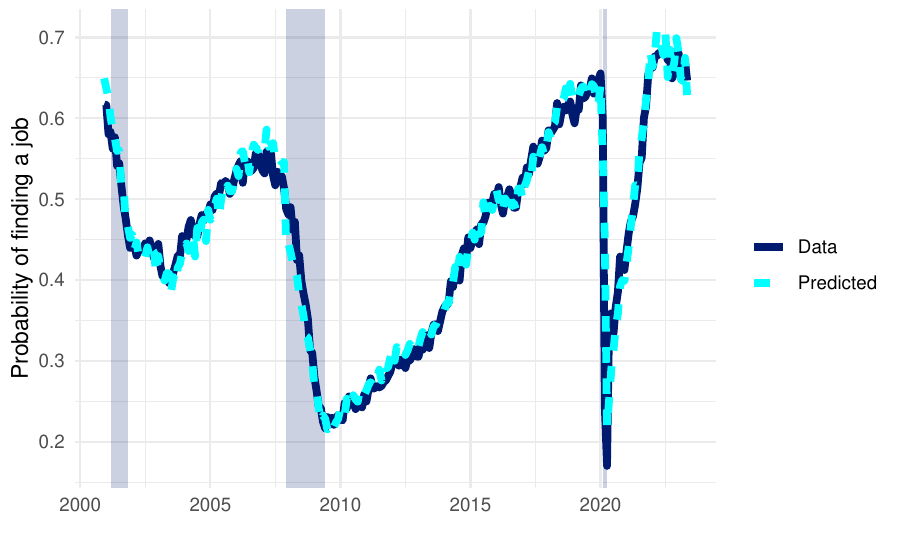}}
\caption[]{\label{fig:estimated-find} Comparison of
  the corrected probability of finding a job and the estimated probability
  of finding a job,
  December 2000 through May 2023.
  Shaded areas indicate US recessions.}
\begin{figurenotes}[Notes]
  Predictions are based on the estimated statistical model found in \eqref{eq:match-estimation}.
  The predicted level (as opposed to the log)
  uses the smear estimator in \eqref{eq:smear-estimate}.
  Predictions are conditional on shifters of matching efficiency.
\end{figurenotes}
\end{figure}

Figure \ref{fig:estimated-find}
depicts the corrected probability of finding a job between December 2000 and May 2023 in blue.
This series is labeled Data and is the same as the corrected series in figure \ref{fig:find-adj}.
It is used to construct the left side of \eqref{eq:match-estimation}.
The broken series is the prediction based on
the estimation of the statistical model in \eqref{eq:match-estimation} and
conditional on dates that shift matching efficiency.
The predicted level of finding a job uses the estimator in \eqref{eq:smear-estimate}.

\begin{figure}[htbp]
\centerline{\includegraphics[width=\textwidth]{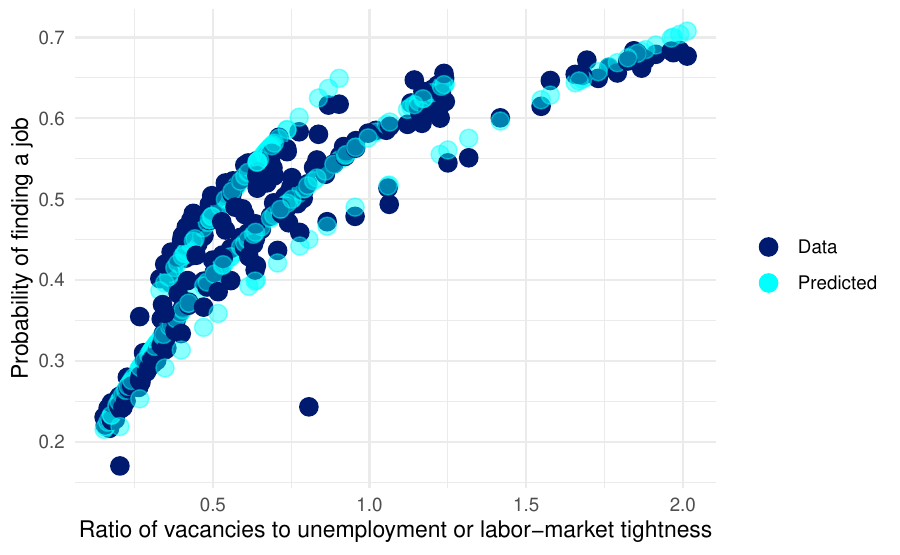}}
\caption[]{\label{fig:tight-vs-find} Tightness, $\theta$, versus the monthly probability of finding a job.}
\begin{figurenotes}[Notes]
  The ordered pairs labeled Data refer to probabilities
  corrected for worker transitions and how the JOLTS program records hires.
  The corrected monthly probabilities of finding a job are reported in figure \ref{fig:find-adj}.
  The ordered pairs labeled Predicted refer to predictions based on
  the estimated statistical model found in \eqref{eq:match-estimation}.
  The predicted level (as opposed to the log)
  uses the smear estimator in \eqref{eq:smear-estimate}.
  Predictions are conditional on shifters of matching efficiency.
\end{figurenotes}
\end{figure}

\begin{figure}[htbp]
\centerline{\includegraphics[width=\textwidth]{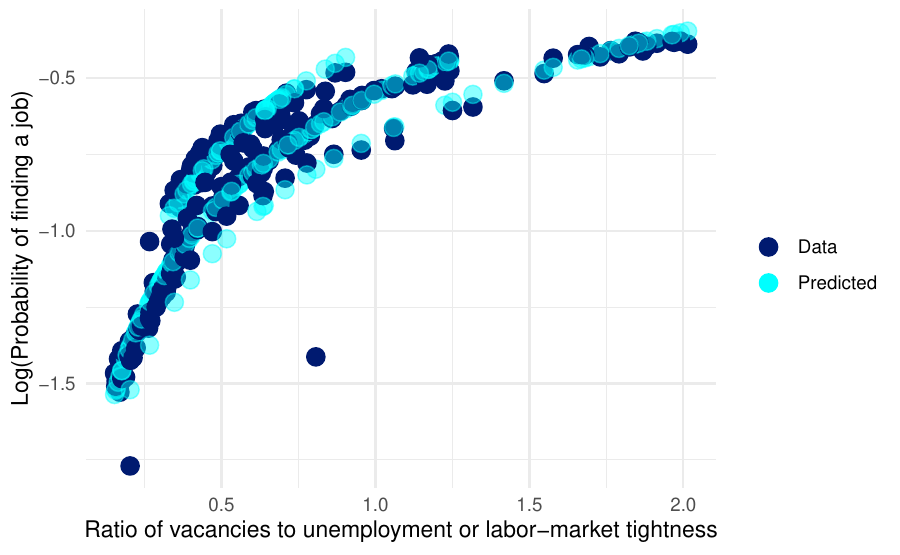}}
\caption[]{\label{fig:tight-vs-lfind} Tightness, $\theta$, versus the log of the monthly job-finding probability.}
\begin{figurenotes}[Notes]
  The ordered pairs labeled Data refer to probabilities
  corrected for worker transitions and how the JOLTS program records hires.
  The corrected monthly probabilities of finding a job are reported in figure \ref{fig:find-adj}.
  The ordered pairs labeled Predicted refer to predictions based on
  the estimated statistical model found in \eqref{eq:match-estimation}.
  Predictions are conditional on shifters of matching efficiency.
\end{figurenotes}
\end{figure}

Figures \ref{fig:tight-vs-find} and \ref{fig:tight-vs-lfind}
show tightness in the labor market
and probabilities of finding a job.
The figures compare the corrected data versus predictions
based on the estimated statistical model in \eqref{eq:match-estimation}.

\subsection{Elasticities of Matching and Bounds}

\begin{figure}[htbp]
\centerline{\includegraphics[width=\textwidth]{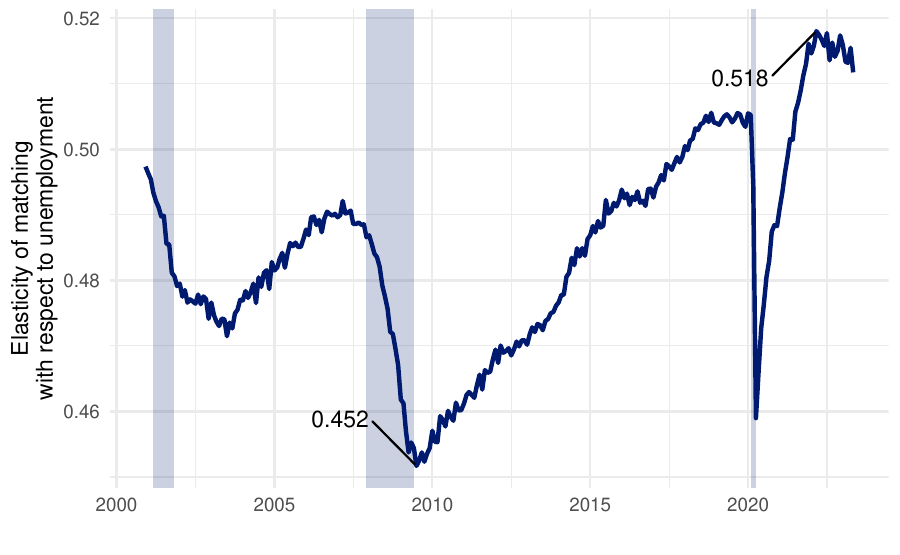}}
\caption[]{\label{fig:elasticity-matching} Elasticity of matching with respect to $u$,
  December 2000 through May 2023.}
\begin{figurenotes}[Notes]
  The elasticity of matching with respect to unemployment
  is computed using the matching function in \eqref{eq:match-fnc} and
  equals $\theta^{\gamma} / \left( 1+\theta^{\gamma} \right)$.
  The elasticity is computed
  for values of $\theta$ observed in the US economy after December 2000.
  The value of $\gamma$ is $\estgamma$,
  the estimate from the statistical model in \eqref{eq:match-estimation}.
\end{figurenotes}
\end{figure}

Figure \ref{fig:elasticity-matching} shows
the elasticity of matching with respect to unemployment
for values of $\theta$ observed in the US economy after December 2000.
The functional form is
based on the matching function in \eqref{eq:match-fnc}.
To compute the elasticities I take
the value of $\gamma$ to be $\estgamma$,
which comes from the estimated statistical model in \eqref{eq:text:match-estimation}.

A remarkable part of figure \ref{fig:elasticity-matching}
is how close the elasticities are to $0.5$.
This value would be the elasticity of
matching with respect to unemployment
if the matching function were Cobb--Douglas and
the exponent was $0.5$.

\begin{figure}[htbp]
\centerline{\includegraphics[width=\textwidth]{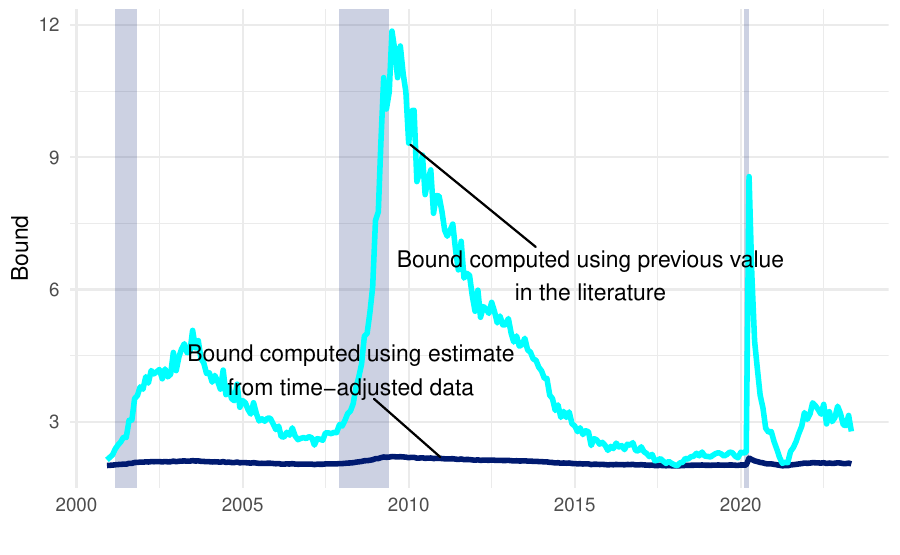}}
\caption[]{\label{fig:bound-hi} Implied bounds for $\Upsilon$
  based on the estimate of $\gamma$ compared to a value found in the literature,
  December 2000 through May 2023.}
\begin{figurenotes}[Notes]
  The bounds are computed using \eqref{eq:Upsilon-bound}.
The elasticity of matching with respect to unemployment,
  $\theta^{\gamma} / \left( 1+\theta^{\gamma} \right)$,
  is based on the matching function in \eqref{eq:match-fnc} and
  uses values of $\theta$ observed in the US economy after December 2000.  
\end{figurenotes}
\end{figure}

\begin{figure}[htbp]
\centerline{\includegraphics[width=\textwidth]{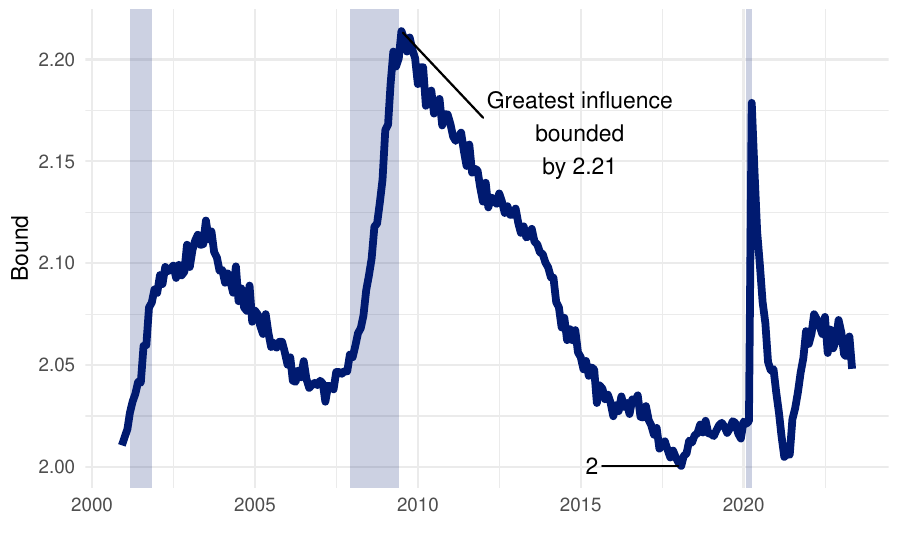}}
\caption[]{\label{fig:bound} Upper bound for $\Upsilon$ computed using $\gamma = \estgamma$ for values of tightness
  observed in the data after December 2000.}
\begin{figurenotes}[Notes]
  The choice of $\gamma$ comes from the estimated statistical model in \eqref{eq:text:match-estimation}.
  The bounds are computed using \eqref{eq:Upsilon-bound}.
\end{figurenotes}
\end{figure}

Figures \ref{fig:bound-hi} and \ref{fig:bound}
show upper bounds for $\Upsilon$,
which are given in \eqref{eq:Upsilon-bound}.
In figure \ref{fig:bound-hi}
the series in bright blue
shows the bound computed using $\gamma = 1.27$,
which is a value found in the literature.
The series in dark blue
shows the bound computed using $\gamma = \estgamma$.
The series from the literature
suggests terms in $\Upsilon$
matter more for dynamics in the labor market
in periods characterized by low tightness.

Figure \ref{fig:bound} removes the series in bright blue from figure \ref{fig:bound-hi}.
The removal facilitates inspection. 

\begin{figure}[htbp]
\centerline{\includegraphics[]{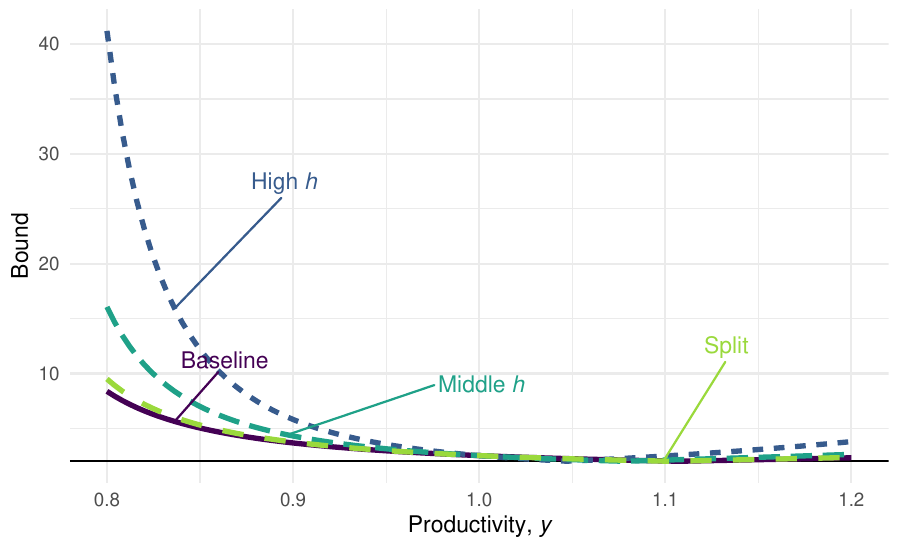}}
\caption[]{\label{fig:bound-indexed-y} Upper bound for $\Upsilon$ computed using $\gamma = 1.27$
  in economies indexed by job-creation costs.
  The upper bound is computed using \eqref{eq:Upsilon-bound}.
  Market tightness varies as productivity is perturbed around $y=1$.}
\end{figure}

Figure \ref{fig:bound-indexed-y} shows the same idea in a different way.
The upper bound for $\Upsilon$
is computed for the economies indexed by costs of job creation in table \ref{tab:model-results}.
These bounds are much different than the bounds in \ref{fig:bound}.

\end{document}

%% file: tbl_04-plot-elasticity-experiment.tex
\begin{table}

\caption{\label{tab:model-results} Model results at different combinations of job-creation costs.}
\centering
\begin{threeparttable}
\begin{tabular}[t]{lSSSSS}
\toprule
Economy & {$c$} & {$h$} & {$\ell$} & {$\eta_{\theta,y}$} & {$\eta_{w,y}$}\\
\midrule
Baseline & 0.369 & 0 & 0 & 3.602 & 0.991\\
Middle $h$ & 0.185 & 4.488 & 0 & 4.846 & 0.928\\
High $h$ & 4.111e-05 & 8.976 & 0 & 7.402 & 0.991\\
Split & 0.354 & 4.488 & 4.488 & 3.760 & 0.884\\
\bottomrule
\end{tabular}
\begin{tablenotes}[para]
\item \textit{Notes: } 
\item The column $\eta_{\theta,y}$ is the elasticity of market tightness with respect to productivity. 
                       The column $\eta_{w,y}$ is the elasticity of wages with respect to productivity.
\end{tablenotes}
\end{threeparttable}
\end{table}